\documentclass[twoside,11pt]{article}

\usepackage{blindtext}
\usepackage{booktabs} 
\usepackage{mathtools}
\usepackage{subcaption}
\usepackage[preprint]{jmlr2e}
\usepackage{graphicx}
\usepackage{microtype}
\usepackage{caption}
\usepackage{xcolor}
\usepackage{graphicx}
\usepackage{microtype}
\usepackage{mathtools,amsfonts,dsfont,amsmath,amssymb}
\usepackage{tabularx,multirow,array}
\usepackage{rotating}
\usepackage[algo2e]{algorithm2e}
\usepackage{algorithm, algpseudocode}
\usepackage{cleveref}

\DeclareMathOperator*{\argmax}{arg\,max}

\newtheorem{assumption}[theorem]{Assumption}

\definecolor{peppecolor}{rgb}{0.1, 0.4, 0.7}

\usepackage{lastpage}
\jmlrheading{26}{2026}{1-\pageref{LastPage}}{11/26; Revised ?/26}{?/26}{21-0000}{Giuseppe Martucci, Janusz M. Meylahn, G\"{u}rdal Arslan, and Serdar Y\"{u}ksel}

\ShortHeadings{Soft Decentralized Expected SARSA}{Soft Decentralized Expected SARSA}
\firstpageno{1}

\begin{document}

\title{Soft Decentralized Expected SARSA: Equilibrium Selection via Gradual Adaptation}
%Soft Decentralized Expected SARSA: Equilibrium Selection via Incremental Policy Revision Dynamics
%Convergence and Equilibrium Selection Beyond Stabilization
\author{\name Giuseppe Martucci \email giuseppe.martucci@utwente.nl \\
       \addr Department of Applied Mathematics,  \\ University of Twente, Netherlands
       \AND
       \name Janusz M. Meylahn \email j.m.meylahn@utwente.nl \\
       \addr Department of Applied Mathematics,  \\ University of Twente, Netherlands
       \AND
        \name G\"{u}rdal Arslan
       \email gurdal@hawaii.edu \\
       \addr Department of Electrical Engineering, \\ University of Hawaii at Manoa, USA
       \AND
       \name Serdar Y\"{u}ksel
       \email yuksel@queensu.ca\\
       \addr Department of Mathematics and Statistics, Queen's University
       }

\editor{-}

\maketitle

\begin{abstract}%   
%We introduce Soft Decentralized Expected SARSA (SDES), a variant of Decentralized Q-learning in which agents update policies gradually through a soft update rate rather than switching directly to newly learned best responses. We prove that SDES preserves the convergence guarantees of DQ in weakly acyclic stochastic games, converging to equilibrium policies under self-play for sufficiently large batch sizes. Beyond stability, we show that soft updating reshapes basins of attraction without changing the equilibrium set: in the deterministic limit, basin boundaries are governed by the spectrum of target Q-value differences $\Delta q_i^{\pi_{-i}}$, and smaller $\alpha$ makes trajectories track that geometry more faithfully. In the repeated Prisoner's Dilemma with memory, we derive analytical basin bounds and show that this mechanism can substantially alter the probability of learning the cooperative equilibrium. Soft updating is thus not merely a stabilization device but a mechanism of equilibrium selection in decentralized learning.
We introduce Soft Decentralized Expected SARSA (SDES), a modification of Decentralized Q-learning in which agents maintain persistent exploration and update their acting Q-values gradually rather than switching directly to newly learned best responses when updating. We prove convergence to $\epsilon$-greedy Nash equilibria in weakly acyclic stochastic games for sufficiently large batch sizes. We then show, in the repeated Prisoner's Dilemma with memory, that soft updating can reshape basins of attraction without changing the equilibrium set. In the deterministic limit, we derive analytical basin bounds governed by the Q-value differences between actions in the target policies and show that reducing the policy learning rate can favor either cooperative or competitive equilibria. These results suggest that soft updating is not merely a stabilization device but also a mechanism of equilibrium selection in decentralized learning.
\end{abstract}


\newpage

\section{Introduction}
\label{sec:introduction}

When implementing multiagent reinforcement learning (MARL) algorithms in real-world environments, two common features are that policy updates are incremental and that exploration persists throughout learning. Incremental updates are typically motivated by the stability they provide, whereas persistent exploration is motivated, in part, by the adaptability it affords in potentially changing environments. 

Incremental policy revision, or soft updating, is a widely used technique in modern reinforcement learning (RL). Rather than immediately replacing a value estimate or policy parameter with its latest learned version, a soft update interpolates between the current estimate and a newly computed target according to 
\[ 
x \leftarrow (1-\alpha)x+\alpha x_{\mathrm{target}}, 
\] 
where $\alpha\in(0,1)$ is a learning rate; the case $\alpha=1$ recovers a hard replacement. In deep RL, this device was introduced for target networks in Deep Deterministic Policy Gradient (DDPG) \citep{lillicrap2016continuous}, in contrast to the periodic hard copies of the target network used by the original Deep Q-Network (DQN) \citep{mnih2015human}, and was retained in Twin Delayed DDPG (TD3) \citep{fujimoto2018addressing} and Soft Actor-Critic (SAC) \citep{haarnoja2018soft}. Modern implementations of DQN also commonly support soft target updates as an alternative to periodic hard copies. Soft updates are typically motivated as a stability mechanism: they slow the movement of bootstrap targets and reduce training variance. Throughout this paper, \emph{soft updating} refers to the incremental interpolation above, applied between batches to each agent's acting Q-values. It is unrelated to $\epsilon$-soft exploration policies \citep{sutton2018reinforcement}, although our agents use $\epsilon$-greedy policies, and the maximum-entropy (``soft'') objective of SAC. 

Persistent exploration is likewise a recurring feature of multiagent learning algorithms. Rather than allowing exploration to vanish, an agent may retain a positive probability of selecting every action, for example, through a fixed $\epsilon$-greedy policy or a smooth best-response rule. Such persistent randomization has been studied most extensively for smooth Q-learning, where it can regularize the learning dynamics and induce convergence to quantal-response equilibria in important classes of games \citep{leslie2005individual,leonardos2021exploration}. More broadly, the exploration rate can affect both proximity to equilibrium and payoff performance in multiagent Q-learning dynamics \citep{hussain2023impact}. The literature on perturbed decentralized learning further shows that continued experimentation can determine the long-run distribution over equilibria and other recurrent outcomes \citep{marden2014achieving,meylahn2026equilibrium}. Persistent exploration is therefore more than a means of collecting samples: it can regularize the learning process, preserve responsiveness to change, and influence which outcomes are observed in the long run. Throughout this paper, \emph{persistent exploration} means that agents use $\epsilon$-greedy policies with a fixed $\epsilon>0$ during and between batches so that every action retains positive probability. 

Establishing convergence guarantees and understanding equilibrium selection are central challenges in decentralized MARL. Although substantial progress has been made on convergence, much less is known about how algorithmic design choices affect which equilibrium is ultimately reached. A major contribution in the former direction is Decentralized Q-learning (DQ) by \citet{arslan2017:1}, which converges to Nash equilibria (NE) in weakly acyclic stochastic games with arbitrarily high probability for sufficiently large batch sizes. DQ, however, uses hard policy updates: whenever an agent revises its policy, it switches immediately to a newly learned best response. 

In this paper, we introduce Soft Decentralized Expected SARSA (SDES), a generalization of DQ in which agents update their acting Q-values gradually, at a policy learning rate $\alpha$, while maintaining persistent exploration and learning within each batch using Expected SARSA. We prove that SDES converges to an $\epsilon$-greedy NE in weakly acyclic stochastic games for every fixed $\epsilon\in(0,1)$, provided that the batch size is sufficiently large. Thus, SDES retains the convergence mechanism of DQ while removing the small-exploration restriction required there. The use of Expected SARSA is central to this extension: it evaluates the exploratory policy that the agent will actually continue to use, rather than learning an unrestricted greedy best response and treating exploration as a perturbation. 

Our second main result is that soft updating can have a consequence beyond stability. Although the policy-revision graph and its equilibrium set are independent of $\alpha$, the policy learning rate can change the basins of attraction of the learning dynamics and, consequently, the probabilities with which different equilibria are learned. For the repeated Prisoner's Dilemma with one-period memory, we show that, in the deterministic soft-update limit, key conditional basins and the resulting bounds on the aggregate basins are determined by the target Q-value differences of competing best responses. We refer to the collection of these differences as the \emph{Q-value spectrum}. In this limit, the soft-update dynamics form a piecewise-linear flow whose basin geometry can be expressed in terms of the relative magnitudes of these differences. Our simulations indicate that this geometry remains predictive for small and intermediate policy learning rates. The analysis therefore suggests a broader hypothesis: in more general settings, the relative magnitudes and configuration of target Q-value differences may help determine the basins of attraction of competing equilibria. Thus, at least in the setting studied here, soft updating is not only a smoothing device but also a mechanism of equilibrium selection.

We establish these results in the repeated Prisoner's Dilemma with one-period memory. We characterize how the payoff parameters, discount factor, and persistent exploration rate determine the $\epsilon$-greedy best-response graph and identify the parameter region in which the cooperative Grim-Trigger equilibrium coexists with the All-Defect equilibrium. Within this region, we derive bounds on their basins of attraction and show how these bounds depend on the policy learning rate and the relevant target Q-value differences. In particular, reducing $\alpha$ can increase the probability of convergence to Grim Trigger when the Q-value spectrum favors that equilibrium, whereas the same reduction increases the probability of convergence to All-Defect when the spectrum favors defection. Whereas previous analyses of MARL dynamics have predominantly focused on stateless repeated games, explicit analytical basin calculations remain scarce for multistate learning dynamics. Our analysis provides such a calculation for a repeated game with endogenous state transitions, yielding explicit conditional basins and bounds on the aggregate basins of the Grim Trigger and All-Defect equilibria.

\subsection{Related Work}
\label{sec:relatedwork}

\paragraph{Provably Convergent Learning in Games.} \citet{qin2024learning} provide a systematic review of learning in normal-form and Markov games. Foundational approaches include fictitious play~\citep{brown1951iterative,robinson1951iterative}, with convergence in zero-sum and potential games~\citep{monderer1996fictitious}, and no-regret methods such as regret matching~\citep{hart2000simple}, online mirror descent~\citep{nemirovski2009robust}, and multiplicative weights for attaining coarse correlated equilibria~\citep{freund1999adaptive}. 

Since the foundational work of~\cite{littman1994markov}, Markov games have become the standard framework for MARL. In zero-sum games, minimax Q-learning~\citep{littman1994markov} converges asymptotically~\citep{szepesvari1999unified}, but requires observed actions, knowledge of the game structure, and the solution of a matrix game at each step. Convergence was subsequently established beyond zero-sum games for several Q-learning variants under coordination assumptions~\citep{littman2001friend,hu2003nash,greenwald2003correlated}, and for a two-timescale risk-sensitive Q-learning algorithm converging to a generalized NE~\citep{borkar2002q}. Independent policy-gradient methods converge at a polynomial rate in Markov potential games~\citep{leonardos2021global}, while continuous-time best-response and discrete-time fictitious-play dynamics converge in zero-sum Markov games~\citep{sayin2020fictitious}; the latter use two timescales to address non-stationarity but require opponents' strategies or actions. By contrast, radically uncoupled learning can achieve convergence and rationality~\citep{sayin2021decentralized}, in line with the information settings of~\cite{foster2006regret,leslie2005individual}. 

More recently, fully decentralized bandit learning was shown to converge almost surely in continuous-action general-sum Markov games under strong monotonicity, using only the public state and each agent's realized rewards~\citep{hofer2026decentralized}. Recent work also extends potential-game theory: \citet{arefizadeh2024characterizations} characterize potential, ordinal-potential, and generalized ordinal-potential games; \citet{guo2025markov} introduce Markov $\alpha$-potential games and algorithms for approximate equilibria; and \citet{guo2025towards} characterize dynamic potential games through value-function decompositions and symmetry of functional derivatives. For linear-quadratic games, \citet{hosseinirad2026linear} show that potential structure is restrictive under full-state feedback but richer with decoupled dynamics and information. Other scalable directions include mean-field learning~\citep{perrin2020fictitious,elie2020convergence} and decentralized MARL~\citep{zhang2021multi}. Overall, this literature studies equilibrium learning under realistic information and interaction constraints.

Within this landscape, the present paper belongs to the class of \emph{graph-theoretic policy-revision} methods: convergence properties are read off a graph on the finite policy space whose edges are induced by the revision dynamics, with weak acyclicity as the key structural condition; see \citet{yongacoglu2024generalizing} for a recent generalization of better-response paths and weakly acyclic games in this framework. Within this class, \citet{altabaa2023decentralized} model the idealized policy updates as an absorbing Markov chain on the joint policy space and obtain closed-form convergence probabilities to each equilibrium as a function of the revision rule and the best-reply graph alone. This is the framework underlying DQ~\citep{arslan2017:1}. Building on~\cite{arslan2017:1}, subsequent work has studied decentralized learning under local control and global state information in stochastic teams and games~\citep{yongacoglu2022decentralized}, introduced satisficing paths to analyze independent MARL dynamics~\citep{yongacoglu2023satisficing}, and provided a two-timescale analysis of unsynchronized DQ under asynchronous updates~\citep{yongacoglu2025unsynchronized}. DQ has also been applied to urgency-constrained workload offloading in edge computing~\citep{fox2025multiagentreinforcementlearningtask}. SDES operates within this framework: its convergence guarantee (Theorem~\ref{thm:convergence}) is a statement about paths on the $\epsilon$-BR graph on the policy space, while the basin analysis of Section~\ref{sec:theory} adds a quantitative selection layer, which path is taken, and with what probability, on top of the same graph.

\paragraph{Dynamical Systems.} From the perspective of dynamical systems, MARL can be interpreted as a high-dimensional, non-equilibrium process in which agents continuously update their policies in response to environmental feedback and each other's evolving behavior, seeking to explain how macroscopic patterns such as cooperation emerge from local interactions among heterogeneous agents~\cite{barfuss2023collective, helbing2012social}. A first family of models, grounded in evolutionary game theory~\cite{nowak2006five}, nonlinear dynamics~\cite{strogatz2018nonlinear}, and network theory~\cite{jackson2008social}, makes minimal behavioral assumptions and offers tractable insight into when cooperative dynamics arise and stabilize~\cite{perc2017statistical}, but abstracts away from the algorithmic complexity and cognitive capacities of real learning agents. MARL closes this gap by instantiating agents with learning algorithms capable of temporal credit assignment, policy optimization, and strategic adaptation in complex, often partially observed environments~\cite{rahwan2019machine}, and a growing body of work integrates this algorithmic perspective with mechanistic insight from dynamical systems~\cite{baker2017rational, kleiman2022value}.

\paragraph{Deterministic Learning Dynamics.} \cite{barfuss2019deterministic} derive deterministic continuous-time limits of temporal-difference learning in stochastic games, covering Q-learning, SARSA, and actor--critic methods and enabling dynamical-systems analysis of learning attractors~\citep{barfuss2019deterministic}. This framework has been linked to a minimum free-energy principle and approximated by an online sample-batch TD algorithm~\citep{barfuss2022dynamical}, extended to infinite-horizon memory~\citep{barfuss2020infinite}, and generalized to partial observability and perceptual uncertainty~\citep{barfuss2022partial}. In the iterated Prisoner's Dilemma with one-period memory, also studied here, finite-batch fluctuations substantially promote cooperation, approximately doubling the frequency of cooperative outcomes relative to the deterministic limit~\citep{barfuss2023intrinsic}. SDES belongs to this class of sample-batch TD algorithms but replaces the hard best response at each batch boundary by moving the acting Q-values toward the valuation Q-values at rate~$\alpha$. Thus, $\alpha$ can reshape the deterministic basins of attraction independently of finite-batch fluctuations~\citep{barfuss2023intrinsic}, while SDES additionally provides almost-sure convergence guarantees.

\paragraph{Algorithmic Collusion.} Tacit collusion among autonomous learning agents is a relatively recent but serious concern in decentralized MARL. Simulations show that independent Q-learners may converge to collusive outcomes in repeated pricing games \citep{Calvano2020AER,Calvano2020}, including sequential pricing environments \citep{Klein2021RAND}. More generally, \citet{askenazi2024bounds} establish \emph{local} convergence of Q-learning and gradient-based dynamics to strict finite-recall equilibria in repeated games, yielding a learning-based analogue of the Folk Theorem that delimits the range of potentially collusive outcomes; notably, they identify the size and geometry of the basins of attraction of these equilibria, the natural equilibrium-selection mechanism, as an open problem. In the same local spirit, \citet{possnig2026monitoring} studies the stability of learned outcomes through a limiting differential equation and derives conditions under which monitoring and market primitives steer learning toward competitive or collusive equilibria in repeated Bertrand competition. The generality of these findings is still debated: \citet{denBoer2022artificial} argue that collusion in \citet{Calvano2020AER} requires implausibly long convergence times, and \citet{den2024mathematical} argue that what constitutes algorithmic collusion needs to be properly defined, so that algorithmic collusion remains an open question regarding generality, empirical relevance, and theoretical foundations. Methods for quantifying collusion show how algorithmic parameters affect its likelihood \citep{meylahn2022limiting,meylahn2025quantifying,meylahn2026intermediate}. The analysis of the present paper is global rather than local: for the repeated Prisoner's Dilemma we compute (bounds on the) basins of attraction in closed form and identify the Q-value spectrum as a factor that governs their size, extending this quantification program, which measures the likelihood of learning collusive strategy equilibria under hard policy updates, to soft policy updates.

\subsection{Contributions}
\label{sec:contributions}
The main contributions of this paper are as follows.
\begin{enumerate}
    \item In Section~\ref{sec:setting}, we introduce Soft Decentralized Expected SARSA (SDES; Algorithm~\ref{alg:SDES}), a decentralized learning rule that combines two modifications of DQ: (i)~a soft update of the acting Q-values at batch boundaries, with policy learning rate $\alpha\in(0,1)$, and (ii)~Expected SARSA within each batch. 
    %Soft updating is defined as the incremental interpolation $q_{\mathrm{act}}\leftarrow(1-\alpha)q_{\mathrm{act}}+\alpha q_{\mathrm{val}}$ and is distinguished from $\epsilon$-soft exploration and from maximum-entropy RL. 
    We also formalize the statewise $\epsilon$-greedy framework in which the algorithm operates: the $\epsilon$-greedy policy spaces (Definition~\ref{def:egreedy}) and the statewise notions of $\epsilon$-BR and $\epsilon$-NE (Definitions~\ref{def:BR} and~\ref{def:NE}).
    \item In Section~\ref{sec:theory}, Theorem~\ref{thm:convergence} establishes that SDES converges to an $\epsilon$-NE in weakly acyclic stochastic games for \emph{all} $\epsilon\in(0,1)$, provided the batch size is sufficiently large. Soft updates realize each unilateral $\epsilon$-BR step over a finite block of batches rather than in one hard jump, preserving the graph-theoretic structure underlying DQ; the use of Expected SARSA removes the small-$\epsilon$ condition required by DQ.
    \item In the same section, for the iterated Prisoner's Dilemma with a two-state abstraction, Proposition~\ref{thm:IBR} characterizes the possible $\epsilon$-IBR graphs: their topology depends on the discount factor $\gamma$ and the exploration rate $\epsilon$ through explicit critical thresholds $\gamma_{c_1}$ and $\gamma_{c_2}$, which partition the parameter space into a unique-defection regime, a coexistence regime, and a unique-cooperation regime. The learning parameters $\alpha$ and $\lambda$ affect equilibrium selection only in the coexistence regime.
    \item Still in Section~\ref{sec:theory}, we provide a geometric analysis of equilibrium selection that goes beyond path existence on the best-response graph. In the deterministic soft-update limit, basin boundaries are determined by the target Q-value differences $\Delta q_i^{\pi_{-i}}$; we derive closed-form conditional basins $\hat{B}$, $\hat{C}$, $\hat{D}$ and aggregate bounds on the basin of attraction of the cooperative equilibrium (Theorem~\ref{thm:BoA}). Soft updating amplifies the equilibrium favored by this Q-value geometry; it does not inherently favor cooperation.
    \item In Section~\ref{sec:simulations}, simulations of the soft inertial dynamics across a range of $(\alpha,\lambda,\gamma,\epsilon)$ validate the theoretical basin bounds for small $\alpha$ and show that the Q-difference geometry remains predictive at intermediate learning rates, while the inertia $\lambda$ affects the basins mainly at intermediate $\alpha$ and not at the extremes.
\end{enumerate}

\section{Setting}
\label{sec:setting}
In this section we introduce the notation and definitions we will need (Section~\ref{sec:notationanddef}), the environment in which we will illustrate the effect of algorithmic hyperparameters on the probability of learning different NE (Section~\ref{sec:environment}) and our novel algorithm (Section~\ref{sec:algorithm}).

\subsection{Notation and Definitions}
\label{sec:notationanddef}

\begin{definition}[Stochastic Games]
\label{def:MG}
A \emph{Stochastic game} for $N$ players is defined by the tuple $(\mathcal{S}, \{\mathcal{A}_i\}_{i=1}^{N}, p, \{r_i\}_{i=1}^{N}, \gamma, \mu)$. Here, $\mathcal{S}$ is a finite set of states, $\mathcal{A}_i$ is the finite set of actions available to player $i$, $p: \mathcal{S} \times \mathcal{A}_1 \times \cdots \times \mathcal{A}_N \rightarrow \Delta(\mathcal{S})$ is the state transition probability function, $r_i: \mathcal{S} \times \mathcal{A}_{1} \times \cdots \times \mathcal{A}_{N} \rightarrow \mathbb{R}$ is the reward function for player $i$, $\gamma \in [0,1)$ is a discount factor shared by all players, and $\mu \in \Delta(\mathcal{S})$ is the initial state distribution. At each time step $t$, all players simultaneously select actions $\boldsymbol{a}_t = (a_{1,t}, \ldots, a_{N,t})$, the environment transitions to the next state $s_{t+1} \sim p(\cdot \mid s_t, \boldsymbol{a}_t)$, and each player receives a reward $r_i(s_t, \boldsymbol{a}_t)$.
\end{definition}

Our convergence result will hold under the following assumption on the environment which is the same as Assumption 1 in \cite{arslan2017:1}.

\begin{assumption}
\label{ass:reachability}
    For all $(s, s')$, there exists a finite integer $H\geq 0$ and a sequence of action profiles $(\tilde{\mathbf{a}}_{0}, \ldots, \tilde{\mathbf{a}}_{H})$ so that 
    \begin{equation}
        p(s_{H+1} = s'| (s_{0}, \mathbf{a}_{0}, \ldots, \mathbf{a}_{H}) = (s, \tilde{\mathbf{a}}_{0}, \ldots, \tilde{\mathbf{a}}_{H})) > 0.
    \end{equation}
\end{assumption}

We will assume that $|r_{i}|<R_{\text{max}}$ for all $i\in\{1, 2, \ldots, N\}$. In general, players can employ policies $\pi_{i}:\mathcal{S}\rightarrow\Delta(\mathcal{A}_{i})$. However, the algorithm we will analyze makes use of the $\epsilon$-greedy action-selection mechanism with a fixed exploration rate $\epsilon\in(0,1)$, which restricts the policies available to player $i$ to the set of $\epsilon$-greedy policies.

\begin{definition}[$\epsilon$-greedy Policies]
\label{def:egreedy}
Fix $\epsilon\in(0,1)$. Given a \emph{base policy} $g_{i}:\mathcal{S}\rightarrow\mathcal{A}_{i}$, i.e., a deterministic stationary policy, the \emph{$\epsilon$-greedy policy induced by $g_{i}$} is
\begin{equation}
\label{eq:egreedybase}
\pi_{i}^{g_{i}}(a\mid s) :=
\begin{cases}
1-\epsilon+\dfrac{\epsilon}{|\mathcal{A}_i|}, & \text{if } a = g_{i}(s),\\[0.8em]
\dfrac{\epsilon}{|\mathcal{A}_i|}, & \text{else.}
\end{cases}
\end{equation}
The set of $\epsilon$-greedy policies available to player $i$ is
\begin{equation}
\label{eq:Piepsilon}
\Pi_{i}^{\epsilon} := \left\{\pi_{i}^{g_{i}} : g_{i}:\mathcal{S}\rightarrow\mathcal{A}_{i}\right\},
\end{equation}
and the joint $\epsilon$-greedy policy space is $\Pi^{\epsilon} := \Pi_{1}^{\epsilon}\times \cdots\times\Pi_{N}^{\epsilon}$.
\end{definition}

Note that $\Pi_{i}^{\epsilon}$ is a finite set with $|\Pi_{i}^{\epsilon}| = |\mathcal{A}_{i}|^{|\mathcal{S}|}$ elements, as each $\epsilon$-greedy policy is identified by its base policy. We denote a policy profile by $\boldsymbol{\pi}\in\Pi^{\epsilon}$.

Given a joint policy profile $\boldsymbol{\pi}\in\Pi^\epsilon$, the statewise expected discounted return of player $i$ is 
\begin{equation} 
\label{eq:br_discret} 
V_i^{\boldsymbol{\pi}}(s) := \mathbb{E}_{\boldsymbol{\pi}} \left[ \left. \sum_{k=0}^{\infty} \gamma^k r_i(s_k,\boldsymbol{a}_k) \,\right|\, s_0=s \right], \quad \text{and}\quad s\in\mathcal{S}, 
\end{equation} 
where $\mathbb{E}_{\boldsymbol{\pi}}$ denotes expectation with respect to the distribution over trajectories induced by $\boldsymbol{\pi}$ and the transition function $p$. When an aggregate performance measure under the initial-state distribution $\mu$ is required, we write \[ J_i^\mu(\boldsymbol{\pi}) := \mathbb{E}_{s_0\sim\mu} \left[ V_i^{\boldsymbol{\pi}}(s_0) \right]. \] The distribution $\mu$ is used only for aggregate evaluation and does not enter the definitions of best response or NE below. In RL, statewise optimality is characterized through the Q-values associated with each state-action pair. Each agent $i$ keeps track of Q-values $q_i(s,a)$, where $s\in\mathcal{S}$ and $a\in\mathcal{A}_i$. Our restriction to the $\epsilon$-greedy policy space $\Pi^\epsilon$ means that the policy employed by player $i$, given $q_i$, is the $\epsilon$-greedy policy induced by the greedy base policy of $q_{i}$:
\begin{equation}
\label{eq:egreedypolicy}
\pi_{i}^{q} := \pi_{i}^{g_{i}^{q}}
\qquad\text{with}\qquad
g_{i}^{q}(s) \in \argmax_{a'\in\mathcal{A}_{i}} q_i(s,a'),
\end{equation}
where ties in the argmax are broken according to a fixed ordering of $\mathcal{A}_{i}$ (any deterministic tie-breaking rule will do). Explicitly, $\pi_{i}^{q}(a\mid s) = 1-\epsilon+\epsilon/|\mathcal{A}_{i}|$ if $a = g_{i}^{q}(s)$ and $\epsilon/|\mathcal{A}_{i}|$ otherwise. The tie-breaking convention guarantees that every Q-table induces exactly one element of $\Pi_{i}^{\epsilon}$. Since tied actions have equal Q-values, the convention does not affect any expected value computed under $\pi_{i}^{q}$; it only fixes which of the tied $\epsilon$-greedy policies is selected.

\begin{definition}[Optimal $\epsilon$-greedy Q-values]
The optimal $\epsilon$-greedy Q-values for player $i$ given that the opponents employ policies $\pi_{-i}$ are given by the fixed point of 
\begin{align*}
    q_{i}(s, a_{i}) = \mathbb{E}_{\pi_{-i}}\Big[r_{i}(s, \boldsymbol{a})+\gamma \sum_{s'\in\mathcal{S}}p(s'|s, \boldsymbol{a})\sum_{a'\in\mathcal{A}_{i}}\pi_{i}^{q}(a'|s')q_{i}(s', a')\Big],
\end{align*}
which we denote by $q_{i}^{\pi_{-i}}$.
\end{definition}

\begin{definition}[$\epsilon$-greedy Best Response]
\label{def:BR}
Given a set of $\epsilon$-greedy policies for all agents other than agent \(i\), denoted by  
$\pi_{-i} = (\pi_1, \dots, \pi_{i-1}, \pi_{i+1}, \dots, \pi_N)$, the set of $\epsilon$-greedy best responses of player $i$ is \begin{equation} \label{eq:epsilon_BR} \operatorname{BR}_i^\epsilon(\pi_{-i}) := \left\{ \pi_i\in\Pi_i^\epsilon: V_i^{(\pi_i,\pi_{-i})}(s) \geq V_i^{(\widetilde{\pi}_i,\pi_{-i})}(s) \quad \text{for every } s\in\mathcal{S} \text{ and } \widetilde{\pi}_i\in\Pi_i^\epsilon \right\}. \end{equation} Any policy  $\pi_i^\star\in\operatorname{BR}_i^\epsilon(\pi_{-i})$  is called an $\epsilon$-greedy best response, or an $\epsilon$-BR, to $\pi_{-i}$. \end{definition} 

Thus, an $\epsilon$-BR is optimal within the restricted policy space $\Pi_i^\epsilon$ from every initial state. This requirement is stronger than maximizing the expected return averaged under $\mu$ when $\mu$ does not assign positive probability to every state. An $\epsilon$-BR may differ from an unrestricted greedy best response.

The set $\operatorname{BR}_i^\epsilon(\pi_{-i})$ is never empty: the $\epsilon$-greedy policy $\pi_{i}^{q}$ induced via \eqref{eq:egreedypolicy} by the optimal $\epsilon$-greedy Q-values $q_{i}^{\pi_{-i}}$ belongs to $\operatorname{BR}_i^\epsilon(\pi_{-i})$. This is the $\epsilon$-greedy analogue of the classical fact that a greedy policy with respect to the optimal Q-values is optimal, established for restricted persistent-exploration policy classes in \cite[Theorem~3]{singh2000convergence} (see also Lemma~\ref{lemma1adapt}). When $q_{i}^{\pi_{-i}}$ has tied maximizers in some state, every $\epsilon$-greedy policy whose base action in that state is one of the tied maximizers is an $\epsilon$-BR; the tie-breaking convention in \eqref{eq:egreedypolicy} merely selects one of them.

\begin{definition}[$\epsilon$-greedy Best-Response Graph]
\label{def:brgraph}
An \emph{$\epsilon$-greedy Best Response Graph} ($\epsilon$-BR graph) is a directed graph with nodes representing all possible strategy profiles, i.e.,  $V = \Pi^{\epsilon}$, and directed edge set 
\begin{align*}
    E=\left\{(\boldsymbol{\pi}, \tilde{\boldsymbol{\pi}}) \in \Pi^{\epsilon}\times\Pi^{\epsilon} : \exists\, i \text{ such that } \tilde{\pi}_{i}\in \operatorname{BR}_{i}^{\epsilon}(\pi_{-i}),\;
    \pi_{i}\notin \operatorname{BR}_{i}^{\epsilon}(\pi_{-i}),\;
    \text{and } \tilde{\pi}_{-i}=\pi_{-i}\right\}.
\end{align*}
\end{definition}
Intuitively, the $\epsilon$-BR graph contains edges from every $\epsilon$-greedy strategy profile to the strategy profiles that can be reached by allowing exactly one player, who is not already playing an $\epsilon$-BR, to switch to some element of its $\epsilon$-BR set, while all other players keep their policies unchanged. A natural learning goal in multiagent systems is that of an NE given for the restriction to $\epsilon$-greedy strategies in Definition~\ref{def:NE}. $\epsilon$-greedy NE appear as sinks in the $\epsilon$-BR graph. 

\begin{definition}[$\epsilon$-greedy Nash equilibrium]
\label{def:NE}
    A strategy profile $\boldsymbol{\pi}^{*}\in\Pi^{\epsilon}$ is an $\epsilon$-greedy Nash equilibrium ($\epsilon$-NE) if
    \begin{align*}
        \forall i\in \{1, 2, \ldots, N\}\; \pi_{i}^{*} \in \operatorname{BR}_{i}^{\epsilon}(\pi_{-i}^{*}),
    \end{align*}
    or, equivalently, if for every player $i$, every state $s\in\mathcal{S}$, and every $\widetilde{\pi}_{i}\in\Pi_{i}^{\epsilon}$,
    \begin{align*}
        V_i^{\boldsymbol{\pi}^{*}}(s) \geq V_i^{(\widetilde{\pi}_i,\pi_{-i}^{*})}(s).
    \end{align*}
\end{definition}

We denote the set of $\epsilon$-greedy NE for a game by $\Pi^{\epsilon}_{\text{eq}}$. Since the best-response correspondence of Definition~\ref{def:BR} is statewise, the $\epsilon$-NE notion is statewise as well. Throughout the paper, best responses and NE are always understood in this restricted $\epsilon$-greedy sense, and we abbreviate them as $\epsilon$-BR and $\epsilon$-NE. Our algorithm will converge to an $\epsilon$-NE in games in which the $\epsilon$-BR graph is weakly acyclic. 

\begin{definition}[$\epsilon$-greedy Weakly Acyclic Games]
\label{def:WAG}
A game is called \emph{$\epsilon$-greedy weakly \\ acyclic} iff, in its $\epsilon$-BR graph, there exists a finite directed path from every node to a sink, i.e., an $\epsilon$-NE. 
\end{definition}

The example we will use to illustrate the effect of changes to the algorithm's hyperparameters is a symmetric, two-player game in which every $\epsilon$-BR set is a singleton, so that the $\epsilon$-BR correspondence reduces to a function. When restricting to this class of games, the structure of the best-response relations is captured more succinctly by the Individual Best-Response Graph \citep{meylahn2025quantifying}.

\begin{definition}[$\epsilon$-greedy Individual Best-Response ($\epsilon$-IBR) Graph]
\label{def:ibrgraph}
The \emph{$\epsilon$-IBR graph} for player $i$ is a directed graph with nodes $V=\Pi_{i}^{\epsilon}$ and edge set
\begin{align*}
    E = \{(\pi_{i}, \tilde{\pi}_{i}): \tilde{\pi}_{i}\in\operatorname{BR}_{i}^{\epsilon}(\pi_{i})\}.
\end{align*}
\end{definition}
Due to the restriction to symmetric games, the $\epsilon$-IBR graph is the same for both players. Such games are weakly acyclic if and only if their $\epsilon$-IBR graphs contain no cycles of length three or more \citep[Theorem III.2]{meylahn2025quantifying}. This follows as self-loops and two-cycles correspond to symmetric and non-symmetric $\epsilon$-NE, respectively, while cycles of length three or more lead to inescapable cycles in the $\epsilon$-BR graph. The $\epsilon$-IBR graph with all cycles removed consists of trees, which cannot lead to cycles in the $\epsilon$-BR graph. 

If the players would have access to the current policies of their opponents as well as their $\epsilon$-greedy Best Response function, they could execute the following idealized dynamics. 

\begin{definition}[Inertial $\epsilon$-greedy Best-Response Dynamics (I$\epsilon$BRD)]
\label{def:IBRD}
    At time $t=0$ each player $i$ initializes their policy uniformly from $\Pi_{i}^{\epsilon}$. At time $t>0$, each player that is not already playing an $\epsilon$-BR to the other players' policies
    \begin{enumerate}
        \item plays an $\epsilon$-BR to the other players' policies at time $t-1$ with probability $1-\lambda$,
        \item plays their policy at time $t-1$ with probability $\lambda$,
    \end{enumerate}
    with $\lambda\in(0, 1)$.
\end{definition}

The inertial best response dynamics (introduced in \cite{stratlearning}) converge to an NE in weakly acyclic games, and the convergence of the DQ algorithm relies on showing that the algorithm mimics these dynamics arbitrarily closely \citep{arslan2017:1}.  

In addition to establishing whether the game is weakly acyclic or not, the IBR graphs also characterize the probability with which the I$\epsilon$BRD reach NE. The probability of reaching an NE is given by the fraction of nodes in the weakly connected component of the graph containing the equilibrium \citep[Theorem III.8]{meylahn2025quantifying}. They will serve as a benchmark against which we can compare the dynamics of our generalized algorithm.

Our algorithm is designed to mimic a generalization of the I$\epsilon$BRD using the optimal Q-values associated with the $\epsilon$-BR given in Definition~\ref{def:softIBRD}.  
\begin{definition}[Soft Inertial $\epsilon$-greedy Best-Response Dynamics (SI$\epsilon$BRD)]
\label{def:softIBRD}
At time $t=0$ all players initialize their Q-values $q_{i}^{0}$ uniformly from $[-\frac{R_{\text{max}}}{1-\gamma}, \frac{R_{\text{max}}}{1-\gamma}]$ and set their policy to be the corresponding $\epsilon$-greedy policy. At time $t>0$ each player 
\begin{enumerate}
    \item updates their Q-values as 
    $$
    q_{i}^{t} =  (1-\alpha)q_{i}^{t-1} + \alpha q_{i}^{\pi_{-i}^{t-1}}
    $$
    with probability $1-\lambda$,
    \item leaves their Q-values unchanged with probability $\lambda$.
\end{enumerate}
\end{definition}

To show how our algorithm differs from DQ, we will analyze the probability with which different NE are learned. For this purpose, we define the Stochastic Basin of Attraction in Definition~\ref{def:BoA}\footnote{Note that this differs from the usual definition of a basin of attraction, as the dynamics we are considering are stochastic due to the inertia parameter. Nevertheless, we use this terminology, as the characterization of this limiting probability strongly resembles the usual use of ``basin of attraction''.}.

\begin{definition}[Stochastic Basin of Attraction]
\label{def:BoA}
    Assume that dynamics $D$ (with a specified initial distribution) are guaranteed to converge to an $\epsilon$-NE. Let $\mathbb{P}_{t}^{D}(\pi^{*})$ denote the probability of observing a process following dynamics $D$ in strategy profile $\boldsymbol{\pi^{*}}\in\Pi^{\epsilon}_{\text{eq}}$ at time $t$. The stochastic basin of attraction of $\boldsymbol{\pi^{*}}$ is given by $$\text{BoA}(\boldsymbol{\pi^{*}}):=\lim_{t\rightarrow\infty}\mathbb{P}_{t}^{D}(\boldsymbol{\pi^{*}}).$$
\end{definition}

We will also make use of conditional stochastic basins of attractions $\text{BoA}(\boldsymbol{\pi^{*}}|\boldsymbol{\pi})$, which are defined as in Definition~\ref{def:BoA} with the addition that the dynamics are conditioned on starting in $\boldsymbol{\pi}$.

\subsection{Illustrative Environment}
\label{sec:environment}

To show how the algorithmic hyperparameters can influence the probability of converging to different NE, we require a simple environment in which multiple NE are possible. The iterated Prisoner’s Dilemma (IPD) with a particular choice of state space is ideal for this purpose. In the IPD, the players have two possible actions $\mathcal{A}=\{C, D\}$ (`cooperating' and `defecting'), and the stage-game payoff matrix is
\[
\begin{array}{c|cc}
 & C & D \\
\hline
C & R,R & S,T \\
D & T,S & P,P \\
\end{array},
\]
where we assume $T>R>P>S\geq 0$. To reduce the dimensionality of the parameter space, and since we are predominantly interested in the effect of algorithmic hyperparameters on the probability of reaching different NE, we will set $T = 1$ and $S = 0$ for most of the analysis. 

Recent work has studied this game when the players make use of RL and a state space consisting of all possible one-period histories, i.e., $$ \mathcal{S}=\{(D, D), (D, C), (C, D), (C, C)\}. $$ 
Here, the first entry of a state tuple represents the action taken by the first player in the previous round, and the second entry represents the action taken by the second player. This enhances the strategy space available to the players to include strategies such as Tit-for-Tat, which repeats the opponent's action in the previous round.  In the full state space, $\mathcal{S}$, there exist three NE, all of which are symmetric: Win–Stay, Lose–Shift (WSLS), Grim Trigger (GT), and All-Defect (AD) \citep{usui, meylahn2022limiting}. Note that GT refers to a one-period memory version of the usual Grim Trigger strategy. That is, the player cooperates when both players cooperated in the previous round and defects otherwise. When the strategies are implemented in an $\epsilon$-greedy fashion, this means that it is possible to re-establish cooperation after defection when both players play GT. 

\begin{figure}[H]
\centering
\includegraphics[width=0.7\textwidth]{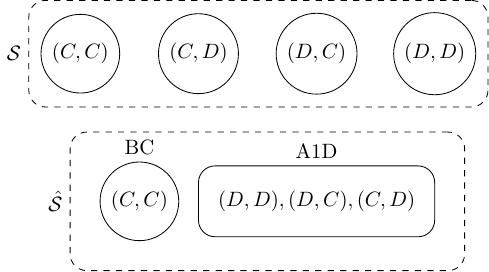}
\caption{Illustration of the state spaces $\mathcal{S}$ and $\hat{\mathcal{S}}$.}
\label{fig:state_abstraction}
\end{figure}

This choice of state space leads to a total of 256 possible strategy profiles with three NE. Our analysis will not be feasible with that many strategy profiles, and since we only require two NE to study the effect of hyperparameters on the probability of reaching different NE, we will consider a lower-dimensional state space. We can achieve this by partitioning the full state space, and we will refer to such a partitioning as a state space abstraction. There are multiple options for reducing the state space dimensionality; however, we require an abstraction that preserves two NE. Since the AD strategy is also an NE in the stateless game, this will always be preserved. We thus have a choice of preserving WSLS or GT. We choose the latter. The resulting two-dimensional state space abstraction is $$ \hat{\mathcal{S}}=\{\{(C, C)\}, \{(D, D), (D, C), (C, D)\}\}. $$
For ease of reference we will refer to the state $\{(C, C)\}$ as BC (for `Both Cooperated') and the state $\{(D, D), (D, C), (C, D)\}$ as A1D (for `At Least One Defected'). The diagram in Figure~\ref{fig:state_abstraction} visualizes how the full state space is abstracted into two states.

Studying the effect of different state space abstractions in games is of independent interest (see \cite{Li_State}). Intuitively, in choosing a state space abstraction in a multiagent environment, the agent wants to balance the expressivity of the policy space with the speed of learning. Smaller state spaces reduce the dimension of the strategy space but facilitate faster learning.

\subsection{Algorithm}
\label{sec:algorithm}
To reduce the notational clutter in the analysis, we define Algorithm~\ref{alg:SDES} for a fixed batch size $K$. This is in contrast to DQ, which is defined for batch size sequences $\{K_{k}\}_{k=1}^{\infty}$ indexed by $k$. Our analysis can be extended to this case straightforwardly. In addition, we omit the subscript for the identity of the agent.

\begin{algorithm}[ht]
\caption{Soft Decentralized Expected SARSA}\label{alg:SDES}
Given learning rate $\alpha$, exploration rate $\epsilon$ and inertia $\lambda$, each in $(0,1)$, discount factor $\gamma \in [0,1)$,\\ batch size $K.$\\
\Begin{
    Initialize $q_{\text{act}}(s, a)=q_{\text{val}}(s, a)$ randomly in $[-\frac{R_{\text{max}}}{1-\gamma}, \frac{R_{\text{max}}}{1-\gamma}]$ for all $(s, a)$, \\
    initialize the visit counter $\nu(s,a) \leftarrow 0$ for all $(s,a)$,\\
    set $\pi_{\text{act}}$ as $\epsilon$-greedy policy from $q_{\text{act}}$,

    set $\pi_{\text{val}}$ as $\epsilon$-greedy policy from $q_{\text{val}}$,
    
    observe current state $s$. 
    
    \Repeat{done}{
        % Per-batch counter reset: the within-batch step-size schedule restarts
        % at every batch, matching the per-phase counters of Arslan-Yuksel and
        % the uniform-over-batches concentration horizon used in the proof.
        %Reset the visit counter $\nu(s,a) \leftarrow 0$ for all $(s,a)$;\\
        \For{$t=1$ \textbf{to} $K$ }{
        
        Execute action $a \sim $ $\pi_{\text{act}}(\cdot|s)$;
        
        observe reward $r$ and next state $s'$;
        
        update $\nu(s,a) \leftarrow \nu(s,a) + 1$; 
        % <- nu(s,a) counts the visits to (s,a) within the current batch
        
        set $\tilde{\alpha} \leftarrow \frac{1}{(\nu(s,a)+1)}$;
        
        set $q_\text{val}(s, a) \leftarrow (1-\tilde\alpha)q_{\text{val}}(s, a) + \tilde \alpha \Big[r +\gamma \sum_b \pi_{\text{val}}(b|s') q_\text{val}(s', b) \Big]$;

        set $\pi_{\text{val}}(\cdot|s)$ as the $\epsilon$-greedy policy from $q_{\text{val}}(s, \cdot)$;
        
        set $s\leftarrow s'$.
        }
        Generate random variable $u$ from $U \sim [0, 1]$;

        \If{$u > \lambda$}{
            % Update q_act only once (not for all (s,a))
            Set $q_{\text{act}}(\hat{s},\hat{a}) \leftarrow (1-\alpha) q_{\text{act}}(\hat{s},\hat{a})
                + \alpha q_{\text{val}}(\hat{s},\hat{a})$ for all $(\hat{s},\hat{a})$.
            % <- Only update the values estimated in the batch
        }
        
        % Update the acting policy based on the newly updated q_act
        Set $\pi_{\text{act}}$ as $\epsilon$-greedy policy from $q_{\text{act}}$.
        
        % Copy the updated acting Q-function into the evaluation Q-function
        Set $q_{\text{val}} \leftarrow q_{\text{act}}$.

        }
    }
\end{algorithm}

During each batch, each agent \(i\) follows a stationary policy \(\pi_i\in\Pi^{\epsilon}\), while the Q-function $q_{\text{val}}$ is updated to learn the optimal Q-values given the opponents' fixed policies. At the end of each batch, the information gathered during the batch (captured by $q_{\text{val}}$) is transferred to the behavioural policy (captured by $q_{\text{act}}$) using a soft update with probability $1-\lambda.$  We refer to $q_{\text{val}}$ as the ``valuation Q-values'' and to $q_{\text{act}}$ as the ``acting Q-values''.

Algorithm~\ref{alg:SDES} differs from DQ in two crucial aspects:
\begin{enumerate}
    \item Instead of updating the behavioural policy at the end of each batch to the best-response given what was learnt during the batch (with probability $1-\lambda$), the Q-values encoding the behavioural policy are updated \emph{towards} the Q-values learnt during the batch. This is reminiscent of soft updating used in deep learning and is typically motivated by noting that the estimates obtained from a batch are noisy. We will see that soft updating affects the probability of learning different NE in multiagent settings even when the estimates obtained from each batch are exact. 
    \item During the batch, $q_{\text{val}}$ is updated using Expected SARSA as introduced by \cite{van2009theoretical} instead of standard Q-learning. This means that the estimates converge to the Q-values corresponding to the optimal $\epsilon$-greedy policy when the batch size grows. We motivate this change in two ways: (a) using Expected SARSA reflects a more rational approach of the agents who realise that they will continue exploring in the next batch, (b) the convergence of DQ requires that the exploration rate $\epsilon$ be smaller than some critical value, which essentially ensures that the greedy and $\epsilon$-greedy best response graphs are identical. In contrast, our result will hold for all $\epsilon\in(0, 1)$.
\end{enumerate}

%[What to do with this text?] Although \Cref{alg:SDES} maintains two Q–functions ($q_{\text{val}}$ and $q_{\text{act}}$) and their associated $\epsilon$–greedy policies, only $q_{\text{val}}$ is updated via stochastic approximation using Expected SARSA. The acting Q–function $q_{\text{act}}$ is updated deterministically by  tracking $q_{\text{val}}$ through a constant step-size update governed by the policy learning rate $\alpha$. Accordingly, in the theoretical development below we denote by $q_i^t \equiv q_{\mathrm{val},i}^t$ the \emph{valuation} Q-function of agent~$i$, and by $\pi_i^t \equiv \pi_{\mathrm{val},i}^t$ the associated $\epsilon$-greedy \emph{valuation policy}.

In the algorithm, we make a particular choice of step size sequence $\tilde{\alpha}_k\in(0,1)$ which satisfies the standard Robbins-Monro conditions, but the results naturally hold for any choice satisfying these conditions.

\section{Results}
\label{sec:result}

We split the results section in two. First we cover the theoretical results (Section \ref{sec:theory}) on (1) the convergence of Algorithm \ref{alg:SDES} in weakly acyclic stochastic games, (2) the equilibrium structure, and (3) the basins of attraction, where (2) and (3) are with respect to the environment introduced in Section \ref{sec:environment}. Then we cover empirical results (Section \ref{sec:simulations}) based on simulations which validate the theoretical results and provide additional insights into the dynamics of Algorithm \ref{alg:SDES}. 

\subsection{Theoretical Results}
\label{sec:theory}

The three main theoretical results concern convergence, equilibrium structure, and the basins of attraction.

\begin{itemize}
    \item \textbf{Proof of Convergence (Theorem~\ref{thm:convergence}):} In weakly acyclic games, Algorithm~\ref{alg:SDES} converges to an $\epsilon$-NE with arbitrarily high probability in finite time when the batch size $K$ is sufficiently large.
    
    \item \textbf{Equilibrium structure (Proposition~\ref{thm:IBR}):} The topology of the induced $\epsilon$-IBR graph depends on discount factor $\gamma$ and exploration rate $\epsilon$. Critical thresholds $\gamma_{c1}$ and $\gamma_{c2}$ partition the parameter space into three regimes: two having a unique defection equilibrium and one with coexistence of multiple equilibria. Learning parameters $\alpha$ and $\lambda$ can only influence equilibrium selection in the coexistence region.
    
    \item \textbf{Basin of attraction (Theorem~\ref{thm:BoA}):} In the deterministic soft-update limit ($\alpha \to 0$, $\lambda \to 0$, $K \to \infty$), basin boundaries are governed by target Q-value differences $\Delta q_i^{\pi_{-i}}$. We derive parameter-independent bounds, refined aggregate bounds via the conditional basins $\hat{B}$, $\hat{C}$, $\hat{D}$ across all $16$ initial profiles, and limiting bounds as $(\gamma,\epsilon)\to(1,0)$.
\end{itemize}

\subsubsection*{Convergence}
\begin{theorem}[Convergence of Algorithm~\ref{alg:SDES}]
\label{thm:convergence}
{\it
 
Consider a weakly acyclic stochastic game satisfying Assumption \ref{ass:reachability} with $N$ agents employing Algorithm~\ref{alg:SDES} with fixed parameters $\alpha,\,\epsilon,\,\lambda \in (0,1)$ and $\gamma\in[0,1)$. In addition, assume that the within-batch step-size schedule $\{\tilde{\alpha}_k\}_{k=1}^{\infty}$ satisfies the Robbins-Monro conditions and is restarted at the beginning of each batch (i.e.\ the visit counters $\nu(s,a)$ are reset at batch boundaries, as in Algorithm~\ref{alg:SDES}). For any tolerance level $\eta>0$, there exist finite constants
$\widetilde{K}<\infty$ and $\widetilde{\ell}<\infty$ such that, if 
$K \ge \widetilde{K}$,
\[
\mathbb{P}\!\left[\pi_\ell \in \Pi^{\epsilon}_{\text{eq}}\right] \;\ge\; 1-\eta,
\qquad \forall\, \ell \ge \widetilde{\ell}.
\]
}
\end{theorem}

\noindent\textit{Proof sketch.} The proof of Theorem \ref{thm:convergence} is provided in Appendix \ref{proof:convergence}. Here we provide a brief sketch of the main ideas.
The argument follows the two-layer template of \cite{arslan2017:1}: (A)~\emph{within each batch}, the valuation Q-values concentrate around the relevant best-response targets; (B)~\emph{across batches}, the soft policy updates realise, with positive probability, the steps of an inertial $\epsilon$-BR process that converges to an $\epsilon$-NE on a weakly acyclic $\epsilon$-BR graph. The essential departure from \cite{arslan2017:1} is that SDES replaces the \emph{hard} end-of-phase best-response jump of DQ (a policy learning rate equal to one) by a \emph{soft} update of the acting Q-values with rate $\alpha\in(0,1)$; consequently a single best-response step of the idealised process is realised not by one batch, but by a finite block of consecutive batches.

\vspace{0.5em}
\noindent\textit{Within each batch.}
The acting policies of all agents are held fixed, and each agent runs Expected SARSA on its valuation Q-values with an $\epsilon$-greedy target. In expectation, each such update applies a Bellman-type operator that maps a Q-table to the expected one-step reward plus the discounted value of the next state under the $\epsilon$-greedy policy of that table. This operator shrinks sup-norm distances between Q-tables by the factor $\gamma$ (it is a $\gamma$-contraction), and therefore has a unique fixed point: the optimal $\epsilon$-greedy Q-values $q_i^{\pi_{-i}}$ against the opponents' fixed acting policies $\pi_{-i}$. By standard stochastic-approximation arguments, as the batch size $K$ grows, the valuation Q-values concentrate uniformly over their batch-initial values around $q_i^{\pi_{-i}}$. Once the estimate lies within a tolerance $\zeta<\delta/2$ of this target, where $\delta>0$ is the minimal separation between distinct best-response action values, its ordering coincides with that of $q_i^{\pi_{-i}}$ and each agent's $\epsilon$-BR is correctly identified. A union bound over the $N$ agents makes this hold jointly with probability close to one.
\vspace{0.5em}
\\
\noindent\textit{Across batches.}
After each batch, an agent either updates its acting Q-values, 
\[ 
q_{\mathrm{act}} \leftarrow (1-\alpha)q_{\mathrm{act}}+\alpha q_{\mathrm{val}}, 
\] 
with probability $1-\lambda$, or leaves them unchanged with probability $\lambda$. When a single agent updates over several consecutive batches while its opponents remain inert, its acting Q-values track the best-response target: the distance to $q_i^{\pi_{-i}}$ contracts by a factor $1-\alpha$ per update, up to the within-batch tolerance $\zeta$. Because $\zeta<\delta/2$, after a finite number $\hat L$ of updates the acting Q-values enter the $\delta/2$-neighbourhood of the target, and the agent's $\epsilon$-greedy acting policy is an $\epsilon$-BR. 

A block of $\hat L$ such batches therefore realises one strict \emph{unilateral} $\epsilon$-BR step of the inertial $\epsilon$-greedy best-response dynamics (Definition~\ref{def:IBRD}). Exactly one agent switches to an $\epsilon$-BR while all other agents retain their policies, with the step now spread over $\hat L$ batches rather than occurring in a single batch. 

Weak acyclicity means that, from every policy profile, there is a finite path of such unilateral $\epsilon$-BR steps, of length at most $L$, ending at an $\epsilon$-NE. The algorithm follows one such path with positive probability: at every batch boundary, the independent inertia draws give a positive probability that the advancing agent performs the soft update while all other agents remain inert, and this pattern repeats at the $\hat L$ consecutive boundaries required for each step. Hence, conditioned on within-batch concentration holding throughout a block of $B=L\hat L$ batches, the joint policy reaches an equilibrium within that block with probability at least $p_{\min}>0$. This probability depends on the inertia $\lambda$, the learning rate $\alpha$, the gap $\delta$ through $\hat L$, and the best-response graph through $L$. 

Removing the conditioning costs only the small block-level concentration failure probability. Tracking the equilibrium probability across successive blocks then yields a recursion of the same form as in \cite{arslan2017:1}, which drives this probability above $1-\eta$ for any target tolerance $\eta>0$ once $K$ is sufficiently large. Finally, an equilibrium is a fixed point of the soft update and, once reached, is therefore not left. This absorption property links the local best-response steps to the weak-acyclicity argument and yields convergence to $\Pi^{\epsilon}_{\mathrm{eq}}$. The full argument, including the concentration and soft-update tracking lemmas, is given in Appendix~\ref{proof:convergence}.

\begin{figure}[H]
    \centering
    \begin{subfigure}[t]{0.45\linewidth}
        \centering
        \includegraphics[width=\linewidth]{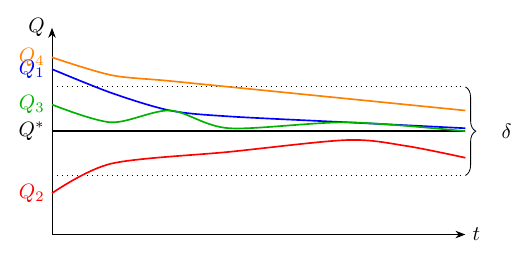}
        \caption{\textbf{Within Each Batch.} Convergence of $q_{i,k}^\ell$ toward its best-response target $q_i^{\pi_{-i}}$, ensuring correct ordering once $\|q_{i,k}^\ell-q_i^{\pi_{-i}}\|_\infty\le\zeta<\delta/2$.}
        \label{fig:within_batch}
    \end{subfigure}
    \hfill
    \begin{subfigure}[t]{0.45\linewidth}
        \centering
        \includegraphics[width=\linewidth]{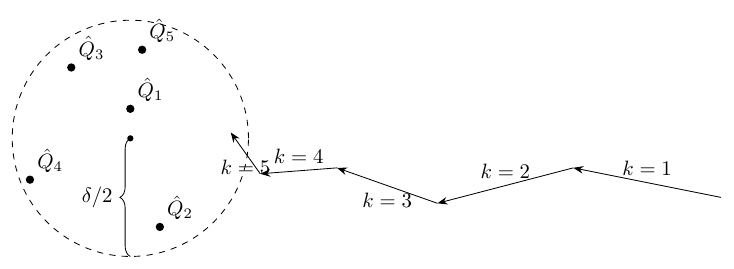}
        \caption{\textbf{Across Batches.} Repeated soft updates pull $q_{\mathrm{act}}$ into the $\delta/2$-neighbourhood of the best-response target; the policy flips after $\hat L$ updates, realising one best-response step.}
        \label{fig:across_batches}
    \end{subfigure}
    \caption{Illustration of \emph{Within} and \emph{Across Batches}.}
    \label{fig:within_across_batches}
\end{figure}

\subsubsection*{Equilibrium Structure and Baseline Basins of Attraction}
\label{sec:ldba}
In this section, we analyse the dynamics of \Cref{alg:SDES} in the large batch size limit, i.e., $K\rightarrow\infty$ with $\alpha = 1$. This corresponds to the I$\epsilon$BRD. We do this in the environment introduced in Section \ref{sec:environment}. We identify which strategy profiles can form an NE, the conditions for their existence, the structure for the IBR graphs and establish baselines for the basins of attraction.

When using $\hat{\mathcal{S}}$, there are four possible base policies $g_{i}:\hat{\mathcal{S}}\rightarrow\mathcal{A}_{i}$, and hence, by \eqref{eq:Piepsilon}, four $\epsilon$-greedy strategies available to each player $i$:
\begin{equation}
    \Pi_{i}^{\epsilon} = \{(D, D), (C, D), (D, C), (C, C)\},
\end{equation}
where the first entry in a tuple encodes the base action taken in state \textit{BC}, the second entry encodes the base action taken in state \textit{A1D}, and each tuple is identified with the $\epsilon$-greedy policy it induces via \eqref{eq:egreedybase}. The joint strategy space $\Pi^{\epsilon} = \Pi_{1}^{\epsilon}\times\Pi_{2}^{\epsilon}$ therefore contains $16$ strategy profiles. The notation used here is overloaded (cf. Figure~\ref{fig:state_abstraction}), so we provide names for the strategies in the table below to avoid confusion. 
\begin{table}[ht]
\centering
\begin{tabular}{lc}
\toprule
\multicolumn{2}{c}{\textbf{Strategy Names}} \\
\midrule
$(D, D)$ & All-Defect (AD) \\
$(C, D)$ & Grim Trigger (GT)\\
$(D, C)$ & Anti-Grim Trigger (AGT)\\
$(C, C)$ & All-Cooperate (AC)\\
\bottomrule
\end{tabular}
%\caption{}
\label{tab:strategynames}
\end{table}

\begin{figure}[ht]
    \centering
    \includegraphics[width=0.3\textwidth]{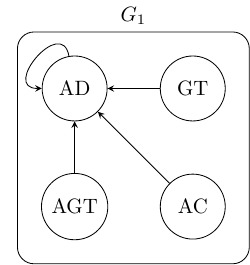}
    \includegraphics[width=0.315\textwidth]{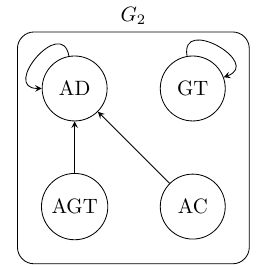}
    \includegraphics[width=0.3\textwidth]{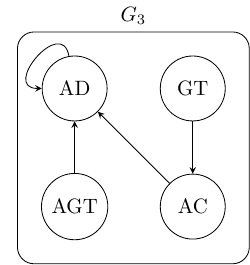}
    \caption{Possible IBR graphs for the IPD, when players use $\epsilon$-greedy strategies and the state space $\hat{\mathcal{S}}$.}
    \label{fig:IBRgraphs}
\end{figure}

If we restrict the players to play $\epsilon$-greedy strategies, the system induces a finite set of IBR graphs that encode the strategic response structure of the game. These are shown in ~\Cref{fig:IBRgraphs}, where nodes represent strategies and edges represent best responses.

\begin{figure}[H]
    \centering
    \includegraphics[scale=0.25]{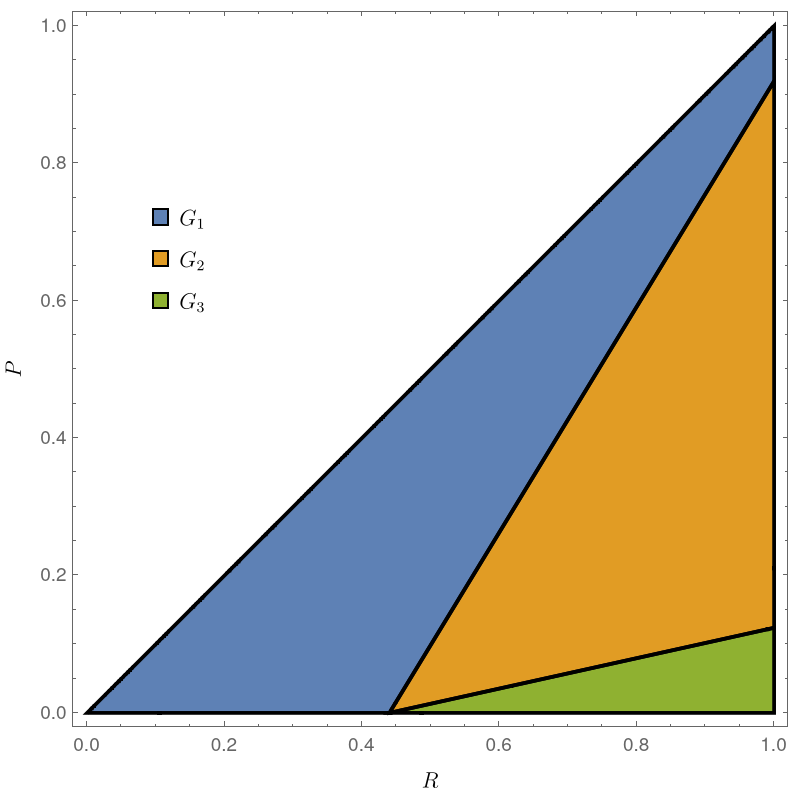}
    \includegraphics[scale=0.25]{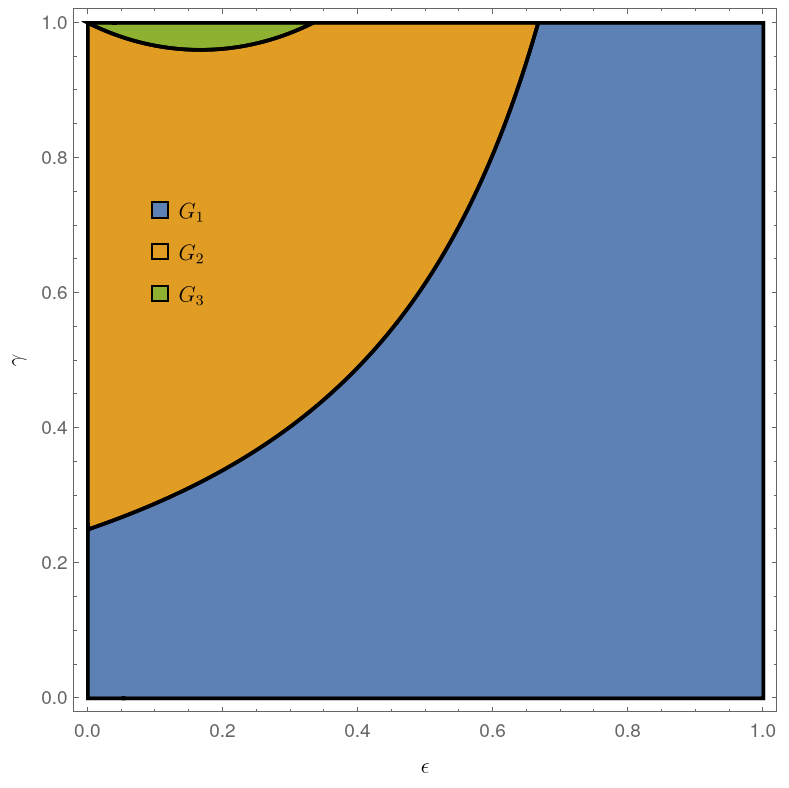}
    \caption{Phase diagrams for the possible IBR graphs in two slices of the parameter space. In both cases, we have fixed $T=1$ and $S=0$. On the left, we set $\gamma=0.7$ and $\epsilon=0.2$ while varying $R$ and $P$. On the right, we fix $R=0.8$ and $P=0.2$ while varying $\gamma$ and $\epsilon$.}
    \label{fig:phasediagrams}
\end{figure}

The structure of these IBR graphs depends not only on the payoff structure of the game but also on learning parameters, namely, the discount factor $\gamma$ and the exploration rate $\epsilon$. At specific combinations of these parameters, the topology of the IBR graph shifts, marking transitions in the strategic dynamics. The critical conditions at which such transitions occur are shown in~\Cref{fig:phasediagrams}.

\begin{proposition}[Possible IBR Graphs]
\label{thm:IBR}
Let 
\begin{equation}
    \gamma_{c_{1}}(\epsilon, T, R, P, S) =\frac{\epsilon (T+S-P-R) -2(T-R)}{(1-\epsilon) (\epsilon(T+S-2P) -2(T-P))}, 
\end{equation}
and 
\begin{equation}
    \gamma_{c_{2}}(\epsilon, T, R, P, S) = \frac{\epsilon (T+S-P-R)+2(P- S)}{(1-\epsilon) (\epsilon (T+S-2P) +2(P-S))}.
\end{equation}
Then for the prisoner's dilemma with state space $\hat{\mathcal{S}}$, 
\begin{enumerate}
    \item if $\gamma < \gamma_{c_{1}}(\epsilon, T, R, P, S)$, the IBR graph is $G_{1}$, with a unique NE at $\boldsymbol{\pi}^{\text{AD}}=(AD,AD)$,
    \item if $\gamma_{c_{1}}(\epsilon, T, R, P, S) < \gamma < \gamma_{c_{2}}(\epsilon, T, R, P, S)$, the IBR graph is $G_{2}$, with $\boldsymbol{\pi}^{\text{AD}}=(AD, AD)$ and $\boldsymbol{\pi}^{\text{GT}}=(GT, GT)$ as NE,
    \item if $\gamma > \gamma_{c_{2}}(\epsilon, T, R, P, S)$, the IBR graph is $G_{3}$, with a unique NE at $\boldsymbol{\pi}^{\text{AD}}=(AD,AD)$.
\end{enumerate}
\end{proposition}

\begin{proof}
    This is a result of brute force algebraic calculations performed using a computer algebra system as described in \cite{meylahn2022limiting} or \cite{barfuss2023intrinsic}. In this case, the calculations were performed using Mathematica for which the files can be found in \cite{repo}.
\end{proof}

Both equilibria can lead to a nonzero level of mutual cooperation as the strategies are implemented in an $\epsilon$-greedy fashion. However, $\boldsymbol{\pi}^{\text{GT}}$ leads to more mutual cooperation, as once the state BC state is reached, the players are likely to continue cooperating for a while before returning to mutual defection. On the other hand, after a mutual cooperation in $\boldsymbol{\pi}^{\text{AD}}$, the players are likely to immediately return to A1D. We therefore refer to $\boldsymbol{\pi}^{\text{GT}}$ as the cooperative equilibrium and to $\boldsymbol{\pi}^{\text{AD}}$ as the competitive equilibrium. 

\begin{corollary}[Convergence]
    The I$\epsilon$BRD and SI$\epsilon$BRD converge to an $\epsilon$-NE in the prisoner's dilemma with state space $\hat{\mathcal{S}}$.
\end{corollary}
\begin{proof}
    Both dynamics converge in weakly acyclic games, so it suffices to exhibit, from every strategy profile, a finite sequence of single-agent $\epsilon$-BR updates ending at an $\epsilon$-NE. In $G_{2}$, the unique best response to every strategy is either $AD$ or $GT$; letting the agents update one at a time thus reaches $(AD,AD)$ or $(GT,GT)$ in at most two updates. For instance, from $(AD,GT)$ a single update by one agent yields $(AD,AD)$ or $(GT,GT)$. In $G_{1}$ and $G_{3}$, the self-loop at $AD$ is the only cycle of the IBR graph, so alternating single-agent updates follow the arrows of the graph and reach $(AD,AD)$ directly. Weak acyclicity also follows from \cite[Theorem III.2]{meylahn2025quantifying}, since the IBR graphs contain no cycles of length three or more.
\end{proof}

Since each strategy has a unique best response, the IBR graph forms a functional relation. By \citet[Theorem II.11]{meylahn2025quantifying}, each weakly connected component therefore consists of a unique cycle (corresponding to an equilibrium) with directed trees pointing toward it. The basin of attraction (in the sense of Definition~\ref{def:BoA}) of an equilibrium under the I$\epsilon$BRD is precisely the fraction of strategies in the weakly connected component containing its associated cycle \citep[Theorem III.8]{meylahn2025quantifying}.

\begin{proposition}[BoA for I$\epsilon$BRD]
\label{thm:BoAIBRD}
    Under the I$\epsilon$BRD in the prisoner's dilemma with state space $\hat{\mathcal{S}}$, if $\gamma_{c_{1}}(\epsilon, T, R, P, S) < \gamma < \gamma_{c_{2}}(\epsilon, T, R, P, S)$, $\text{BoA}(\boldsymbol{\pi}^{\text{GT}}) = \frac{1}{4}$, $\text{BoA}(\boldsymbol{\pi}^{\text{AD}}) = \frac{3}{4}$ and $\text{BoA}(\boldsymbol{\pi}^{\text{GT}}|(AD, GT)) = \frac{1}{2}$ .
\end{proposition}
\begin{proof}
    The unconditional basins of attraction follow immediately from applying \cite[Theorem III.8]{meylahn2025quantifying} to $G_{2}$. For the conditional basin of attraction, note that four transitions are possible from $(AD, GT)$: (1) no player updates their strategy, which occurs with probability $\lambda^{2}$, (2) + (3) exactly one player updates their strategy, which occurs with probability $\lambda(1-\lambda)$, or (4) both players update their strategy, which occurs with probability $(1-\lambda)^{2}$. In cases (1) and (4), the players remain in essentially the same situation. In cases (2) and (3) the players transition to one of the NE. Since (2) and (3) occur with equal probability, the probability of reaching either equilibrium is a half. 
\end{proof}

Crucially, only within the orange region of Figure~\ref{fig:phasediagrams} does the game admit multiple equilibria. It is only in this region that the hyperparameters of the algorithm can influence the basins of attraction. Outside this region the dynamics will always converge to $\boldsymbol{\pi}^{\text{AD}}$. In the rest of the article we thus focus on this region. Our bounds on the basin of attraction of the cooperative equilibrium are collected in Theorem~\ref{thm:BoA} below. Part~(i) holds for every learning rate and inertia: whenever $\gamma_{c_{1}}(\epsilon, T, R, P, S) < \gamma < \gamma_{c_{2}}(\epsilon, T, R, P, S)$,
$$
\frac{1}{16} \;\;\leq\;\; \text{BoA}(\boldsymbol{\pi}^{\text{GT}}) \;\;\leq\;\; \frac{3}{4}.
$$
This reveals a fundamental limitation: in the best-case scenario, no more than \(75\%\) of total strategy profiles can converge to $\boldsymbol{\pi}^{\text{GT}}$ under the SI$\epsilon$BRD. Comparing Proposition~\ref{thm:BoAIBRD} with these bounds, we see that introducing the soft updates has the potential to change the BoAs drastically. In the worst case, the BoA could be reduced from $\frac{1}{4}$ to $\frac{1}{16}$ and in the best case, it could increase from $\frac{1}{4}$ to $\frac{3}{4}$. Parts~(ii) and~(iii) of Theorem~\ref{thm:BoA} refine these bounds in the small learning rate and inertia limit, using conditional basins that we derive next. 

\subsubsection*{Influence of hyperparameters on basin of attraction}
\label{sec:rationalsubsec}

Here we analyze the SI$\epsilon$BRD in the environment of Section \ref{sec:environment} to investigate how the introduction of the learning rate $\alpha$ influences the basins of attraction of $\boldsymbol{\pi}^{\text{GT}}$ and $\boldsymbol{\pi}^{\text{AD}}$. Furthermore, we focus on the limit as $\lambda$ approaches zero, where the Q-value updates become deterministic. This allows us to visualize the learning dynamics as a vector field in the eight-dimensional Q-value space given by 
\begin{equation}
    q_{i}^{t}(s, a) \;=\; (1-\alpha)\,q_{i}^{t-1}(s, a) \;+\; \alpha\,q^{\pi^{t-1}_{-i}}_{i}(s, a).
    \label{eq:Q_deterministic_update}
\end{equation}
Note that the dynamics in \eqref{eq:Q_deterministic_update} partition the Q-value space into regions in which the dynamics are linear with the boundaries of the regions given by $q_{i}(s, a) = q_{i}(s, a')$ with $a\neq a'$. This is because the target Q-values $q^{\pi^{t-1}_{-i}}_{i}$ potentially change at these hyperplanes. Using this, we define the Q-difference between defection and cooperation in state $s \in \mathcal{S}$ as
\begin{equation}
\label{eq:Qdiffdef}
    \Delta q_i^{t}(s) = q_i^{t}(s, D) - q_i^{t}(s, C).
\end{equation} 
By substituting \eqref{eq:Q_deterministic_update} into \eqref{eq:Qdiffdef} we obtain 
\begin{equation}
    \Delta q_i^{t+1}(s) = (1-\alpha)\,\Delta q_i^{t}(s) + \alpha\,\Delta q^{\pi^{t}_{-i}}_{i}(s),
    \label{eq:difference_update}
\end{equation}
where $\Delta q_{i}^{\pi^{t}_{-i}}$ is the Q-value difference in the target policy.
  
The advantage of analysing the evolution of the Q-difference rather than the full Q-function lies in the dimensionality reduction it provides: the state of the system is now captured by a single scalar per player in each state, and policy changes correspond to the Q-difference crossing zero. 

\begin{figure}[ht]
    \centering
    \includegraphics[width=0.9\textwidth]{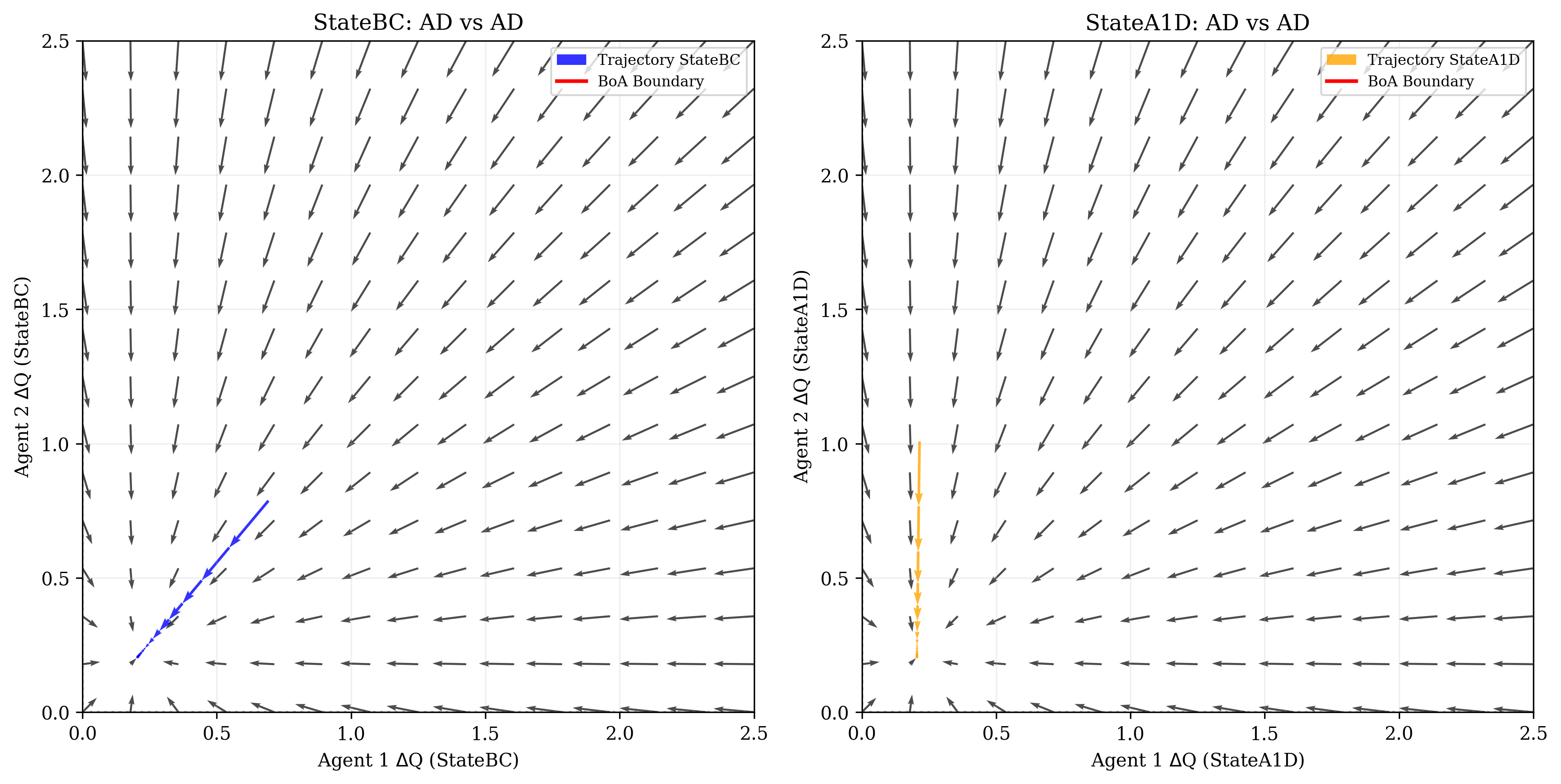}
    \caption{The vector fields corresponding to the dynamics of \eqref{eq:difference_update} in the two states BC (left) and (A1D) given that $\Delta q_i^{0}(s) > 0$ for all $i\in\{1, 2\}, s\in\hat{\mathcal{S}}$, together with a sample trajectory. }
    \label{fig:equilibria}
\end{figure}

To illustrate the above, consider the case where the players are initialized with Q-values having the ordering corresponding to $\boldsymbol{\pi}^{\text{AD}}$ so that $\Delta q_i^{0}(s) > 0$ for all $i\in\{1, 2\}, s\in\hat{\mathcal{S}}$. In this case, the target Q-value differences are in the same quadrant as the initialized Q-value differences, and the dynamics in \eqref{eq:difference_update} will converge to the targets. This is illustrated in Figure \ref{fig:equilibria}. Similarly, initializations with $\Delta q_i^{0}(A1D) > 0$ and $\Delta q_i^{0}(BC) < 0$, i.e., in the quadrant corresponding to $\boldsymbol{\pi}^{\text{GT}}$, will converge to $\boldsymbol{\pi}^{\text{GT}}$.

%Paragraph on limit as $\alpha$ goes to zero.
We will analyze the dynamics in the limit that $\alpha$ goes to zero. In this limit (under rescaling time as $\tau = \alpha t$), we can consider the evolution of the Q-value differences to be 
\begin{equation}
    \frac{\partial \Delta q_i^{\tau}(s)}{\partial \tau} = -(\Delta q_i^{\tau}(s) - \Delta q^{\pi^{\tau}_{-i}}_{i}(s)).
    \label{eq:continuousevolution}
\end{equation}

\begin{figure}[ht]
    \centering
    \includegraphics[width=0.9\textwidth]{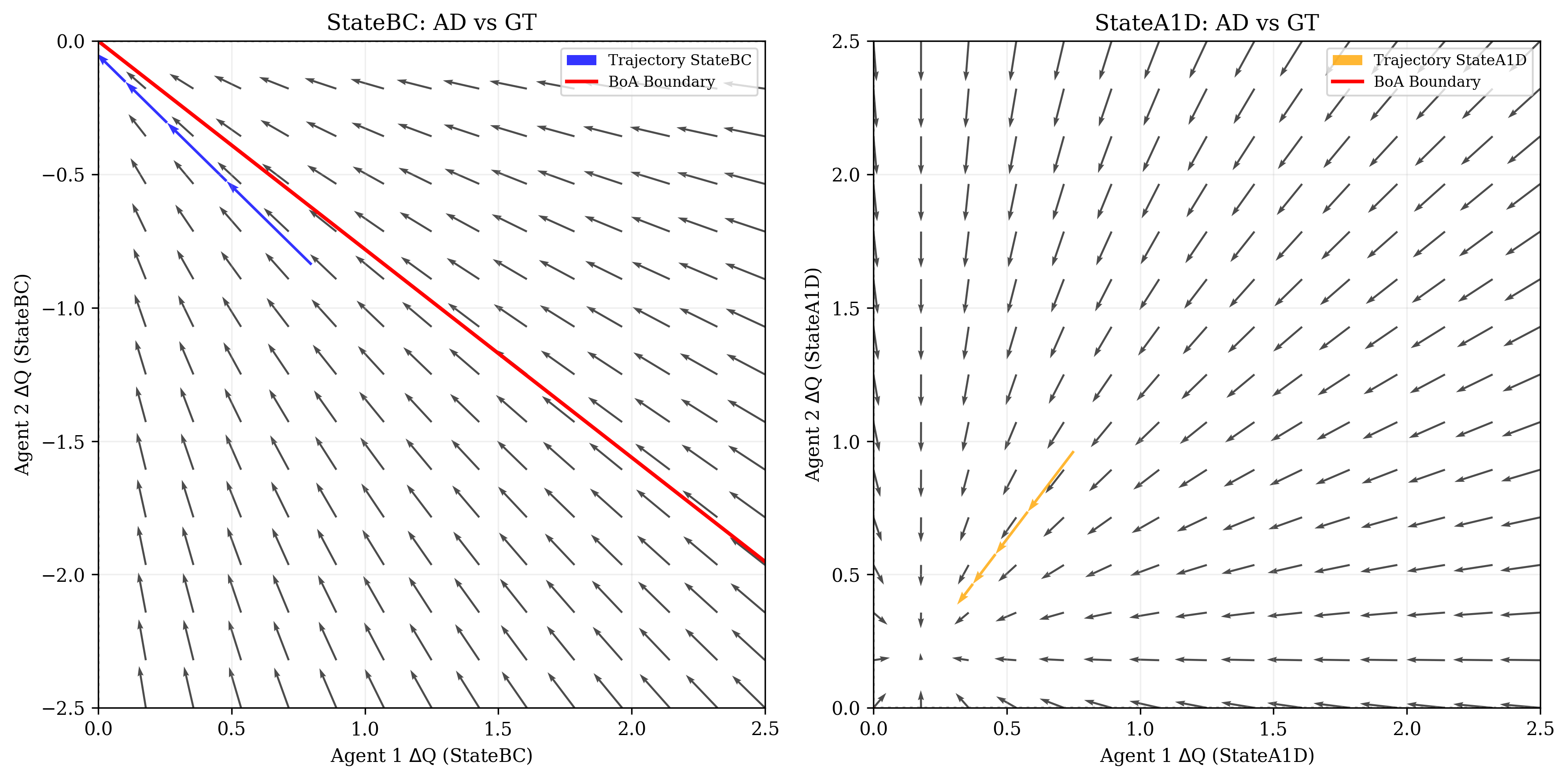}
    \caption{The vector fields corresponding to the dynamics of \eqref{eq:difference_update} in the two states BC (left) and A1D (right) given that $\Delta q_1^{0}(BC) > 0$ , $\Delta q_1^{0}(A1D) > 0$, $\Delta q_2^{0}(BC) < 0$, and $\Delta q_2^{0}(A1D) > 0$, together with a sample trajectory.}
    \label{fig:adgt}
\end{figure}

A central example in our analysis will be the case where the Q-value initializations are such that one agent is playing GT and the other is playing AD, i.e., the profiles $(AD, GT)$ and $(GT, AD)$. For our illustration, we assume that player one is playing AD and player two is playing GT. This corresponds to 
\begin{equation}
    \Delta q_1^{0}(BC) > 0 , \; \Delta q_1^{0}(A1D) > 0, \; \Delta q_2^{0}(BC) < 0, \text{ and } \Delta q_2^{0}(A1D) > 0.
\end{equation}
The corresponding vector field, and a sample trajectory are shown in Figure \ref{fig:adgt}. In this case, the target Q-value differences in the A1D state are in the same quadrant as the initial values, but the target Q-value differences in the BC state are in the opposite quadrant to the initial values. For the player playing AD (whose opponent plays GT) the best response is to play GT, while for the player playing GT (whose opponent plays AD) the best response is to play AD. Which equilibrium will be reached depends on which player changes their strategy first, which is determined by which side of the basin of attraction boundary (BoAB) the initial values are. The BoAB is defined by the straight line in the quadrant being considered intersecting the origin and the target Q-value differences. Such a boundary, therefore, only arises in quadrants where the Q-value difference target is in the opposite quadrant (as opposed to being in the same or an adjacent quadrant). 

In the case that we condition on the players starting in $(AD, GT )$, the BoAB allows us to calculate the probability of reaching either of the two equilibria. In particular, we can calculate the probability of reaching $\boldsymbol{\pi}^{\text{GT}}$ by calculating the area below the BoAB in Figure \ref{fig:adgt} and dividing this by the area of the entire quadrant. The latter is given by $\frac{1}{(1-\gamma)^{2}}$. The calculation of the former depends on whether the area below the BoAB is smaller or larger than half the area of the quadrant. This is determined by comparing $|\Delta q_{1}^{GT}(BC)|$ to $|\Delta q_{2}^{AD}(BC)|$. If $|\Delta q_{2}^{AD}(BC)| > |\Delta q_{1}^{GT}(BC)|$, the area below the BoAB forms a triangle with area 
\begin{equation}
    \frac{1}{2} \frac{1}{(1-\gamma)^{2}}\frac{|\Delta q_{1}^{GT}(BC)|}{|\Delta q_{2}^{AD}(BC)|}.
\end{equation}
A similar calculation for the case that $|\Delta q_{2}^{AD}(BC)| < |\Delta q_{1}^{GT}(BC)|$ gives
\begin{align}
\label{eq:Bhat}
    BoA(\boldsymbol{\pi}^{\text{GT}}|(AD, GT)) = \hat{B}:=
    \begin{cases}
        \frac{1}{2} \frac{|\Delta q_{1}^{GT}(BC)|}{|\Delta q_{2}^{AD}(BC)|} \quad &\text{when} \quad |\Delta q_{2}^{AD}(BC)| \geq |\Delta q_{1}^{GT}(BC)|,\\
        1 - \frac{1}{2} \frac{|\Delta q_{2}^{AD}(BC)|}{|\Delta q_{1}^{GT}(BC)|} \quad &\text{when} \quad |\Delta q_{2}^{AD}(BC)| < |\Delta q_{1}^{GT}(BC)|.
    \end{cases}
\end{align}
Naturally, $BoA(\boldsymbol{\pi}^{\text{AD}}|(AD, GT)) = 1 - BoA(\boldsymbol{\pi}^{\text{GT}}|(AD, GT))$.

Due to the symmetry of the game, the basin of attraction calculation is invariant under permutation of player indices. Specifically, for any initial configuration $(S_1, S_2)$, with $S_{i}\in\Pi^{\epsilon}$, the probability of reaching $\boldsymbol{\pi}^{\text{GT}}$ equals that from $(S_2, S_1)$. We thus only need to consider 10 distinct cases. In \Cref{appendix:case_by_case} we give the vector fields and possible paths to an equilibrium for all 10 distinct initial Q-value configurations.

\begin{theorem}[Basins of attraction under the SI$\epsilon$BRD]
\label{thm:BoA}
Assume agents follow SI$\epsilon$BRD in the IPD with state space $\hat{\mathcal{S}}$, with inertia parameter $\lambda\in(0, 1)$ and learning rate $\alpha\in(0, 1)$, and let $\gamma_{c_{1}}(\epsilon, T, R, P, S) < \gamma < \gamma_{c_{2}}(\epsilon, T, R, P, S)$. Then the basin of attraction of $\boldsymbol{\pi}^{\text{GT}}$ satisfies the following.
\begin{enumerate}
    \item[(i)] For all $\lambda, \alpha\in(0, 1)$,
    $$
    \frac{1}{16} \;\;\leq\;\; \text{BoA}(\boldsymbol{\pi}^{\text{GT}}) \;\;\leq\;\; \frac{3}{4}.
    $$
    \item[(ii)] In the limit as $\alpha, \lambda\rightarrow 0$, with payoffs normalized so that $T=1$ and $S=0$,
    $$
    \frac{1}{16}\left(1 + 2 \hat{B} + 2\hat{C} + 2\hat{D}\right) \;\;\leq\;\; \text{BoA}(\boldsymbol{\pi}^{\text{GT}}) \;\;\leq\;\; \frac{1}{16}\left(8 + 2\hat{B} + 2\hat{D}\right).
    $$
    \item[(iii)] If, in addition, $3P+R>2$, then in the $(\gamma, \epsilon) \rightarrow (1, 0)$ limit,
    \begin{align}
        \frac{1}{16}\left(4+2\frac{3P-1}{3P-R}\right) \leq \lim_{\gamma\rightarrow 1, \epsilon\rightarrow 0}BoA(\boldsymbol{\pi}^{\text{GT}}) \leq \frac{11}{16}.
    \end{align}
\end{enumerate}
\end{theorem}
The proofs of parts (i), (ii), and (iii) are given in Appendices~\ref{proof:baselineBoASIBRD}, \ref{proof:PropBoA}, and~\ref{proof:limitingBoA}, respectively. The proof of part~(iii) relies on showing that 
\begin{align}
    \lim_{\gamma\rightarrow 1, \epsilon\rightarrow 0}\hat{B} = 1,\quad \lim_{\gamma\rightarrow 1, \epsilon\rightarrow 0} \hat{D} = \frac{1}{2}, \quad \text{and} \quad \lim_{\gamma\rightarrow 1, \epsilon\rightarrow 0} \hat{C} = \frac{3P-1}{3P-R}.
\end{align}

Part~(iii) of Theorem~\ref{thm:BoA} implies that introducing a learning rate $\alpha$ can strictly increase the basin of attraction of $\boldsymbol{\pi}^{\text{GT}}$ relative to the case with $\alpha$ equal to one (see Proposition~\ref{thm:BoAIBRD}) when the discount factor is large enough and the exploration rate is small enough. With favorable instances of $R$ and $P$ the lower bound can reach values up to $\frac{3}{8}$, which is an increase of 50\% from the baseline. 

\subsection{Simulation Results}
\label{sec:simulations}
In this section, we validate our theoretical results with simulations and show that the insights gained through the theory in the small learning rate and inertia limit are also valid for larger values of $\alpha$ and $\lambda$. In addition, the simulations provide insights beyond those gained from the theoretical analysis. We conduct simulations of the SI$\epsilon$BRD using value iteration (Algorithm~\ref{alg:value_iteration}) to mimic the infinite batch size limit. The use of value iteration also confirms our results on the IBR graph structure and optimal Q-value targets. 

%\paragraph{Simulation Parameters.}
We select the algorithmic hyperparameters for the simulation so as to ensure that we stay in the region of $G_{2}$ and that the simulations include instances that can validate our theoretical results. In particular, we take $\alpha \in\Theta_{\alpha}:= \{0.01, 0.25, 0.5, 0.75, 1.0\}$,  $\lambda \in \Theta_{\lambda}:=\{0.01, 0.1, 0.2, 0.3, 0.4\}$, $\epsilon \in \Theta_{\epsilon}:=\{0.05, 0.1, 0.2\}$, and $\gamma \in \Theta_{\gamma}:=\{0.5, 0.75, 0.95\}$. For each parameter combination we simulate $M=2000$ independent learning trajectories. Convergence in the value iteration is established when the difference between Q-value updates falls below tolerance $\tau = 10^{-5}$. We approximate the BoA of $\boldsymbol{\pi}^{\text{GT}}$ as the fraction of trajectories that eventually reach $\boldsymbol{\pi}^{\text{GT}}$.

We select two instances of the environmental parameters so that both cases in \eqref{eq:Bhat} will be represented in the simulations when varying over the algorithmic hyperparameters. To achieve this, we select (1) $T=1,\ S=0,\ R=0.7,\ P=0.2$, which leads to $\hat{B}>1/2$ for most parameter instances, and (2) $T=1,\ S=0,\ R=0.8,\ P=0.5$, which leads to $\hat{B}<1/2$ for most parameter instances. In addition, (2) satisfies the condition for Theorem~\ref{thm:BoA}(iii) while (1) does not. 

\begin{figure}[ht]
    \centering
    \begin{subfigure}[b]{0.48\textwidth}
        \centering
        \includegraphics[width=\textwidth]{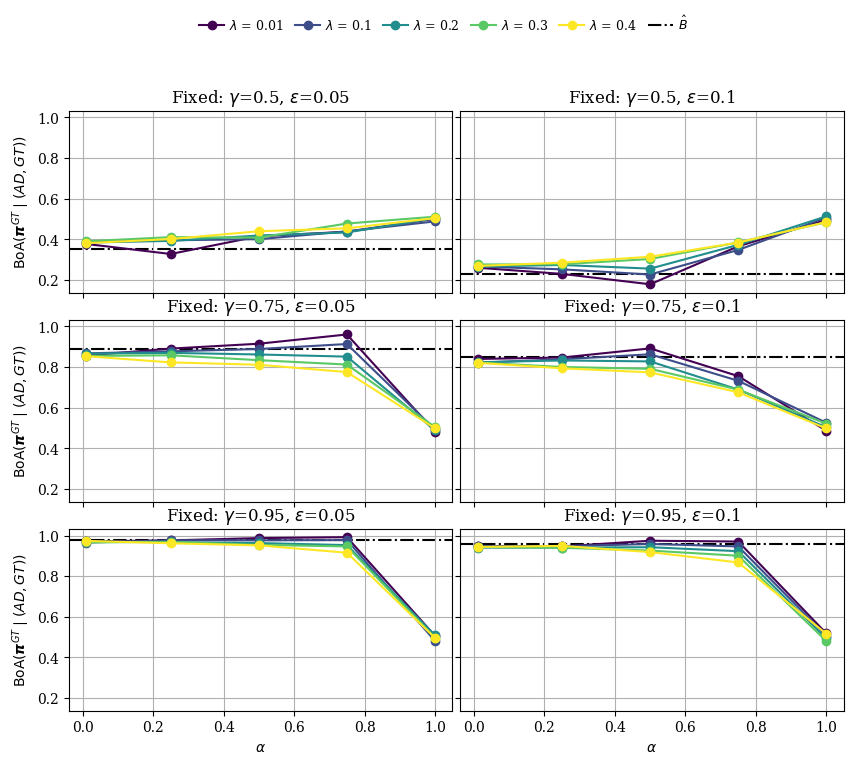}
        \caption{$R = 0.7,\ P = 0.2$}
        \label{fig:alpha_effect_ADGT_R07}
    \end{subfigure}
    \hfill
    \begin{subfigure}[b]{0.48\textwidth}
        \centering
        \includegraphics[width=\textwidth]{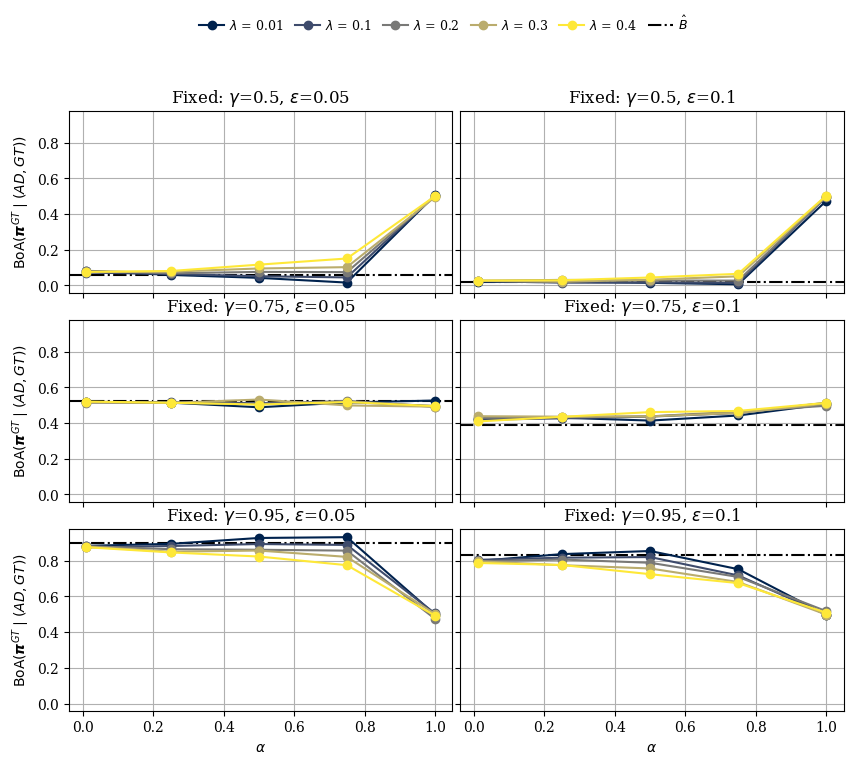}
        \caption{$R = 0.8,\ P = 0.5$}
        \label{fig:alpha_effect_ADGT_R08}
    \end{subfigure}
    \caption{Comparison of the empirical BoA of $\boldsymbol{\pi}^{\text{GT}}$ under the SI$\epsilon$BRD conditioned on starting in the $(\text{AD},\text{GT})$ strategy profile with the theoretical basin $\hat{B}$ (dash-dotted line), defined in \eqref{eq:Bhat}, as a function of the learning rate $\alpha$ for different values of $\lambda$. Each subplot fixes $(\gamma, \varepsilon)$.}
    \label{fig:alpha_effect_ADGT}
\end{figure}

In our first simulation, shown in Figure \ref{fig:alpha_effect_ADGT}, we condition on the initial Q-values being such that the players start in $(AD, GT)$. This allows us to validate the expression for $BoA(\boldsymbol{\pi}^{\text{GT}}|(AD, GT))$ given in \eqref{eq:Bhat}. First, note that the BoA of $\boldsymbol{\pi}^{\text{GT}}$ given that the dynamics start in $(AD, GT)$ is always equal to $1/2$ when $\alpha=1$ independent of $\lambda$ as predicted in Proposition~\ref{thm:BoAIBRD}. When moving to values of $\alpha$ less than one, we see that the simulations match the theory well when $\alpha$ and $\lambda$ are small, but also that the theory is a good predictor of the BoA for values of $\alpha$ less than $1/2$. Regarding the inertia, we see that the influence of $\lambda$ on the BoA of $\boldsymbol{\pi}^{\text{GT}}$ is relatively small for all values of $\alpha$, but especially as $\alpha$ approaches zero and one. Finally, we see that the BoA of $\boldsymbol{\pi}^{\text{GT}}$ conditioned on starting in $(AD, GT)$ increases to one as $\gamma$ approaches one and $\epsilon$ approaches zero.  
%with random Q-value initializations drawn uniformly from $[0, 1/(1-\gamma)]$.

\begin{figure}[ht]
    \centering
    \begin{subfigure}[b]{0.48\textwidth}
        \centering
        \includegraphics[width=\textwidth]{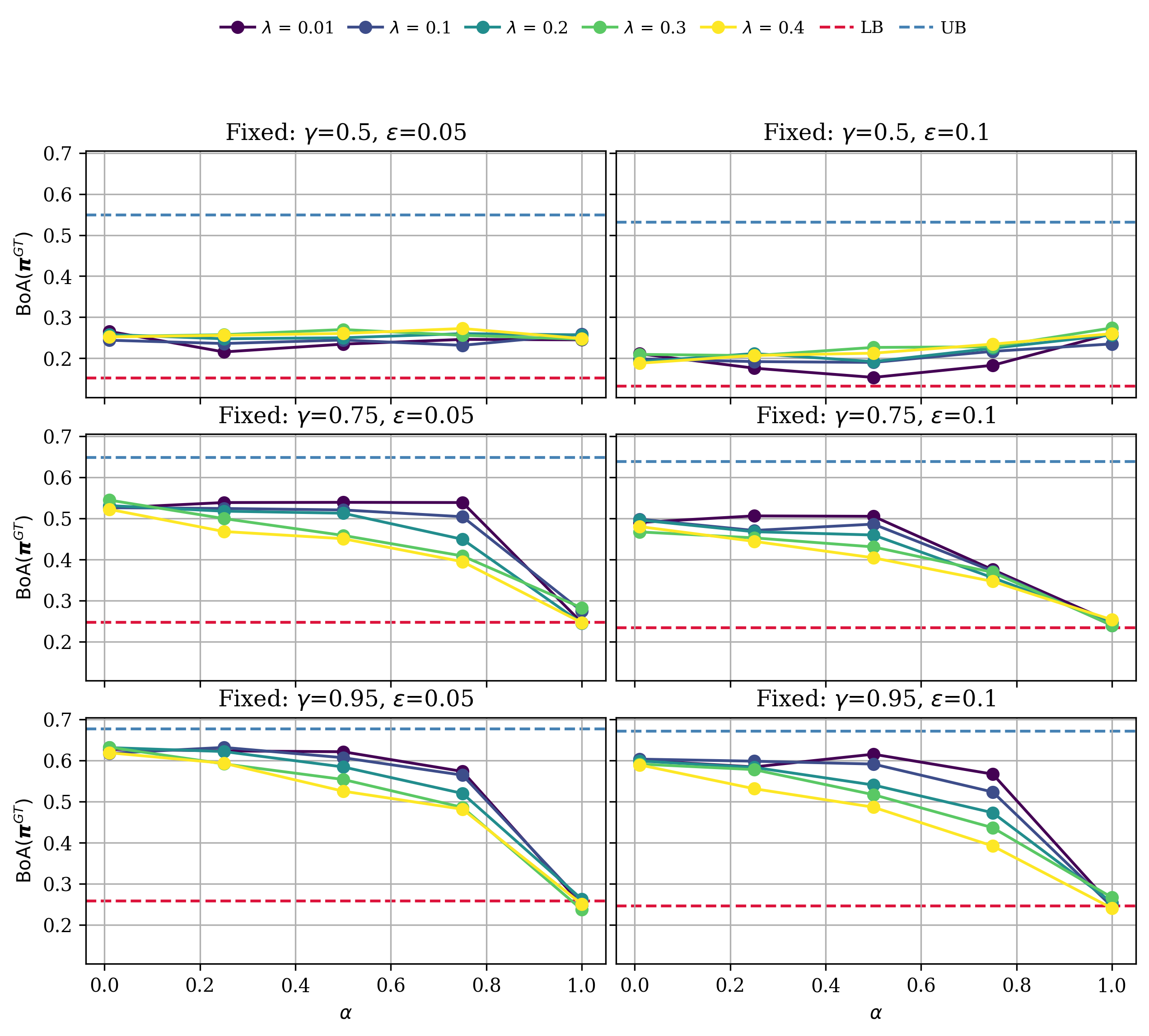}
        \caption{$R = 0.7,\ P = 0.2$}
        \label{fig:alpha_effect_R07}
    \end{subfigure}
    \hfill
    \begin{subfigure}[b]{0.48\textwidth}
        \centering
        \includegraphics[width=\textwidth]{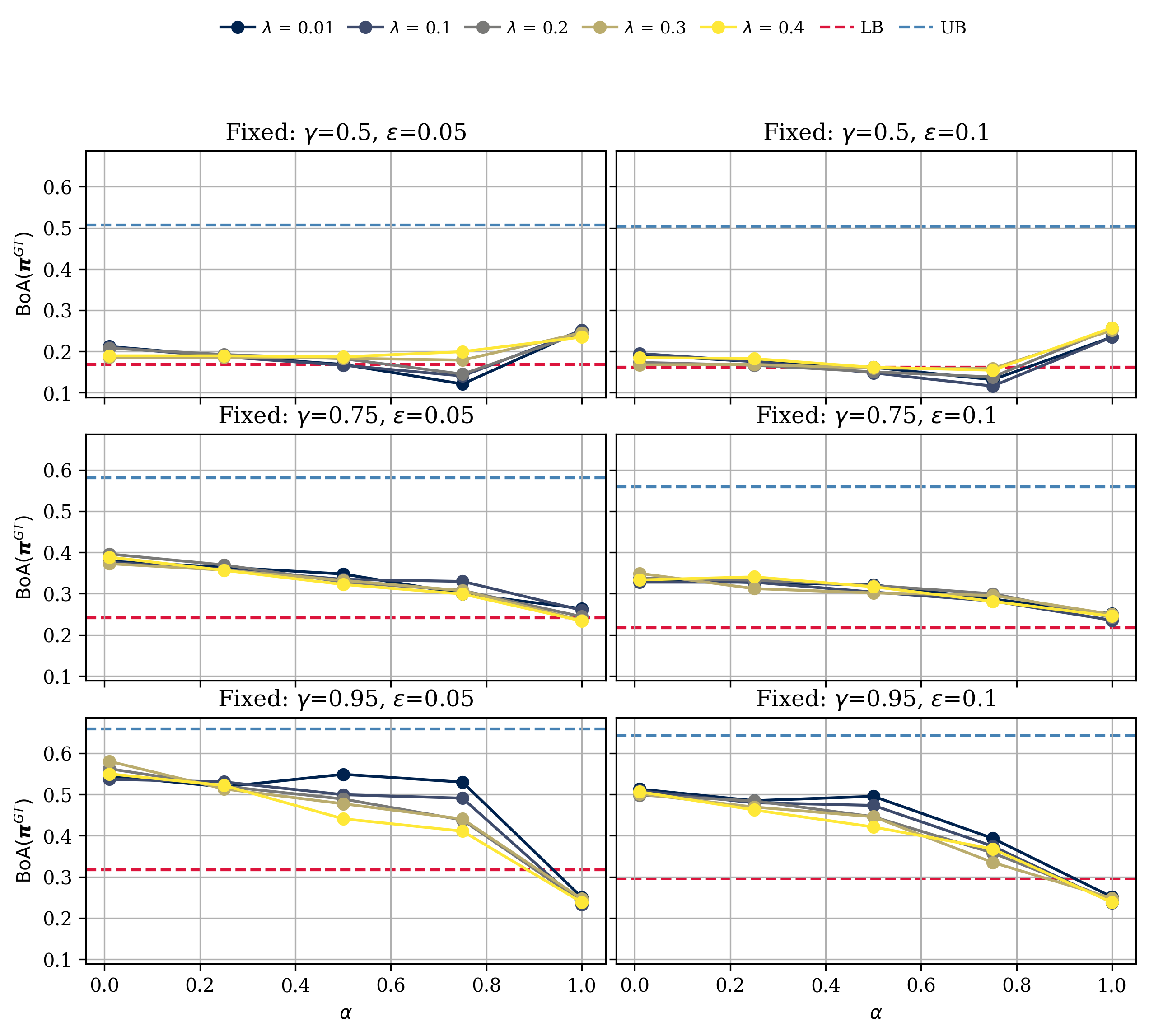}
        \caption{$R = 0.8,\ P = 0.5$}
        \label{fig:alpha_effect_R08}
    \end{subfigure}
    \caption{Comparison of the empirical BoA of $\boldsymbol{\pi}^{\text{GT}}$ under the SI$\epsilon$BRD with the theoretical bounds as a function of the learning rate $\alpha$ for different values of $\lambda$. Each subplot fixes $(\gamma, \epsilon)$. Dashed horizontal lines indicate the theoretical lower (red) and upper (blue) bounds from Theorem~\ref{thm:BoA}(ii).}
    \label{fig:alpha_effect}
\end{figure}

In our second simulation, shown in Figure~\ref{fig:alpha_effect}, we initialize the Q-values uniformly from $[0, 1/(1-\gamma)]$. This allows us to validate Proposition~\ref{thm:BoAIBRD} and parts (ii) and (iii) of Theorem~\ref{thm:BoA}. We find that the theoretical bounds derived in Theorem~\ref{thm:BoA}(ii) in the limit as $\alpha$ and $\lambda$ tend to zero always hold for the smallest values of $\alpha$ and $\lambda$ that we simulate. We also find that the theoretical basin of Proposition~\ref{thm:BoAIBRD} matches the simulations where $\alpha=1$. In addition, we find that the empirical BoA falls within the theoretical bounds for almost all values of $\lambda$ and $\alpha$ less than one. Regarding Theorem~\ref{thm:BoA}(iii), we see that the BoA of $\boldsymbol{\pi}^{\text{GT}}$ falls in the predicted range. This is the case for both configurations, even though configuration (1) does not satisfy the condition for Theorem~\ref{thm:BoA}(iii). The simulations, however, show a more specific effect, namely, that the BoA of $\boldsymbol{\pi}^{\text{GT}}$ increases as $\gamma$ approaches one and $\epsilon$ approaches zero. Finally, we see that $\lambda$ plays a larger role in determining the BoA for intermediate values of $\alpha$ compared to the role it played in Figure~\ref{fig:alpha_effect_ADGT}, but we still see that it has no effect for the extremal values of $\alpha$.

\section{Discussion}
\label{sec:discussion}
In this section we interpret the theoretical and empirical results and draw out the structural message they carry. The starting observation is an invariance. The $\epsilon$-BR graph is built entirely from the stage payoffs, the discount factor $\gamma$, and the exploration rate $\epsilon$: its vertices and edges encode which unilateral policy revisions are available from each strategy profile, and the learning parameters $\alpha$ and $\lambda$ do not enter it. At the level of the graph, changing $\alpha$ therefore changes nothing; the equilibria, the admissible revision paths, and the convergence guarantee of Theorem~\ref{thm:convergence} are all unaffected. Indeed, for hard updates the selection probabilities are fully determined by the revision rule and the best-reply graph \citep{altabaa2023decentralized}; a soft update is not expressible at that level, since the dynamics leave the discrete policy space. Yet Theorem~\ref{thm:BoA}, together with the simulations of Section~\ref{sec:simulations}, show that $\alpha$ is a primary determinant of \emph{which} equilibrium is selected. Equilibrium selection is thus a layer of structure that the best-response graph alone does not resolve. In the two-state Prisoner's Dilemma studied here, the additional information needed to describe this selection is contained in the relative target Q-value differences $\Delta q_i^{\pi_{-i}}$ separating actions. We refer to the indexed collection of these differences as the \emph{Q-value spectrum}.

In the soft-update limit, policy revision becomes a deterministic, piecewise-linear flow in Q-space whose basins of attraction are explicit functions of these gaps (Section~\ref{sec:rationalsubsec}). As $\alpha\to0$ the discrete soft-update dynamics approach the deterministic piecewise-linear flow analyzed in Section~\ref{sec:theory}. The simulations show that the basin geometry of this limiting flow is predictive for sufficiently small $\alpha$ and often remains informative at intermediate values. The structure of the soft-update dynamics suggests that this mechanism may extend beyond the two-state IPD: whenever several $\epsilon$-BR targets coexist and updates interpolate in Q-space, the relative target Q-value differences may contain information about equilibrium selection that is absent from the best-response graph. Establishing when this interpretation extends to more general stochastic games remains an open question. A core message of our paper, supported by both the theory and the simulations, can be stated as follows: the best-response graph determines which equilibria decentralized learners can reach, but not which equilibrium they select. Selection is governed by the spectrum of target Q-value differences $\Delta q_i^{\pi_{-i}}$: the policy learning rate $\alpha$, invisible to the graph, sets how faithfully the dynamics follow this spectrum, and in the soft-update limit, the basins of attraction are computable from it in closed form.

Our focus in the simulations is on the learning rate $\alpha$ and the inertia $\lambda$. Both dampen the responsiveness of policy updates, but they act differently. The parameter $\alpha$ interpolates continuously in Q-space and therefore reshapes basins; $\lambda$ only decides whether an update occurs at a batch boundary. When $\alpha=1$, the BoA of $\boldsymbol{\pi}^{\text{GT}}$ (in the coexistence region $\gamma_{c_{1}}<\gamma<\gamma_{c_{2}}$) equals $1/4$ for all $\lambda$, as predicted by the IBR graph. In the limit $\alpha,\lambda\to 0$, the BoA is determined by the target differences $\Delta q_i^{\pi_{-i}}$, which depend on payoffs, $\gamma$, and $\epsilon$.

\emph{Inertia:} The inertia parameter was introduced in DQ to guarantee escape from cycles in weakly acyclic games. Empirically, the BoA is independent of $\lambda$ at the extremes $\alpha\in\{0.01,1\}$, while at intermediate $\alpha$ the effect of $\lambda$ is visible but secondary (Figure~\ref{fig:alpha_effect}). A natural explanation is that $\lambda$ slows updates uniformly without biasing direction in Q-space, whereas $\alpha$ changes how far each update moves toward the target. Smaller $\alpha$ also increases the number of batches $\hat{L}$ needed to realize one $\epsilon$-BR step, which parallels a larger $\lambda$, but $\alpha$ has an additional geometric effect that $\lambda$ does not share. We still require $\lambda>0$ in the general convergence proof. The continuous-time analysis nevertheless suggests that, as $\alpha\to0$, simultaneous policy changes become unlikely because they require simultaneous boundary-crossing times. Whether this observation permits the inertia requirement to be weakened or removed for suitable classes of games is an open question.

\emph{Policy learning rate and the Q-value spectrum:} Soft updating between batches is typically motivated by stability, yet we find that it changes which equilibrium is selected. In our environment the effect is not monotone toward cooperation: the favored attractor is the one favored by the limiting vector field of $\Delta q$ differences. Conditional basins such as $\hat{B}$, $\hat{C}$, and $\hat{D}$ are explicit functions of those differences (Section~\ref{sec:rationalsubsec}). For favorable payoff configurations satisfying the conditions of Theorem~\ref{thm:BoA}(iii), the limiting lower bound on the basin of $\boldsymbol{\pi}^{\text{GT}}$ can reach $3/8$, compared with the hard-update value $1/4$. Thus, even the guaranteed lower bound can be 50\% larger than the hard-update basin. Simulations indicate still larger shifts in some regimes (Figure~\ref{fig:alpha_effect}), always in the direction predicted by the Q-difference geometry. For example, the bottom left plot of Figure~\ref{fig:alpha_effect} shows and increase of close to 150\% from the baseline as $\alpha$ approaches zero.

\emph{Implications for algorithmic collusion:} Our results indicate how soft updating could influence the learning of collusive strategy equilibria. The basin of attraction of an equilibrium is the probability that independent learners converge to it, the quantity proposed as an equilibrium-selection mechanism, and left open, by \citet{askenazi2024bounds}, and the quantity measured for hard policy updates by \citet{meylahn2025quantifying}. In pricing environments, equilibria supported by reward-punishment strategies analogous to $\boldsymbol{\pi}^{\text{GT}}$ are natural candidates for collusive strategy equilibria. Our analysis suggests that the probability of learning such equilibria is governed not only by market primitives and monitoring, which enter through the payoffs, $\gamma$, and $\epsilon$ and fix the revision graph, but also by an internal algorithmic parameter, the policy learning rate, which is invisible to equilibrium analysis. Slow policy adjustment could substantially increase the probability of learning a collusive strategy equilibrium when the Q-value spectrum favors it, and equally suppress it when the spectrum favors competition. A seemingly innocuous stabilization choice is therefore relevant to the ongoing debate on the plausibility and regulation of algorithmic collusion \citep{denBoer2022artificial,den2024mathematical}.

\emph{Persistent exploration and the equilibrium concept:} SDES uses Expected SARSA to evaluate the exploratory policy that will actually continue to be played, so its learning target and equilibrium concept are both defined within the fixed $\epsilon$-greedy policy class. This alignment removes the small-$\epsilon$ condition needed for convergence when a greedy best response is learned while an exploratory policy generates the data. The resulting guarantee is convergence to an $\epsilon$-greedy NE, not to an NE. Persistent exploration therefore enters not only as a sampling device but also as part of the solution concept. As $\epsilon$ changes, both the target Q-values and the best-response graph may change, as illustrated by the critical thresholds in Proposition~\ref{thm:IBR}.

\emph{Limitations.} The convergence and basin results operate at different levels of generality. Theorem~\ref{thm:convergence} applies to weakly acyclic stochastic games under the stated communicating and non-degeneracy assumptions, but provides an arbitrarily-high-probability guarantee for sufficiently large finite batches; it does not assert almost-sure convergence for every fixed batch size. The explicit basin analysis is considerably more specialized. It uses a two-player, two-action Prisoner's Dilemma with a two-state abstraction, uniform Q-value initialization, and the deterministic limit $K\to\infty$ and $\alpha,\lambda\to0$. Moreover, parts (ii) and (iii) of Theorem~\ref{thm:BoA} provide bounds on the aggregate basin rather than a complete closed-form characterization. The simulations examine the idealized SI$\epsilon$BRD through value iteration and show that the limiting geometry remains informative away from the strict limit, but they do not yet quantify how finite-batch estimation noise modifies the basins of SDES itself.

\section{Conclusion}
\label{sec:conclusion}

We introduced Soft Decentralized Expected SARSA (SDES), a generalization of Decentralized Q-learning in which agents maintain persistent exploration and revise their acting Q-values gradually at a policy learning rate $\alpha$. We showed that SDES converges to an $\epsilon$-greedy NE in weakly acyclic stochastic games for sufficiently large batch sizes. 

Our analysis of the repeated Prisoner's Dilemma with one-period memory shows that soft updating can also affect equilibrium selection. Although $\alpha$ does not alter the $\epsilon$-best-response graph or its equilibria, it changes the basins of attraction. In the deterministic soft-update limit, key conditional basins and bounds on the aggregate basins are determined by the relative target Q-value differences between competing actions. Consequently, reducing $\alpha$ can enlarge the basin of either Grim Trigger or All-Defect, depending on the underlying Q-value geometry; soft updating does not intrinsically favor cooperation. 

These results show that soft updating can be more than a stabilization device: in the setting studied here, it is also a mechanism of equilibrium selection. They further suggest that, more generally, the Q-value information omitted by the best-response graph may help determine which equilibria decentralized learners select. Establishing the scope of this mechanism in larger stochastic games with more agents and actions is an important direction for future work.

\section*{Data availability}
Full computational implementations, including BoA calculations for all configuration pairs and simulation code is available in~\cite{repo}.
%\acks{}

\vskip 0.2in
\bibliography{references}

\newpage

\appendix

\section{Q-value Difference Values}
\label{app:QvalDiff}
Here we provide the algebraic expressions for the target Q-value differences when 
$$
\gamma_{c_{1}}(\epsilon, T, R, P, S) < \gamma < \gamma_{c_{2}}(\epsilon, T, R, P, S),
$$ 
i.e, for $G_{2}$. If the opponent is playing the AD strategy, the target Q-value differences are
\begin{align}
    \Delta q_{i}^{AD}(BC) = \Delta q_{i}^{AD}(A1D) = \frac{1}{2} \left(2 P - 2 S + \epsilon (T + S - R - P)\right).
\end{align}
If the opponent is playing the AC strategy, the target Q-value differences are
\begin{align}
    \Delta q_{i}^{AC}(BC) = \Delta q_{i}^{AC}(A1D) = T - R + \frac{1}{2} \epsilon \left(P + R - S - T\right).
\end{align}
If the opponent is playing GT, the target Q-value differences are 
\begin{align}
    \Delta q_{i}^{GT}(BC) = \frac{2T -2 R + \epsilon (P + R - S - T) 
 -\gamma (1 - \epsilon) (2 T - 2P - \epsilon (S + T-2P))}{2 - 2 \gamma (1 - \epsilon)}
\end{align}
and 
\begin{align}
    \Delta q_{i}^{GT}(A1D) = \frac{(2 - 2 \gamma (1 - \epsilon)^2) P - 2 (1 - \gamma) S + \epsilon (T+S-R-P - \gamma (3 - \epsilon) S  - \gamma (1 - \epsilon) T)}{2 - 2 \gamma (1 - \epsilon)}.
\end{align}
Finally, if the opponent is playing AGT, the target Q-value differences are 
\begin{align}
    \Delta q_{i}^{AGT}(BC) = \frac{2 (P - S) + \epsilon (T+S-R -P + \gamma (1 - \epsilon)(T- S))}{2 + \gamma (1 - \epsilon) \epsilon}
\end{align}
and
\begin{align}
    \Delta q_{i}^{AGT}(A1D) = \frac{2 (T - R) + \epsilon (P + R - S - T) + \gamma (1 - \epsilon) (2T -2 (1 - \epsilon) P - \epsilon (S + T))}{2 + \gamma (1 - \epsilon) \epsilon }.
\end{align}

\section{Proofs}
\subsection{Proof of Theorem~\ref{thm:convergence}}
\label{proof:convergence}
We establish convergence in two parts, mirroring the proof sketch. \emph{Part~A} (within each batch) shows that the valuation Q-values concentrate around the best-response targets, so that on a high-probability event $E_\ell$ each agent's best response is correctly identified. \emph{Part~B} (across batches) shows that the soft policy update turns these correctly identified best responses into actual policy flips, each best response step being realized over a finite block of batches, and runs the resulting block recursion to equilibrium. The two parts are linked through the concentration event $E_\ell$ of Part~A, which is the standing hypothesis of both the soft-update tracking lemma (Lemma~\ref{lem:flip}) and the block-reaching lemma (Lemma~\ref{lem:block}) of Part~B.
\\
\noindent
Indexing convention: $\ell$ is the batch index; $k\in\{0,\dots,K\}$ the
within-batch iteration index; $i\in\{1,\dots,N\}$ the agent index;
$n$ the block counter of the across-batch recursion (Part~B); $\nu(s,a)$ the
within-batch visit counter of Algorithm~\ref{alg:SDES} (reset at every batch).

\paragraph{Part A: Within-batch learning.}
For each batch $\ell$, let $\pi_\ell=(\pi_{1,\ell},\dots,\pi_{N,\ell})$ denote the joint acting policy applied during the batch, held fixed throughout the batch. During the batch each agent~$i$ runs Expected SARSA on its valuation Q-values $q_{i,k}^\ell$ with an $\epsilon$-greedy target. We first identify the limit of this within-batch recursion and quantify the concentration around it.
% REVISED (2nd pass): the per-batch restart of the step-size schedule is
% ESSENTIAL here. In AY the visit counters are per-phase ("the number of visits
% to (x_t,u_t) in the k-th exploration phase"); with a global, never-reset
% counter the step sizes at the start of batch ell are O(1/ell), and for fixed K
% the iterate cannot reach the zeta-neighbourhood: eq:batch_concentration with a
% T uniform over ell would be FALSE. Algorithm 1 resets nu(s,a) at batch
% boundaries to match this analysis.
Throughout, the within-batch step sizes are $\tilde\alpha_{\nu}=1/(\nu+1)$ with $\nu=\nu(s,a_i)$ the number of visits to the current pair $(s,a_i)$ \emph{within the current batch}, so that each batch runs a fresh Robbins--Monro schedule, exactly as in the per-phase counters of \cite[Algorithm~1]{arslan2017:1}. We stress that the analogy with \cite{arslan2017:1} concerns only the per-phase step-size schedule: the within-batch update itself is Expected SARSA \cite{van2009theoretical}, whereas \cite{arslan2017:1} run standard Q-learning with the max-target (their Eq.~(7)); this is why the fixed point here is the optimal $\epsilon$-greedy Q-value $q_i^{\pi_{-i}}$ rather than the optimal Q-factor of \cite[Eq.~(2)]{arslan2017:1}, and why no exploration-bias correction (\cite[Lemma~3, Assumption~2]{arslan2017:1}) is needed.
The within-batch recursion falls squarely in the \emph{persistent-exploration}
framework of \cite{singh2000convergence}: $\epsilon$-greedy action selection
with fixed $\epsilon$ is one of their restricted rank-based randomized (RRR)
learning policies \cite[Section~2.2]{singh2000convergence}, and their
Theorem~2 shows that the on-policy SARSA(0) recursion under such a policy
converges to the fixed point of the corresponding \emph{restricted}
optimality equation, their Eq.~(7), which -- specialised to $\epsilon$-greedy
weights and to the induced MDP in which the opponents' fixed policies
$\pi_{-i}$ are marginalised into transitions and rewards -- is exactly the
target $q_i^{\pi_{-i}}$ below. Their Theorem~3 further shows that the greedy
restricted policy of this fixed point is optimal \emph{within the restricted
class}, the single-agent counterpart of the $\epsilon$-greedy best response
of Definition~\ref{def:BR}. Expected SARSA differs from SARSA(0) only in that
the expectation over the target policy at $s'$ is computed analytically
rather than sampled; as the proof below makes precise, this removes the
correction term that the SARSA(0) analysis must control
(\cite[Theorem~1]{singh2000convergence}) and strictly reduces the conditional
variance of the update \cite{van2009theoretical}.

% REVISED: the within-batch concentration lemma is now STATED HERE, before it is
% used, so that q_i^{pi_{-i}} is defined as its fixed point (no forward reference).
\begin{lemma}[Within-batch concentration]
\label{lemma1adapt}
Suppose Assumption~\ref{ass:reachability} holds and the data are generated by the $\epsilon$-greedy acting policies of Algorithm~\ref{alg:SDES}. For the operator $\mathcal F^i_{\pi_{-i}}:\mathbb{R}^{|\mathcal S\times\mathcal A_i|}\to\mathbb{R}^{|\mathcal S\times\mathcal A_i|}$, 
\begin{equation}
\label{eq:exp_sarsa_operator}
\mathcal F^{i}_{\pi_{-i}}(q)(s,a_i)
\coloneqq \mathbb{E}_{a_{-i}\sim\pi_{-i}(s)}\!\left[
r_i(s,a_i,a_{-i})
+\gamma\sum_{s'}p(s'|s,a_i,a_{-i})\sum_{b\in\mathcal A_i}\pi^{q}_i(b\mid s')\,q(s',b)\right],
\end{equation}
where $\pi^{q}_i$ is the $\epsilon$-greedy policy of $q$, the following hold. $\mathcal F^i_{\pi_{-i}}$ is a $\gamma$-contraction in $\|\cdot\|_\infty$ with unique fixed point $q_i^{\pi_{-i}}$. Moreover, for any tolerance $\zeta>0$ and any failure level $\delta'>0$, there exists a deterministic horizon $T(\zeta,\delta')<\infty$, uniform over the compact set of initial conditions $q_{i,0}$, such that the Expected SARSA recursion \[ q_{i,k+1}(s,a_i)=(1-\tilde\alpha_k)q_{i,k}(s,a_i) +\tilde\alpha_k\Big[r_i+\gamma\textstyle\sum_b\pi^{q_{i,k}}_i(b\mid s')\,q_{i,k}(s',b)\Big] \]
satisfies
\begin{equation}
\mathbb{P}\!\left[\sup_{k\ge T(\zeta,\delta')}\bigl\|q_{i,k}-q_i^{\pi_{-i}}\bigr\|_\infty\le\zeta\right]
\;\ge\;1-\delta'.
\label{eq:concentration_lemma}
\end{equation}
\end{lemma}
\begin{proof}
\emph{Contraction.} We emphasise at the outset that the fixed-point argument
takes place on the space of Q-tables
$\bigl(\mathbb{R}^{|\mathcal S\times\mathcal A_i|},\|\cdot\|_\infty\bigr)$,
a finite-dimensional normed space, hence a complete metric space on which
Banach's fixed-point theorem applies; no structure on the (finite) set of
$\epsilon$-greedy policies is used, since the policy $\pi^{q}_i$ enters the
operator \eqref{eq:exp_sarsa_operator} only through the weights it assigns in
the backup. Write this backup as
$M(q)(s')=\sum_{b\in\mathcal A_i}\pi^{q}_i(b\mid s')\,q(s',b)$; it is the
Expected SARSA backup of \cite{van2009theoretical} and, for $\epsilon$-greedy
weights, admits the closed form
\begin{equation}
\label{eq:M_closed_form}
M(q)(s')=(1-\epsilon)\max_{b\in\mathcal A_i} q(s',b)
+\frac{\epsilon}{|\mathcal A_i|}\sum_{b\in\mathcal A_i} q(s',b),
\end{equation}
valid also when the maximiser is not unique (the tie-breaking rule of
\eqref{eq:egreedypolicy} places the $(1-\epsilon)$ mass on a single action of
maximal value, so the expectation equals \eqref{eq:M_closed_form} regardless of
which maximiser is selected). As
a convex combination of the two non-expansions $\max_b(\cdot)$ and the
uniform average, $M$ is a non-expansion in $\|\cdot\|_\infty$; this is the
non-expansion property of rank-based averaging established in
\cite[Appendix~C, Lemmas~5 and~7]{singh2000convergence} within the
generalized-MDP framework of \cite{szepesvari1999unified}. For completeness
we also verify it directly, separating the \emph{value change} (same weights,
different values) from the \emph{weight change} (same values, different
weights). For any $q,q'$ and any $s'$,
\begin{align*}
M(q)(s')\hspace{-0.05cm}-\hspace{-0.05cm}M(q')(s') &=\underbrace{\sum_{b}\pi^{q}_i(b\mid s')\big(q(s',b)-q'(s',b)\big)}_{\text{value change}}
+\underbrace{\sum_{b}\big(\pi^{q}_i(b\mid s')-\pi^{q'}_i(b\mid s')\big)q'(s',b)}_{\text{weight change}}.
\end{align*}
The value-change term is bounded by $\|q-q'\|_\infty$, since $\pi^{q}_i(\cdot\mid s')$ is a probability distribution. The weight-change term is non-positive: by construction $\pi^{q'}_i$ is the $\epsilon$-greedy (hence value-maximising within the $\epsilon$-greedy class) policy for $q'$, so
$\sum_{b}\pi^{q}_i(b\mid s')\,q'(s',b)\le\sum_{b}\pi^{q'}_i(b\mid s')\,q'(s',b)$. Hence $M(q)(s')-M(q')(s')\le\|q-q'\|_\infty$, and exchanging the roles of $q,q'$ gives $|M(q)(s')-M(q')(s')|\le\|q-q'\|_\infty$, i.e.\ $M$ is a non-expansion. Composing with the averaging maps $\mathbb{E}_{a_{-i}\sim\pi_{-i}}$ and $\sum_{s'}p(\cdot\mid s,a_i,a_{-i})$ (both non-expansions) and the discount factor $\gamma$ yields, for all $q,q'$, \begin{equation}
    \|\mathcal F^i_{\pi_{-i}}(q)-\mathcal F^i_{\pi_{-i}}(q')\|_\infty\le\gamma\|q-q'\|_\infty.
\end{equation}
By the Banach fixed-point theorem, $\mathcal F^i_{\pi_{-i}}$ has a unique fixed point $q_i^{\pi_{-i}}$. Equivalently, $q_i^{\pi_{-i}}$ is the restricted-optimality Q-function of the induced MDP, whose existence and uniqueness is established by the same argument in \cite[Lemma~6]{singh2000convergence}.
\\
\emph{Step 1: infinite visitation.} Assumption~\ref{ass:reachability}
together with the $\epsilon$-greedy acting policies guarantees that each pair
$(s,a_i)$ is visited infinitely often w.p.~$1$: under the $\epsilon$-greedy
policies every action profile has probability at least
$(\epsilon/\max_j|\mathcal A_j|)^N$ in every state, so by
Assumption~\ref{ass:reachability} every state is reached from every other
with positive probability, and each own action $a_i$ is then played in it
with probability at least $\epsilon/|\mathcal A_i|$; the claim follows from
the extended Borel--Cantelli argument of
\cite[Appendix~B, Lemmas~3 and~4]{singh2000convergence}. In particular the
per-pair step sizes $\tilde\alpha_{\nu}=1/(\nu+1)$ traverse a full
Robbins--Monro sequence on every pair. We note that the data-generating
(acting) policy is \emph{fixed} during the batch while the target in
\eqref{eq:exp_sarsa_operator} uses the $\epsilon$-greedy policy of the
\emph{current} valuation estimate; the scheme is therefore off-policy in the
data sense, and -- unlike the on-policy analysis of
\cite[Theorem~2, Condition~1]{singh2000convergence} -- no condition linking
the sampled next action to the current estimate is needed, because no next
action is sampled in the target.

\emph{Step 2: almost-sure convergence.} Set
$\Delta_k\coloneq q_{i,k}-q_i^{\pi_{-i}}$ and, for the visited pair
$(s_k,a_{i,k})$,
\[
F_k(s_k,a_{i,k})\coloneq r_{i,k}+\gamma\,M(q_{i,k})(s_{k+1})
-q_i^{\pi_{-i}}(s_k,a_{i,k}),
\]
with $F_k(s,a_i)\coloneq 0$ elsewhere, so that the Expected SARSA recursion
takes the form
$\Delta_{k+1}(x)=(1-\tilde\alpha_k(x))\Delta_k(x)+\tilde\alpha_k(x)F_k(x)$ on
$X=\mathcal S\times\mathcal A_i$, the setting of
\cite[Lemma~1]{singh2000convergence} (an extension of
\cite[Theorem~1]{jaakkola1994convergence} and of
\cite[Lemma~12]{szepesvari1999unified}, proved there via martingale
arguments). Its four conditions hold: (i) $X$ is finite; (ii) the step sizes
are Robbins--Monro on every pair by Step~1; (iii) taking conditional
expectations and using that $r_{i,k}$ and $s_{k+1}$ are sampled from the
induced MDP,
\[
\mathbb E[F_k\mid\mathcal P_k]
=\mathcal F^i_{\pi_{-i}}(q_{i,k})-\mathcal F^i_{\pi_{-i}}\bigl(q_i^{\pi_{-i}}\bigr)
\quad\Longrightarrow\quad
\bigl\|\mathbb E[F_k\mid\mathcal P_k]\bigr\|_\infty\le\gamma\|\Delta_k\|_\infty ,
\]
i.e.\ the contraction condition holds \emph{exactly}, with vanishing
correction term $c_k\equiv 0$: because Expected SARSA computes the
expectation over the target policy analytically, the term
$C_k=\gamma[q_{i,k}(s_{k+1},a_{k+1})-M(q_{i,k})(s_{k+1})]$ that the SARSA(0)
analysis must drive to zero (\cite[Theorem~1]{singh2000convergence}, where
the GLIE condition enters) is identically absent; (iv) the noise
$w_k=F_k-\mathbb E[F_k\mid\mathcal P_k]$ is a uniformly bounded
martingale-difference sequence, since $|r_i|<R_{\max}$ and the iterates
remain in $[-R_{\max}/(1-\gamma),R_{\max}/(1-\gamma)]$ (each update is a
convex combination of values in this range), so
$\mathrm{Var}\{F_k(x)\mid\mathcal P_k\}\le K$ for a finite constant $K$.
Hence $\Delta_k\to0$ w.p.~$1$ for every fixed initial condition; the same
conclusion follows from asynchronous stochastic approximation for
$\|\cdot\|_\infty$-contractions \cite[Theorems~2 and~4]{tsitsiklis1994}
(DQ's Remark~2 records the analogous extension of their Lemma~1 beyond the
max-target).

\emph{Step 3: uniform deterministic horizon.} Almost-sure convergence is
asymptotic; the statement of the lemma requires a \emph{deterministic}
horizon $T(\zeta,\delta')$ valid uniformly over initial conditions, which
does not follow from Step~2 alone and is obtained by the
contraction--coupling argument of \cite[Appendix~A, Lemma~1]{arslan2017:1}:
two copies of the recursion started from different initial conditions and
driven by the \emph{same} sample path have a coupled error that contracts,
the coupling step using only that the backup applied to the two trajectories
is a $\gamma$-contraction in the coupled error, which holds here because $M$
is a non-expansion (established in the contraction part above). Two further
ingredients of
\cite[Lemma~2]{arslan2017:1} carry over unchanged: the batch-initial state
lies in the finite $\mathcal S$, and the joint acting policies during a batch
range over the \emph{finite} set of $\epsilon$-greedy policies, so the
horizon can be taken uniform over them. Uniformity over initial conditions
is essential because each batch re-initialises
$q_{\mathrm{val}}\leftarrow q_{\mathrm{act}}$, so the batch-initial condition
is data-dependent but lies in a fixed compact set.
\end{proof}

\noindent
We write $q_i^{\pi_{-i}}$ for the fixed point of Lemma~\ref{lemma1adapt}: the
optimal $\epsilon$-greedy Q-value of agent~$i$ against the opponents' fixed acting
policies $\pi_{-i}$. % REVISED: symbol fixed (was gamma_i, clashing with discount).
Define the minimal separation between distinct best-response action values as
\begin{equation}
\delta_i \;\coloneq\;
\min_{\substack{\pi_{-i}\in\Pi_{-i},\\ s\in\mathcal{S},\, a_i\neq a_i'\in\mathcal{A}_i}}
\Big\{
\big| q^{\pi_{-i}}_i(s,a_i) - q^{\pi_{-i}}_i(s,a_i') \big|
\;\Big|\;
q^{\pi_{-i}}_i(s,a_i)\neq q^{\pi_{-i}}_i(s,a_i')
\Big\},
\label{eq:Di_new}
\end{equation}
and set $\delta\coloneq\min_i\delta_i$. Here $\Pi_{-i}$ denotes the \emph{finite}
set of joint $\epsilon$-greedy policies of the opponents (each induced by a
deterministic base policy, Definition~\ref{def:egreedy}), which is the set of
acting policies realisable by
Algorithm~\ref{alg:SDES}; the minimum is therefore over a finite set and is
attained. As in \cite[Footnote~4]{arslan2017:1} we assume the game is
non-degenerate: for every agent $i$, every opponents' policy
$\pi_{-i}\in\Pi_{-i}$ and every state $s$, the entries of
$q_i^{\pi_{-i}}(s,\cdot)$ are pairwise distinct across actions. This holds for
generic payoffs; it implies $\delta>0$, that every argmax of $q_i^{\pi_{-i}}$
is a singleton, and in particular that every $\epsilon$-BR set is a singleton.
Uniqueness of the maximiser is used in Lemma~\ref{lem:flip}(iii): with tied
maximisers, repeated soft updates could in principle alternate the acting
argmax between tied actions, so that the acting policy could move between
distinct (tied) $\epsilon$-BRs and the absorption argument of
\eqref{eq:stay_eq} would fail. Fix any within-batch tolerance
\begin{equation}
\label{eq:zeta_choice}
\zeta \;<\; \tfrac{\delta}{2}.
\end{equation}
If $\|q_{i,k}^{\ell}-q^{\pi_{-i}}_i\|_\infty\le\zeta$, then, since distinct entries
of $q^{\pi_{-i}}_i$ differ by at least $\delta>2\zeta$, the ordering of
$q_{i,k}^\ell$ coincides with that of $q^{\pi_{-i}}_i$ and the best response is
correctly identified. (Within a single batch any $\zeta\le\delta/2$ already orders
the iterate; the \emph{strict} inequality $\zeta<\delta/2$ is what will be needed
in Part~B, Step~B1, to keep the soft-update tracking bound strictly inside the
$\delta/2$-neighbourhood. We return to this point there.)

Applying Lemma~\ref{lemma1adapt} with this $\zeta$: for any $\delta'>0$ there is a
deterministic $T(\zeta,\delta')<\infty$ (uniform over the compact set of initial
conditions) such that, for $K\ge T(\zeta,\delta')$,
\begin{equation}
\mathbb{P}\!\left[
\sup_{k \ge T(\zeta,\delta')}
\bigl\| q_{i,k}^{\ell} - q_i^{\pi_{-i}} \bigr\|_{\infty}\le \zeta
\right]\;\ge\; 1-\delta'.
\label{eq:batch_concentration}
\end{equation}
In particular $\delta'\to0$ as $K\to\infty$, so in the sequel $K$ is a free
parameter chosen large enough to achieve any desired $\delta'>0$.

\begin{remark}[Persistent exploration versus GLIE]
\label{rem:persistent}
Convergence of on-policy methods to the \emph{unrestricted} optimum $Q^*$
requires the exploration to decay (GLIE policies,
\cite[Theorem~1]{singh2000convergence}). No such condition appears here, and
none is needed: the target of the within-batch recursion is the
$\epsilon$-greedy \emph{restricted} optimum $q_i^{\pi_{-i}}$, the regime of
persistent exploration covered by \cite[Theorem~2]{singh2000convergence},
and the equilibrium notion of Definition~\ref{def:NE} is restricted
accordingly. We emphasise that persistent exploration is \emph{not} what
delivers the Robbins--Monro condition: any policy that visits each
state--action pair infinitely often suffices, including GLIE policies with
decaying exploration. What is specific to our setting is rather the
combination of (i)~a fixed $\epsilon$-greedy acting policy, which under
Assumption~\ref{ass:reachability} yields infinite visitation (Step~1 above),
and (ii)~a within-batch target that is the $\epsilon$-greedy restricted
optimum rather than $Q^*$, so that no exploration decay is needed to align
the learning target with the equilibrium notion. This is also why
Theorem~\ref{thm:convergence} holds for all $\epsilon\in(0,1)$, with no
small-$\epsilon$ condition and no exploration-bias correction as in
\cite[Lemma~3, Assumption~2]{arslan2017:1}.
\end{remark}

Consider the event that, after any given batch, exactly one agent updates while
all others remain inert. With $U_i\sim\mathrm{Unif}[0,1]$ i.i.d.\ and inertia
$\lambda\in(0,1)$, this \emph{unilateral-update} probability is
\begin{equation}
\rho_{\min}\;\coloneq\;\min_i\mathbb{P}\big(U_i>\lambda,\;U_j<\lambda,\;\forall j\neq i\big)
\;=\;(1-\lambda)\lambda^{N-1}\;>\;0.
\label{eq:qmin}
\end{equation}
% REVISED: joint concentration via UNION BOUND (the per-agent events share one
% trajectory and are NOT independent; the previous (1-delta')^N step is removed).
For the per-batch joint concentration event $E_\ell:=\bigcap_{i=1}^N E_\ell^i$ with
\begin{equation}
    E_\ell^i:=\{\sup_{k\ge T(\zeta,\delta')}\|q_{i,k}^\ell-q_i^{\pi_{-i,\ell}}\|_\infty\le\zeta\}
\end{equation}
(the target is taken against the opponents' acting policies $\pi_{-i,\ell}$
\emph{during batch~$\ell$}, so it is batch-dependent),
the union bound over the $N$ agents gives
\begin{equation}
\mathbb{P}(E_\ell)\;=\;\mathbb{P}\Big(\bigcap_{i=1}^N E_\ell^i\Big)
\;\ge\;1-\sum_{i=1}^N\mathbb{P}\big((E_\ell^i)^{c}\big)\;\ge\;1-N\delta',
\label{eq:union_bound}
\end{equation}
where the per-batch failure $N\delta'$ can be made arbitrarily small by increasing
$K$ (Lemma~\ref{lemma1adapt}). The block-level failure is bounded in Part~B once
the block length $B$ is fixed.

\paragraph{Part B: Across batches.}
We now leave the within-batch analysis and study the soft policy update across batch boundaries. The argument has two steps: first, a single agent's soft update realises one best response step over a finite block of batches (Lemma~\ref{lem:flip}); second, chaining such steps along a best response path drives the joint policy to equilibrium (Lemma~\ref{lem:block}) and the resulting recursion yields the theorem. Both steps take the within-batch concentration event $E_\ell^i$ of Part~A as their standing hypothesis, which is precisely how the two parts are joined.
We record the structural fact that drives the whole of Part~B: over any
stretch of batches during which the opponents' acting policies $\pi_{-i}$
are unchanged, the within-batch limits $q_i^{\pi_{-i}}$ of Part~A are
\emph{fixed} targets. Consequently, on the concentration events every policy
revision moves the joint policy along an edge of the $\epsilon$-greedy
best-response graph of Definition~\ref{def:brgraph} -- the same
graph-preservation property that underlies the policy-revision dynamics of
\cite{arslan2017:1} -- with the sole difference that SDES traverses an edge
over a finite block of batches (Lemma~\ref{lem:flip}) rather than in one.

\smallskip
\noindent\emph{Step B1: one best response step per block.}
The following lemma is the mechanism that distinguishes SDES from DQ: with a
soft update of rate $\alpha\in(0,1)$, one best-response step is realised over
$\hat L_i$ batches rather than one. This is also where the strict choice
$\zeta<\delta/2$ from Part~A is used: the positive gap $\delta/2-\zeta>0$ keeps the
tracking residual strictly inside the $\delta/2$-neighbourhood, which is precisely
what makes $\hat L_i$ finite (part~(ii) below); with $\zeta=\delta/2$ the residual
would saturate the neighbourhood and $\hat L_i$ would be infinite.

Throughout Part~B we assume the Q-tables are initialised in the natural range
$[-R_{\max}/(1-\gamma),\,R_{\max}/(1-\gamma)]$ (recall $|r_i|<R_{\max}$), as in
Algorithm~\ref{alg:SDES} and Definition~\ref{def:softIBRD}. This range is
preserved by both the Expected SARSA update and the soft update, since each is a
convex combination of values in the range. Consequently, for any agent~$i$ the
initial tracking error is bounded by
\begin{equation}
\label{eq:Dbound}
D\;\coloneq\;\bigl\|q_{\mathrm{act},i}^{(0)}-q_i^{\pi_{-i}}\bigr\|_\infty
\;\le\;\frac{2R_{\max}}{1-\gamma}\;<\;\infty .
\end{equation}

% NOTE TO AUTHORS: Lemma~\ref{lem:flip} also supplies the only nontrivial content
% of the SI-epsilon-BRD convergence Corollary (the assertion that soft updates
% realise strict best response steps). Per the chosen scope, that Corollary is left
% unchanged in the main text; it should cite Lemma~\ref{lem:flip} for its proof.
\begin{lemma}[Soft-update tracking and policy flip]
\label{lem:flip}
Fix agent~$i$. Suppose the opponents' acting policies $\pi_{-i}$ are held fixed over batches $\ell,\ell+1,\dots$, that on each such batch the concentration event $E_\ell^i$ holds with within-batch tolerance $\zeta<\delta/2$, and that agent~$i$ performs the soft update at $m$ consecutive batch boundaries. Then:
\begin{enumerate}
\item[(i)] $\displaystyle
\bigl\|q_{\mathrm{act},i}^{(m)}-q_i^{\pi_{-i}}\bigr\|_\infty
\;\le\;(1-\alpha)^m\,D+\zeta$;
\item[(ii)] for every
$m\ge \hat L_i\coloneq
\Big\lfloor \dfrac{\log\!\big(D/(\tfrac{\delta}{2}-\zeta)\big)}{\log\!\big(1/(1-\alpha)\big)}\Big\rfloor+1<\infty$,
one has $\|q_{\mathrm{act},i}^{(m)}-q_i^{\pi_{-i}}\|_\infty<\delta/2$, hence the $\epsilon$-greedy policy of $q_{\mathrm{act},i}^{(m)}$ is an $\epsilon$-BR, i.e.\ an element of $\operatorname{BR}_i^\epsilon(\pi_{-i})$;
\item[(iii)] if $\pi_{\mathrm{act},i}\in\operatorname{BR}_i^\epsilon(\pi_{-i})$ (equilibrium coordinate),
then $q_{\mathrm{act},i}^{(m)}$ retains the ordering of $q_i^{\pi_{-i}}$ for all
$m\ge0$, so the policy is unchanged (absorption).
\end{enumerate}
\end{lemma}
\begin{proof}
Write $q^\star=q_i^{\pi_{-i}}$ and $e_m=q_{\mathrm{act},i}^{(m)}-q^\star$. The soft
update gives
$e_m=(1-\alpha)e_{m-1}+\alpha\,(q_{\mathrm{val},i}^{(m-1)}-q^\star)$, and on
$E_\ell^i$, $\|q_{\mathrm{val},i}^{(m-1)}-q^\star\|_\infty\le\zeta$. Hence
$\|e_m\|_\infty\le(1-\alpha)\|e_{m-1}\|_\infty+\alpha\zeta$, and iterating,
$\|e_m\|_\infty\le(1-\alpha)^m\|e_0\|_\infty+\zeta\sum_{j=0}^{m-1}\alpha(1-\alpha)^j
=(1-\alpha)^m\|e_0\|_\infty+\zeta\,(1-(1-\alpha)^m)\le(1-\alpha)^m\|e_0\|_\infty+\zeta$,
which is~(i). For~(ii): the strict inequality $\zeta<\delta/2$ leaves a positive
gap $\delta/2-\zeta>0$; combined with $\|e_0\|_\infty\le D\le 2R_{\max}/(1-\gamma)$
(Eq.~\eqref{eq:Dbound}), the requirement
$(1-\alpha)^m D<\delta/2-\zeta$ is met for $m\ge\hat L_i$, since
$\hat L_i=\lfloor\cdot\rfloor+1$ is \emph{strictly} larger than
$\log\!\big(D/(\tfrac{\delta}{2}-\zeta)\big)/\log\!\big(1/(1-\alpha)\big)$ even
when that ratio is an integer (a ceiling would give only the non-strict
inequality in that case); $\hat L_i$ is finite
precisely because $\delta/2-\zeta>0$ (were $\zeta=\delta/2$, the residual $\zeta$
in~(i) would already saturate the $\delta/2$-neighbourhood and $\hat L_i$ would be
infinite). This gives $\|e_m\|_\infty<\delta/2$. Because distinct entries of $q^\star$ differ by at
least $\delta$ (Eq.~\eqref{eq:Di_new}) and $\|e_m\|_\infty<\delta/2$, the ordering
of $q_{\mathrm{act},i}^{(m)}$ matches that of $q^\star$ on entries with distinct
values; in particular the greedy actions of $q_{\mathrm{act},i}^{(m)}$ lie in the
argmax set of $q^\star$ (if $q^\star$ has tied maximisers, the greedy action of
$q_{\mathrm{act},i}^{(m)}$ is one of them), so the $\epsilon$-greedy policy of
$q_{\mathrm{act},i}^{(m)}$ is an $\epsilon$-BR to $\pi_{-i}$.
For~(iii): if $\pi_{\mathrm{act},i}\in\operatorname{BR}_i^\epsilon(\pi_{-i})$ then in
every state the greedy action of $q_{\mathrm{act},i}^{(0)}$ is the unique
maximiser of $q^\star$ in that state (uniqueness by the non-degeneracy
assumption below \eqref{eq:Di_new}); as $q_{\mathrm{act},i}^{(m)}$
is a convex combination of $q_{\mathrm{act},i}^{(0)}$ and Q-values within $\zeta$
of $q^\star$ (with margin $\delta>2\zeta$), that action remains the strict
argmax of $q_{\mathrm{act},i}^{(m)}$ for all $m$, so the acting policy is
unchanged.
\end{proof}

\noindent
Define $\hat L\coloneq\max_{i}\hat L_i<\infty$. By the explicit
bound~\eqref{eq:Dbound},
\[
\hat L_i\;\le\;
\Big\lfloor \dfrac{\log\!\big(2R_{\max}/((1-\gamma)(\tfrac{\delta}{2}-\zeta))\big)}{\log\!\big(1/(1-\alpha)\big)}\Big\rfloor+1,
\]
so $\hat L$ is a finite, computable constant. Note that $\hat L$ depends on the
policy learning rate $\alpha$ and the gap $\delta$; as $\alpha\to0$,
$\hat L\to\infty$, but $\hat L<\infty$ for every fixed $\alpha\in(0,1)$.

\smallskip
\noindent\emph{Step B2: chaining $\epsilon$-BR steps to equilibrium.} Given weak acyclicity, from any joint policy $\pi\in\Pi^{\epsilon}$ there is a strict $\epsilon$-BR path in the $\epsilon$-BR graph of minimum length $L_\pi<\infty$ ending at an $\epsilon$-NE; let $L=\max_{\pi\in\Pi^{\epsilon}}L_\pi<\infty$ and define the block length (in batches)
\begin{equation}
\label{eq:Lhat}
B\;=\;L\,\hat L.
\end{equation}

\begin{lemma}[Block-wise equilibrium reaching]
\label{lem:block}
Under the concentration events \\$E_\ell,\dots,E_{\ell+B-1}$ and the asynchronous
soft-update mechanism, there exists $p_{\min}\in(0,1)$ such that, for all $\ell$,
\begin{equation}
\label{eq:pmin_block}
\mathbb{P}\!\left(\pi_{\ell+B}\in\Pi^{\epsilon}_{\text{eq}}\,\middle|\,
\pi_\ell\notin\Pi^{\epsilon}_{\text{eq}},\;E_\ell,\dots,E_{\ell+B-1}\right)\;\ge\;p_{\min}.
\end{equation}
The constant $p_{\min}$ depends on the inertia $\lambda$, on $\alpha$ and $\delta$
(through $\hat L$), and on the best-response graph (through $L$).
\end{lemma}
\begin{proof}
Fix a shortest strict $\epsilon$-BR path $\pi=\pi^{(0)}\to\pi^{(1)}\to\dots\to\pi^{(r)}\in\Pi^{\epsilon}_{\text{eq}}$
with $r\le L$. Each step $\pi^{(t)}\to\pi^{(t+1)}$ changes exactly one agent, say
$i_t$. By Lemma~\ref{lem:flip}, this step is realised once $i_t$ performs the soft
update on $\hat L_{i_t}\le\hat L$ consecutive batches while all other agents remain
\emph{inert} and concentration holds. Here ``inert'' means $U_j<\lambda$ for every
$j\ne i_t$, i.e.\ agent~$j$ does \emph{not} perform the soft update at that batch
boundary, so its acting policy $\pi_{\mathrm{act},j}$ (and hence its acting
Q-function $q_{\mathrm{act},j}$) is unchanged; this does \emph{not} freeze
agent~$j$'s valuation Q-function $q_{\mathrm{val},j}$, which continues to be
learned within each batch but does not feed back into $\pi_{\mathrm{act},j}$ until
$j$ updates. The probability that, after a given batch,
$i_t$ updates and the others are inert is at least $\rho_{\min}>0$
(Eq.~\eqref{eq:qmin}); over $\hat L$ batches this pattern occurs with probability at least $\rho_{\min}^{\hat L}$. Chaining $r\le L$ such steps within the block of $B=L\hat L$ batches, and using that concentration holds on each batch by hypothesis, gives $\mathbb{P}(\pi_{\ell+B}\in\Pi^{\epsilon}_{\text{eq}}\mid\cdots)\ge \rho_{\min}^{B}\eqqcolon p_{\min}>0$. (A larger $p_{\min}$ can be obtained by accounting for all admissible paths; the positive lower bound suffices.) Since $\hat L=\hat L(\alpha,\delta)$ and $\rho_{\min}=(1-\lambda)\lambda^{N-1}$, the constant $p_{\min}$ depends on $\lambda,\alpha,\delta,L$ as stated.
\end{proof}

\noindent
% REVISED: block-level failure via union bound over the B batches of a block
% (matches AY Eq.~(12), where the failure is taken over the L phases of a block).
A block spans $B=L\hat L$ batches, on each of which concentration must hold. By a
further union bound over these $B$ batches and the per-batch bound~\eqref{eq:union_bound},
the block-level concentration event satisfies
\begin{equation}
\label{eq:block_union}
\mathbb{P}\Big(\textstyle\bigcap_{j=0}^{B-1}E_{\ell+j}\Big)\;\ge\;1-B\,N\delta'\;=:\;1-\tilde\delta,
\qquad \tilde\delta\coloneq B\,N\delta'.
\end{equation}
Since $B=L\hat L<\infty$ is a fixed constant and $\delta'\to0$ as $K\to\infty$, the
block failure $\tilde\delta$ can be made arbitrarily small by increasing $K$.

% REVISED: equilibrium absorption is Lemma~\ref{lem:flip}(iii).
% REVISED (2nd pass): eq:stay_eq strengthened to equality along the WHOLE block
% (= AY Eq.~(13)); endpoint equality alone does not support the within-block
% conclusion p_{nB+j} >= 1-eta at the end of the proof.
By Lemma~\ref{lem:flip}(iii), under concentration no agent's soft update changes
its policy at an equilibrium, at \emph{any} batch boundary of the block; hence
\begin{align}
\mathbb{P}\!\left(\pi_{\ell+B}\in\Pi^{\epsilon}_{\text{eq}}\mid\pi_\ell\notin\Pi^{\epsilon}_{\text{eq}}\right)
&\ge p_{\min}(1-\tilde{\delta}),\label{eq:reach_eq}\\[1mm]
\mathbb{P}\!\left(\pi_{\ell+B}=\pi_{\ell+B-1}=\dots=\pi_\ell\mid\pi_\ell\in\Pi^{\epsilon}_{\text{eq}}\right)
&\ge (1-\tilde{\delta}),\label{eq:stay_eq}
\end{align}
where~\eqref{eq:reach_eq} combines Lemma~\ref{lem:block} with the block-level
concentration~\eqref{eq:block_union} via the law of total probability, and
\eqref{eq:stay_eq} is the analogue of \cite[Eq.~(13)]{arslan2017:1}.

Let $p_\ell\coloneq\mathbb{P}(\pi_\ell\in\Pi^{\epsilon}_{\text{eq}})$. From
\eqref{eq:reach_eq}--\eqref{eq:stay_eq},
\begin{equation}
\label{eq:AY_style_recursion}
p_{(n+1)B}\;\ge\;(1-\tilde{\delta})\big(p_{nB}+p_{\min}(1-p_{nB})\big),
\end{equation}
which is identical in form to \cite[Eq.~(14)]{arslan2017:1}.

% REVISED: target tolerance eta is kept DISTINCT from the per-block failure
% tilde-delta. The previous single-symbol condition (RHS 1 - tilde-delta) was
% unsatisfiable; the correct AY condition has RHS 1 - eta.
Fix the target tolerance $\eta>0$. Choose $K$ (hence $\tilde\delta=BN\delta'$)
small enough that
\begin{equation}
\label{eq:delta_tilde_condition}
\left(\frac{(1-\tilde{\delta})p_{\min}}{\tilde{\delta}+(1-\tilde{\delta})p_{\min}}
-\tilde{\delta}\right)(1-\tilde{\delta})\;\ge\;1-\eta .
\end{equation}
Such a $\tilde\delta>0$ exists for any $p_{\min}>0$ and $\eta>0$, because the
left-hand side tends to $1$ as $\tilde\delta\to0$. Since $\tilde\delta=BN\delta'$
with $B<\infty$ fixed and $\delta'\to0$ as $K\to\infty$,
\eqref{eq:delta_tilde_condition} holds for $K$ sufficiently large.

\textit{Recursion analysis (following \cite[Eqs.~(14)--(16)]{arslan2017:1}).}
Rewrite~\eqref{eq:AY_style_recursion} as
\begin{equation}
\label{eq:step3}
p_{(n+1)B}-p_{nB}\;\ge\;\bigl[\tilde{\delta}+(1-\tilde{\delta})p_{\min}\bigr]
\left[\frac{(1-\tilde{\delta})p_{\min}}{\tilde{\delta}+(1-\tilde{\delta})p_{\min}}-p_{nB}\right],
\end{equation}
and set $p^\star\coloneq\dfrac{(1-\tilde{\delta})p_{\min}}{\tilde{\delta}+(1-\tilde{\delta})p_{\min}}$.
If $p_{nB}\le p^\star-\tilde{\delta}$ then, by~\eqref{eq:step3},
$p_{(n+1)B}-p_{nB}\ge[\tilde{\delta}+(1-\tilde{\delta})p_{\min}]\tilde{\delta}>0$,
so $p_{nB}$ increases until it exceeds $p^\star-\tilde{\delta}$. If
$p_{nB}\ge p^\star$ then, by~\eqref{eq:AY_style_recursion},
$p_{(n+1)B}\ge(1-\tilde{\delta})p_{nB}\ge p_{nB}-\tilde{\delta}$, so $p_{nB}$ can
decrease by at most $\tilde{\delta}$ and remains above $p^\star-\tilde{\delta}$.
Hence there is $\tilde n<\infty$ with $p_{nB}\ge p^\star-\tilde{\delta}$
for all $n\ge\tilde n$.
Finally, by~\eqref{eq:stay_eq} the policy remains at the equilibrium throughout
the block, so for all $n\ge\tilde n$ and $j\in\{0,\dots,B-1\}$,
$p_{nB+j}\ge p_{nB}\,(1-\tilde\delta)\ge(p^\star-\tilde{\delta})(1-\tilde{\delta})\ge1-\eta$,
the last inequality being again~\eqref{eq:delta_tilde_condition}.

Therefore, for any $\eta>0$ there exist finite $\widetilde{K}<\infty$ and
$\widetilde{\ell}<\infty$ (namely $\widetilde\ell=\tilde n B$) such that, if
$K\ge\widetilde{K}$,
\[
p_\ell=\mathbb{P}(\pi_\ell\in\Pi^{\epsilon}_{\text{eq}})\;\ge\;1-\eta,
\qquad\forall\,\ell\ge\widetilde{\ell}.
\]
This establishes Theorem~\ref{thm:convergence}.

% =====================================================================
% NOTE: The within-batch concentration lemma (Lemma~\ref{lemma1adapt}) is stated
% and proved at the START of Part A, before it is used (no forward reference).
% NOTE: The proofs of parts (i), (ii), and (iii) of
% Theorem~\ref{thm:BoA} (formerly three separate propositions)
% are UNCHANGED from the original
% proofsections.tex and are not reproduced in this revised file.
% =====================================================================
\subsection{Proof of Theorem~\ref{thm:BoA}(i)}
\label{proof:baselineBoASIBRD}
\begin{figure}
    \centering
    \includegraphics[width=1.0\linewidth]{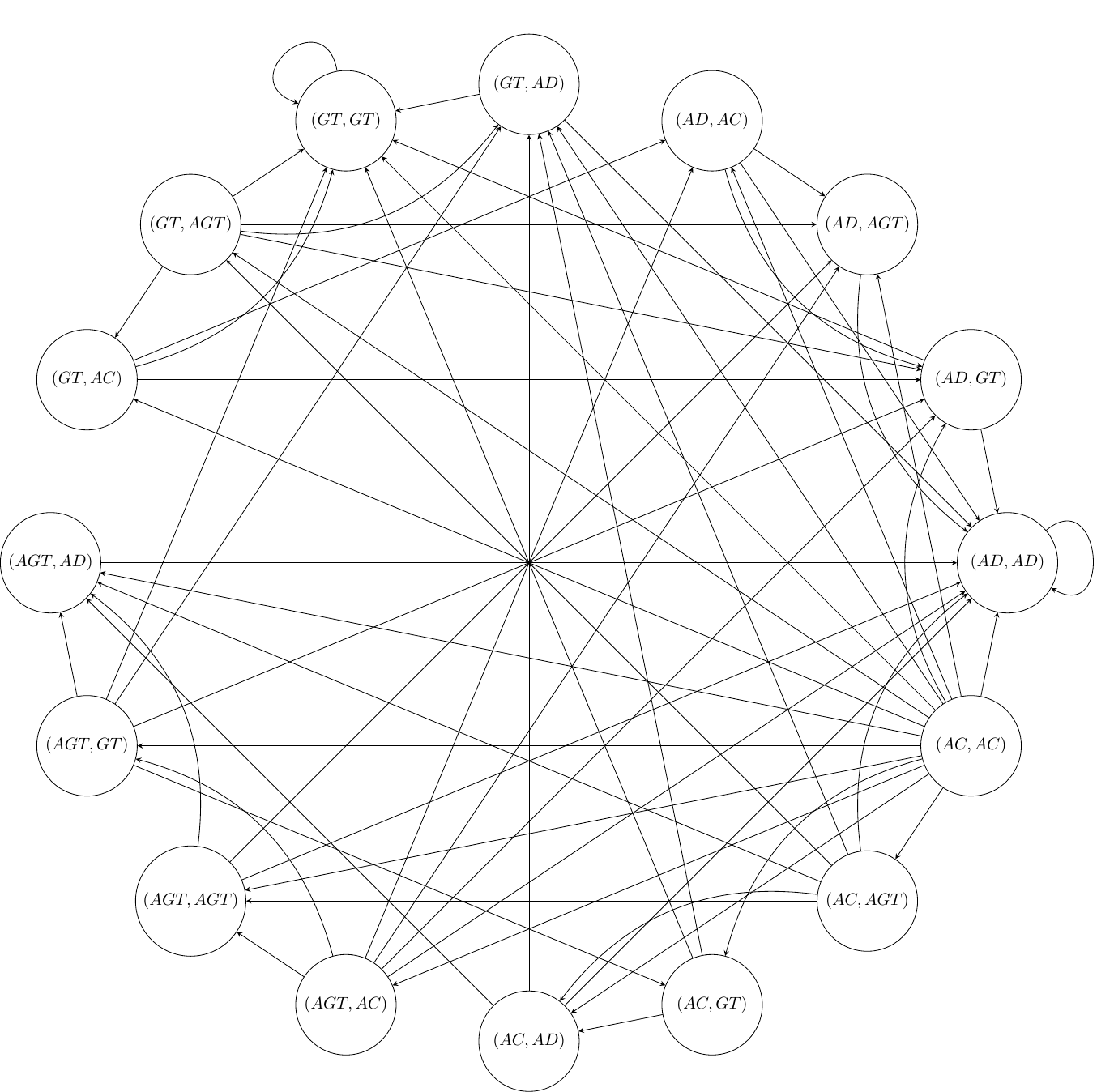}
    \caption{Asynchronous best-response graph.}
    \label{fig:asynchronousBR}
\end{figure}

We proceed by constructing what we call the asynchronous best-response graph corresponding to IBR graph $G_{2}$ in Figure \ref{fig:asynchronousBR}. This graph contains nodes representing all possible policy profiles. The directed edges represent the transitions that can occur if one or both players changes their strategy towards the best-response strategy given their opponent's strategy in one or both components. For example, consider the profile $(AC, AD)$. Given IBR graph $G_{2}$, we see that the best response for the first player is to switch to playing $AD$ while the first player is already best responding. Switching from $AC=(C, C)$ to $AD=(D, D)$ requires changing the greedy action in both states. In the asynchronous BR graph, we thus add directed edges to the profiles reached by changing the strategy $AC$ towards $AD$ in one or both states. That is we add the edges from $(AC, AD)$ to $(GT, AD), (AGT, AD)$, and $\boldsymbol{\pi}^{\text{AD}}$. 
If this graph contains an edge from one strategy profile to another, it is possible to observe that transition in the SI$\epsilon$BRD. To see this, note that the SI$\epsilon$BRD allow for transitions that change a profile towards the profile in which both players are playing a best response to their opponent's previous strategy in any subset of agent-state pairs. These are precisely the edges present in the graph. Therefore, we can calculate a lower bound for the BoA of $\boldsymbol{\pi}^{\text{GT}}$ by taking the fraction of nodes from which it is \emph{only} possible to reach $\boldsymbol{\pi}^{\text{GT}}$ and an upper bound by taking the fraction of nodes from which it is \emph{at all} possible to reach $\boldsymbol{\pi}^{\text{GT}}$. 

\textbf{Lower bound:} The $\boldsymbol{\pi}^{\text{GT}}$ node has eight incoming edges (including the self-loop), coming from seven other strategy profiles. From all seven of these there also exists a path to $\boldsymbol{\pi}^{\text{AD}}$. Therefore, the fraction of nodes from which only $\boldsymbol{\pi}^{\text{GT}}$ can be reached is $\frac{1}{16}$. 

\textbf{Upper bound:} Since the game is weakly acyclic, the fraction of nodes from which it is at all possible to reach $\boldsymbol{\pi}^{\text{GT}}$ is one minus the fraction of nodes from which it is only possible to reach $\boldsymbol{\pi}^{\text{AD}}$. Now the node $\boldsymbol{\pi}^{\text{AD}}$ has eleven incoming edges (including the self-loop), coming from ten other strategy profiles. Of these, it is possible to reach $\boldsymbol{\pi}^{\text{GT}}$ from all nodes except from $(AGT, AD)$, $(AD, AGT)$ and $(AGT, AGT)$. These in turn are reachable from $(AC, AC), (AC, AGT), (AGT, AC), (AC, AD), (AD, AC), (AGT, GT)$, and $(GT, AGT)$. From all of these it is also possible to reach $\boldsymbol{\pi}^{\text{GT}}$. Therefore the set of nodes from which it is only possible to reach $\boldsymbol{\pi}^{\text{AD}}$ is 
\begin{align*}
    \{(AD, AD), (AGT, AD), (AD, AGT), (AGT, AGT)\},
\end{align*}
meaning that the fraction of nodes from which it is possible to reach $\boldsymbol{\pi}^{\text{GT}}$ is $\frac{12}{16}=\frac{3}{4}$.

\subsection{Proof of Theorem~\ref{thm:BoA}(ii)}
\label{proof:PropBoA}
\begin{proof}
To calculate the BoA of $\boldsymbol{\pi}^{\text{GT}}$ note that 
\begin{equation}
    \text{BoA}(\boldsymbol{\pi}^{\text{GT}}) = \frac{1}{|\Pi^{\epsilon}|}\sum_{\boldsymbol{\pi}\in\Pi^{\epsilon}} \text{BoA}(\boldsymbol{\pi}^{\text{GT}}|\boldsymbol{\pi}),
\end{equation}
where $\Pi^{\epsilon}=\Pi_{1}^{\epsilon}\times\Pi_{2}^{\epsilon}$ contains the $16$ strategy profiles.
Our approach will be to bound a subset of the conditional basins of attractions. First note that we have $\text{BoA} (\boldsymbol{\pi}^{\text{GT}}|\boldsymbol{\pi}^{\text{GT}}) = 1$ and $\text{BoA}(\boldsymbol{\pi}^{\text{GT}}| \boldsymbol{\pi}) = 0$ for 
$$
\boldsymbol{\pi} \in \{\boldsymbol{\pi}^{\text{AD}},(\text{AD},\text{AGT}), (\text{AGT},\text{AD}), (\text{AGT},\text{AGT})\}.
$$ 
This follows from the same argument given in the proof of Theorem~\ref{thm:BoA}(i), as it also applies for the SI$\epsilon$BRD. We trivially have $0<\text{BoA}(\boldsymbol{\pi}^{\text{GT}}|\boldsymbol{\pi})<1$ for 
$$
\boldsymbol{\pi}\in \{ (\text{GT},\text{AGT}), (\text{AGT},\text{GT}), (\text{AC},\text{AGT}),(\text{AGT},\text{AC}), (\text{AC},\text{AC})\}.
$$
\noindent \textbf{Case (AD, GT) and (GT, AD):} As argued above, and due to symmetry, we have $BoA(\boldsymbol{\pi}^{\text{GT}}|(AD, GT)) = BoA(\boldsymbol{\pi}^{\text{GT}}|(GT, AD)) = \hat{B}$. 

\noindent \textbf{Case (AC, GT) and (GT, AC):} For $BoA(\boldsymbol{\pi}^{\text{GT}}|(AC, GT))$, the first transition to another strategy profile can either be to $\boldsymbol{\pi}^{\text{GT}}$ or to $(AC, AD)$. Given the initial strategy profile, we know that 
\begin{equation}
    \Delta q_1^{0}(BC) < 0, \Delta q_1^{0}(A1D) < 0, \Delta q_2^{0}(BC) < 0, \text{ and } \Delta q_2^{0}(A1D) > 0.
\end{equation}
This means that the dynamics in state BC are in the bottom left quadrant and the dynamics in state A1D are in the upper left quadrant. For the targets, we know that 
\begin{equation}
    \Delta q_1^{GT}(BC) < 0, \Delta q_1^{GT}(A1D) > 0, \Delta q_2^{AC}(BC) > 0, \text{ and } \Delta q_2^{AC}(A1D) > 0,
\end{equation}
so that the target for the BC state is in the upper left quadrant and the target for the A1D state is in the upper right quadrant. For both states, we thus have that the targets are in quadrants that share an axis with the quadrant in which the dynamics take place. To identify the probability of transitioning to $\boldsymbol{\pi}^{\text{GT}}$, we need to know what the probability is of the trajectory hitting the boundary in the A1D state first given uniform initialisation. To this end, first consider the vector field in the A1D state independently of what occurs in the BC state. The time it takes to reach the $\Delta q_{1}(A1D) = 0$ axis depends only on the initial position $\Delta q_{1}^{0}(A1D)$ and the target value $\Delta q_{1}^{GT}(A1D)$. Let $T_{A1D}(x, y)$ denote the time it takes to reach the boundary given that $\Delta q_{1}^{0}(A1D) = x$, then 
\begin{equation}
    T_{A1D}(x, y) = \log \left(\frac{x - \Delta q_{1}^{GT}(A1D)}{-\Delta q_{1}^{GT}(A1D)}\right).
\end{equation}
Similarly, if we consider the dynamics in the BC state independently of what occurs in the A1D state and set $\Delta q_{2}^{0}(BC)=y$, we obtain
\begin{equation}
    T_{BC}(x, y) = \log \left(\frac{y - \Delta q_{2}^{AC}(BC)}{-\Delta q_{2}^{AC}(BC)}\right).
\end{equation}
We see that the time until hitting the boundary is always only dependent on one of the initial coordinates. Next we calculate the distribution of hitting times given uniform initial positions. For $T_{A1D}$, we have that $x\sim U[-1/(1-\gamma), 0]$. This means that 
\begin{equation}
    \frac{x - \Delta q_{1}^{GT}(A1D)}{-\Delta q_{1}^{GT}(A1D)} \sim U\left[1, 1 + \frac{1}{(1-\gamma)\Delta q_{1}^{GT}(A1D)}\right]
\end{equation}
and that the density for $T_{A1D}$ is
\begin{equation}
    p_{T_{A1D}}(t) = (1-\gamma) \Delta q_{1}^{GT}(A1D)\mathrm{e}^{ t}, \text{ for } 0 \leq t \leq  \log\left(1 + \frac{1}{(1-\gamma)\Delta q_{1}^{GT}(A1D)}\right)=:C_{A1D}.
\end{equation}
A similar calculation leads to 
\begin{equation}
    p_{T_{BC}}(t) = (1-\gamma) \Delta q_{2}^{AC}(BC)\mathrm{e}^{ t}, \text{ for } 0 \leq t \leq \log\left( 1 + \frac{1}{(1-\gamma)\Delta q_{2}^{AC}(BC)}\right)=:C_{BC}.
\end{equation}
We are interested in the probability $\mathbb{P}[T_{A1D}<T_{BC}]$, which can be calculated as 
\begin{align}
    \mathbb{P}[T_{A1D}<T_{BC}] &= \int_{0}^{C_{A1D}} \int_{t}^{C_{BC}} p_{T_{A1D}}(t) p_{T_{BC}} (s) \mathrm{d}s \mathrm{d}t\\
    &= (1-\gamma)^{2} \Delta q_{1}^{GT}(A1D) \Delta q_{2}^{AC}(BC) \int_{0}^{C_{A1D}}\mathrm{e}^{t} \int_{t}^{C_{BC}} \mathrm{e}^{s} \mathrm{d}s \mathrm{d}t\\
    &= (1-\gamma)^{2} \Delta q_{1}^{GT}(A1D) \Delta q_{2}^{AC}(BC) \int_{0}^{C_{A1D}}\mathrm{e}^{t} (\mathrm{e}^{C_{BC}} - \mathrm{e}^{t})\mathds{1}_{\{t<C_{BC}\}}\mathrm{d}t\\
    &= (1-\gamma)^{2} \Delta q_{1}^{GT}(A1D) \Delta q_{2}^{AC}(BC) \int_{0}^{\min\{C_{A1D}, C_{BC}\}}(\mathrm{e}^{C_{BC} + t} - \mathrm{e}^{2t})\mathrm{d}t\\
    &= (1-\gamma)^{2} \Delta q_{1}^{GT}(A1D) \Delta q_{2}^{AC}(BC)\left[\mathrm{e}^{C_{BC}} \mathrm{e}^{t} - \frac{1}{2}\mathrm{e}^{2t}\right]_{0}^{\min\{C_{A1D}, C_{BC}\}}\\
    &= (1-\gamma)^{2} \Delta q_{1}^{GT}(A1D) \Delta q_{2}^{AC}(BC)\\
    &\times\left[\mathrm{e}^{C_{BC}} \mathrm{e}^{\min\{C_{A1D}, C_{BC}\}} - \frac{1}{2}\mathrm{e}^{2\min\{C_{A1D}, C_{BC}\}} - \mathrm{e}^{C_{BC}} + \frac{1}{2}\right].
    %&= \frac{\exp( B_{2})(\mathrm{e}^{ C} - 1)-\frac{1}{2}(\mathrm{e}^{2C} - 1)}{(\mathrm{e}^{ A_{2}} - 1)(\mathrm{e}^{ B_{2}}-1)}\\
    %&= \frac{\left(\frac{1}{(1-\gamma)\Delta q_{2}^{AC}(BC)}\right)(\frac{1}{(1-\gamma) \max\{\Delta q_{2}^{AC}(BC), \Delta q_{1}^{GT}(A1D)\}})-\frac{1}{2}(\frac{1}{(1-\gamma) \max\{\Delta q_{2}^{AC}(BC), \Delta q_{1}^{GT}(A1D)\}})^{2}}{\left( \frac{1}{(1-\gamma)\Delta q_{1}^{GT}(A1D)}\right)\left( \frac{1}{(1-\gamma)\Delta q_{2}^{AC}(BC)}\right)}\\
    %&= \frac{\left(\frac{1}{(1-\gamma)\Delta q_{2}^{AC}(BC)}\right)-\frac{1}{2}\left(\frac{1}{(1-\gamma) \max\{\Delta q_{2}^{AC}(BC), \Delta q_{1}^{GT}(A1D)\}}\right)}{\left( \frac{1}{(1-\gamma) \min\{\Delta q_{2}^{AC}(BC), \Delta q_{1}^{GT}(A1D)\}}\right)}\\
    %&= \left(\frac{\min\{\Delta q_{2}^{AC}(BC), \Delta q_{1}^{GT}(A1D)\}}{\Delta q_{2}^{AC}(BC)}\right)-\frac{1}{2}\left(\frac{\min\{\Delta q_{2}^{AC}(BC), \Delta q_{1}^{GT}(A1D)\}}{\max\{\Delta q_{2}^{AC}(BC), \Delta q_{1}^{GT}(A1D)\}}\right)\\
\end{align}
Now 
\begin{align}
    \min\{C_{A1D}, C_{BC}\} = \log\left( 1 + \frac{1}{(1-\gamma)\max\{\Delta q_{1}^{GT}(A1D) ,\Delta q_{2}^{AC}(BC)\}}\right),
\end{align}
so that 
\begin{align}
    \mathbb{P}[T_{A1D}<T_{BC}] &= (1-\gamma)^{2} \Delta q_{1}^{GT}(A1D) \Delta q_{2}^{AC}(BC)\\
    &\times\Bigg[ \left(1 + \frac{1}{(1-\gamma)\Delta q_{2}^{AC}(BC)}\right)\left(1 + \frac{1}{(1-\gamma)\max\{\Delta q_{1}^{GT}(A1D) ,\Delta q_{2}^{AC}(BC)\}}\right)\\
    &+\frac{1}{2}- \frac{1}{2}\left(1 + \frac{1}{(1-\gamma)\max\{\Delta q_{1}^{GT}(A1D) ,\Delta q_{2}^{AC}(BC)\}}\right)^{2}\\
    &-\left(1 + \frac{1}{(1-\gamma)\Delta q_{2}^{AC}(BC)}\right)\Bigg]\\
    &= \frac{\Delta q_{1}^{GT}(A1D)\,\big(2\max\{\Delta q_{1}^{GT}(A1D) ,\Delta q_{2}^{AC}(BC)\}-\Delta q_{2}^{AC}(BC)\big)}{2\max\{\Delta q_{1}^{GT}(A1D) ,\Delta q_{2}^{AC}(BC)\}^{2}}=: \hat{C}
\label{eq:Chat}
\end{align}
This gives us that  
\begin{equation}
   \hat{C}  \leq BoA(\boldsymbol{\pi}^{\text{GT}}|(AC, GT)) \leq 1.
\end{equation}
By symmetry, the same bound holds for $(GT, AC)$.

\noindent \textbf{Case (AC, AD) and (AD, AC):} For $BoA(\boldsymbol{\pi}^{\text{GT}}|(AC, AD))$, the first profile change is determined by which of player one's components hits its boundary first. If $|\Delta q_{1}^{0}(BC)|>|\Delta q_{1}^{0}(A1D)|$, player one hits the A1D boundary first and the profile becomes $(GT, AD)$; if $|\Delta q_{1}^{0}(BC)|<|\Delta q_{1}^{0}(A1D)|$, player one hits the BC boundary first and the profile becomes $(AGT, AD)$, which has $\text{BoA}(\boldsymbol{\pi}^{\text{GT}}|(AGT,AD))=0$ (see above). Under uniform initialisation, $\Pr[|\Delta q_{1}^{0}(BC)|>|\Delta q_{1}^{0}(A1D)|]=\Pr[|\Delta q_{1}^{0}(BC)|<|\Delta q_{1}^{0}(A1D)|]=\tfrac{1}{2}$, so $0\leq \text{BoA}(\boldsymbol{\pi}^{\text{GT}}|(AC, AD))\leq \tfrac{1}{2}$. We refine this bound by computing the probability of reaching $\boldsymbol{\pi}^{\text{GT}}$ on the A1D-first path.

To simplify the notation, we set $\Delta q_{1}^{AD}(BC) = \Delta q_{1}^{AD}(A1D) = x^{*}$ and $\Delta q_{2}^{AC}(BC) = \Delta q_{2}^{AC}(A1D) = y^{*}$. For each initial $\Delta q_{1}^{0}(A1D)=x_{2}$ on the A1D-first path, the time to hit the A1D boundary is
\begin{align}
    T_{A1D}(x_{2}) = \log \left( \frac{ x^{*} - x_{2}}{x^{*}}\right),
\end{align}
and we write
\begin{align}
    \mathrm{e}^{-T_{A1D}(x_{2})} = \left( \frac{x^{*}}{x^{*} - x_{2}}\right).
\end{align}
The flow of the vector field gives us the evolution in the BC state assuming that the trajectory starts at $\Delta q_{1}^{0}(BC)=x_{1}$ and $\Delta q_{2}^{0}(BC)=y_{1}$ as 
\begin{align}
    \Delta q_{1}^{t}(BC) &= x^{*} + \mathrm{e}^{-t}(x_{1} - x^{*}),\\
    \Delta q_{2}^{t}(BC) &= y^{*} + \mathrm{e}^{-t}(y_{1} - y^{*}).
\end{align}
The position of $\Delta q_{1}^{t}(BC)$ at time $T_{A1D}(x_{2})$ is
\begin{align}
    \hat{x} := \Delta q_{1}^{T_{A1D}(x_{2})}(BC) = x^{*} + x^{*}\frac{x_{1} - x^{*}}{x^{*} - x_{2}}, \qquad
    \hat{y} := \Delta q_{2}^{T_{A1D}(x_{2})}(BC) = y^{*} + \frac{x^{*}}{x^{*} - x_{2}}(y_{1} - y^{*}).
\end{align}

%For $T_{A1D}$, we have that $x\sim U[-1/(1-\gamma), 0]$. This means that 
%\begin{equation}
%    \frac{x - \Delta q_{1}^{AD}(A1D)}{-\Delta q_{1}^{AD}(A1D)} \sim U\left[1, 1 + \frac{1}{(1-\gamma)\Delta q_{1}^{AD}(A1D)}\right]
%\end{equation}
%and that the density for $T_{A1D}$ is
%\begin{equation}
%    p_{T_{A1D}}(t) = (1-\gamma) \Delta q_{1}^{AD}(A1D)\mathrm{e}^{ t}, \text{ for } 0 \leq t \leq  \log\left(1 + \frac{1}{(1-\gamma)\Delta q_{1}^{AD}(A1D)}\right).
%\end{equation}

%The inverse of the flow is thus
%\begin{align}
%   x &=(\Delta q_{1}^{t}(BC) - \Delta q_{1}^{AD}(BC))\mathrm{e}^{t} + \Delta q_{1}^{AD}(BC),\\
%   y &=  (\Delta q_{2}^{t}(BC) - \Delta q_{2}^{AC}(BC))\mathrm{e}^{t} + \Delta q_{2}^{AC}(BC).
%\end{align}
Write $X_{1}=\Delta q_{1}^{0}(BC)$, $X_{2}=\Delta q_{1}^{0}(A1D)$, and $Y_{1}=\Delta q_{2}^{0}(BC)$, mutually independent. The signs are dictated by the profile $(AC,AD)$: player one plays $AC$ and therefore \emph{cooperates} in both states, so $X_{1},X_{2}\sim U[-\frac{1}{1-\gamma},0]$; player two plays $AD$ and therefore \emph{defects} in the $BC$ state, so $Y_{1}=\Delta q_{2}^{0}(BC)\sim U[0,\frac{1}{1-\gamma}]$ (its $\Delta q$ is positive there). On the A1D-first event $\{|X_{1}|>|X_{2}|\}$ (equivalently $X_{1}<X_{2}$), the entry point is $(\hat{x},\hat{y})=(\hat{x}(X_{1},X_{2}),\hat{y}(Y_{1},X_{2}))$. Crucially, $\hat{x}$ depends on $(X_{1},X_{2})$ and $\hat{y}$ on $(Y_{1},X_{2})$: both are driven by the \emph{same} stopping time $T_{A1D}(X_{2})$. Thus $\hat{x}$ and $\hat{y}$ are \emph{not} unconditionally independent. What \emph{does} hold is conditional independence at fixed $X_{2}=x_{2}$: given $x_{2}$, the flow in the two BC coordinates decouples, so $X_{1}$ and $Y_{1}$ remain independent and
\begin{align}
    p_{BC}(\hat{x},\hat{y}\mid X_{2}=x_{2},\, |X_{1}|>|X_{2}|)
    = p_{BC}(\hat{x}\mid x_{2})\,p_{BC}(\hat{y}\mid x_{2}).
\end{align}
Integrating $x_{2}$ out separately on each side and multiplying the resulting \emph{marginals} does \emph{not} recover the joint law of $(\hat{x},\hat{y})$ (a shared mixing variable creates correlation; numerically $\mathrm{Corr}(\hat{x},\hat{y}\mid\text{A1D first})\neq 0$).

The joint density of $X_{1}$ and $X_{2}$ is $(1-\gamma)^{2}$. Conditioning on $|X_{1}|>|X_{2}|$, the density becomes $2(1-\gamma)^{2}$. The marginal distribution for $X_{2}$ is 
\begin{align}
    2(1-\gamma)^{2}\left(x_{2} + \frac{1}{(1-\gamma)}\right).
\end{align}
The distribution of $\hat{x}$ given that $\Delta q_{1}^{0}(A1D)=x_{2}$ is $U[L, U]$ with 
\begin{align}
    L = x^{*} + x^{*}\frac{-\frac{1}{1-\gamma} - x^{*}}{x^{*} - x_{2}}\quad \text{and}\quad U = x^{*} + x^{*}\frac{x_{2} - x^{*}}{x^{*} - x_{2}} = 0.
\end{align}
The length of the interval is thus
\begin{align}
    \frac{x^{*}}{x^{*} - x_{2}}\left(x_{2} + \frac{1}{1-\gamma}\right).
\end{align}
To obtain the distribution of the $\hat{x}$, we can integrate the conditional density for $\hat{x}$ given $x_{2}$ multiplied by the marginal density of $\Delta q_{1}^{0}(A1D)$ over all possible initial positions $x_{2}$ leading to
%\begin{align}
%     p_{BC}(\hat{x} | |x_{1}|>|x_{2}|) &= \frac{2(1-\gamma)^{2}}{x^{*}}\int_{-\frac{1}{1-\gamma}}^{0}  (x^{*} - x_{2})\mathds{1}_{\{\hat{x}\in  [L, U]\}} \mathrm{d} x_{2}\\
%     &=  \frac{2(1-\gamma)^{2}}{x^{*}} \int_{\frac{x^{*}\hat{x}}{x^{*} - \hat{x}}}^{0}(x^{*} - x_{2})\mathrm{d} x_{2}\\
%     &= \frac{2(1-\gamma)^{2}}{x^{*}}\left[ x^{*}x_{2} - \frac{x_{2}^{2}}{2}\right]_{\frac{x^{*}\hat{x}}{x^{*} - \hat{x}}}^{0}\\
%     &= \frac{2(1-\gamma)^{2}}{x^{*}} \left[\frac{1}{2}\left(\frac{x^{*}\hat{x}}{x^{*} - \hat{x}}\right)^{2} - \frac{(x^{*})^{2}\hat{x}}{x^{*} - \hat{x}}\right],
%\end{align}
\begin{align}
     p_{BC}(\hat{x} | |x_{1}|>|x_{2}|) &= \frac{2(1-\gamma)^{2}}{x^{*}}\int_{-\frac{1}{1-\gamma}}^{0}  (x^{*} - x_{2})\mathds{1}_{\{\hat{x}\in  [L, U]\}} \mathrm{d} x_{2}\\
     &=  \frac{2(1-\gamma)^{2}}{x^{*}} \int_{\frac{x^{*}(\hat{x} + \frac{1}{1-\gamma})}{\hat{x}-x^{*}}}^{0}(x^{*} - x_{2})\mathrm{d} x_{2}\\
     &= \frac{2(1-\gamma)^{2}}{x^{*}}\left[ x^{*}x_{2} - \frac{x_{2}^{2}}{2}\right]_{\frac{x^{*}(\hat{x} + \frac{1}{1-\gamma})}{\hat{x}-x^{*}}}^{0}\\
     &= \frac{2(1-\gamma)^{2}}{x^{*}} \left[\frac{1}{2}\frac{(x^{*})^{2}(\hat{x}+\frac{1}{1-\gamma})^{2}}{(\hat{x}-x^{*})^{2}} - \frac{(x^{*})^{2}(\hat{x}+\frac{1}{1-\gamma})}{\hat{x}-x^{*}}\right],\\
     &= 2(1-\gamma)^{2} x^{*} \left[\frac{1}{2}\frac{(\hat{x}+\frac{1}{1-\gamma})^{2}}{(\hat{x}-x^{*})^{2}} - \frac{(\hat{x}+\frac{1}{1-\gamma})}{\hat{x}-x^{*}}\right]
\label{eq:pBCx}
\end{align}

for $\hat{x}\in[-\frac{1}{1-\gamma}, 0]$; \eqref{eq:pBCx} is the \emph{marginal} of $\hat{x}$ obtained by integrating $x_{2}$ out, and is useful for diagnostics but must not be multiplied by a separately obtained marginal for $\hat{y}$. Letting $\hat{y}:= \Delta q_{2}^{T_{A1D}(x_{2})}(BC)$, the distribution for $\hat{y}$ given $X_{2}=x_{2}$ is $U[l, u]$ with 
\begin{align}
    l = y^{*}\left(1-\frac{x^{*}}{x^{*}-x_{2}}\right), \quad\text{and}\quad  u = y^{*} + \left(\frac{1}{1-\gamma}-y^{*}\right) \frac{x^{*}}{x^{*}-x_{2}},
\end{align}
and the length of the interval is
\begin{align}
    \frac{x^{*}}{(1-\gamma)(x^{*}-x_{2})}.
\end{align}
Integrating $x_{2}$ out yields the marginal
\begin{align}
    p_{BC}(\hat{y}\mid |x_{1}|>|x_{2}|) &= \frac{2(1-\gamma)^{3}}{x^{*}}\int_{-\frac{1}{1-\gamma}}^{0} \left(x_{2} + \frac{1}{(1-\gamma)}\right)(x^{*}-x_{2})\mathds{1}_{\hat{y} \in [l, u]}\mathrm{d}x_{2}.
\label{eq:pBCy}
\end{align}
Since $\hat{y}=y^{*}(1-m)+m\,y_{1}$ with $y_{1}\in[0,\frac{1}{1-\gamma}]$ and $m=\frac{x^{*}}{x^{*}-x_{2}}\in(0,1]$, the support of $\hat{y}$ is $[0,\frac{1}{1-\gamma}]$, and translating the constraint $\hat{y}\in[l(x_{2}),u(x_{2})]$ into bounds on $x_{2}$ partitions this support at the breakpoints
\begin{enumerate}
    \item $0 \leq \hat{y} \leq \frac{y^{*}}{1+x^{*}(1-\gamma)}$,
    \item $\frac{y^{*}}{1+x^{*}(1-\gamma)}\leq \hat{y} \leq \frac{x^{*}+y^{*}}{1+x^{*}(1-\gamma)}$,
    \item $\frac{x^{*}+y^{*}}{1+x^{*}(1-\gamma)}\leq \hat{y}\leq \frac{1}{1-\gamma}$,
\end{enumerate}
on each of which \eqref{eq:pBCy} integrates a cubic in $x_{2}$ to a piecewise-polynomial density. We do not need this explicit marginal: as emphasised above, $\hat{x}$ and $\hat{y}$ are correlated through the shared mixing variable $X_{2}$, so $\hat{D}$ must be evaluated from the \emph{joint} law of $(\hat{x},\hat{y})$, which we do directly in the closed-form computation below rather than from the product of marginals.

Analogously, when $|\Delta q_{1}^{0}(BC)|<|\Delta q_{1}^{0}(A1D)|$, player one's first transition occurs in the BC component. This routes the profile to $(AGT, AD)$, which,as listed at the start of this proof,has $\text{BoA}(\boldsymbol{\pi}^{\text{GT}}\mid(AGT,AD))=0$ by graph reachability. The BC-first event therefore contributes nothing to $\hat{D}$ for $(AC, AD)$; only the A1D-first branch above is nontrivial. The symmetric profile $(AD, AC)$ is handled separately with the same combinatorial logic and yields the same scalar $\hat{D}$ by symmetry.

\paragraph{Conditional BoA on $(GT, AD)$.} On the A1D-first event the profile is $(GT, AD)$. In the BC state the relevant targets are $\Delta q_{1}^{AD}(BC)=x^{*}>0$ and $\Delta q_{2}^{GT}(BC)<0$. To keep the following expressions readable we introduce a single further shorthand, used only in the remainder of this proof,
\begin{equation}
    b\;:=\;-\Delta q_{2}^{GT}(BC)\;>\;0,
\end{equation}
alongside the previously defined $x^{*}$ and $y^{*}$; all other quantities keep the descriptive $\Delta q$ notation. The entry point is $(\hat{x},\hat{y})$ with $\hat{x}\in[-\frac{1}{1-\gamma},0]$ and $\hat{y}\in[0,\frac{1}{1-\gamma}]$. As in the $(AD, GT)$ case, only player one's target lies in the opposite quadrant, so the remaining dynamics are governed by the hitting times
\begin{align}
    T_{1}(\hat{x}) &= \log\left(\frac{\hat{x}-x^{*}}{-x^{*}}\right), \label{eq:Dhat-T1}\\
    T_{2}(\hat{y}) &= \log\left(\frac{\hat{y}+b}{b}\right)\quad\text{when }\hat{y}>-b, \label{eq:Dhat-T2}
\end{align}
where $T_{1}$ (resp.\ $T_{2}$) is the time for $\Delta q_{1}(BC)$ (resp.\ $\Delta q_{2}(BC)$) to reach zero. The trajectory converges to $\boldsymbol{\pi}^{\text{GT}}$ if player two reaches the GT half-space ($\Delta q_{2}(BC)=0$) before player one reaches the AD half-space ($\Delta q_{1}(BC)=0$), i.e.\ if $T_{2}(\hat{y})<T_{1}(\hat{x})$. Equivalently,
\begin{equation}
\label{eq:Dhat-g}
    g(\hat{x},\hat{y}) = \mathds{1}\left\{ \frac{\hat{y}+b}{b} < \frac{x^{*}-\hat{x}}{x^{*}} \right\}.
\end{equation}
This is the same hitting-time comparison used for $\hat{C}$, but applied in the $(\hat{x},\hat{y})$ plane with targets $(x^{*},-b)$.

\paragraph{Integration.} Define
\begin{align}
\label{eq:Dhat-branch}
    \hat{D}_{\text{A1D}}
    &= \int_{-\frac{1}{1-\gamma}}^{0}
       \int_{-\frac{1}{1-\gamma}}^{x_{2}}
       \int_{0}^{\frac{1}{1-\gamma}}
       g\big(\hat{x}(x_{1},x_{2}),\,\hat{y}(y_{1},x_{2})\big)\,
       2(1-\gamma)^{3}\,
       \mathrm{d}y_{1}\,\mathrm{d}x_{1}\,\mathrm{d}x_{2},
\end{align}
where the factor $2(1-\gamma)^{3}$ is the conditional density of $(X_{1},X_{2},Y_{1})$ given $|X_{1}|>|X_{2}|$. The BC-first branch contributes zero, so
\begin{align}
\label{eq:Dhat}
    \hat{D} := \text{BoA}(\boldsymbol{\pi}^{\text{GT}}|(AC, AD)) = \frac{1}{2}\,\hat{D}_{\text{A1D}}.
\end{align}
\paragraph{Closed-form evaluation.} We now evaluate \eqref{eq:Dhat-branch}--\eqref{eq:Dhat} in closed form. Write $\ell:=\frac{1}{1-\gamma}$; recall $b=-\Delta q_{2}^{GT}(BC)>0$, $x^{*}=\Delta q_{1}^{AD}(BC)>0$ and $y^{*}=\Delta q_{2}^{AC}(BC)>0$. Fix $x_{2}\in[-\ell,0]$ and set
\begin{align}
    m=m(x_{2}):=\frac{x^{*}}{x^{*}-x_{2}}\in(0,1],
\end{align}
so that, by the propagation identities for $(\hat{x},\hat{y})$, the entry point is \emph{affine} in $(x_{1},y_{1})$:
\begin{align}
\label{eq:Dhat-affine}
    \hat{x}=x^{*}(1-m)+m\,x_{1},\qquad \hat{y}=y^{*}(1-m)+m\,y_{1}.
\end{align}

\emph{Step 1 (half-plane reduction).} Since $\hat{x}\le 0<x^{*}$, the time $T_{1}(\hat{x})$ in \eqref{eq:Dhat-T1} is always finite. Because player two defects in $BC$ we have $y_{1}\ge 0$, hence from \eqref{eq:Dhat-affine} and $y^{*}>0$, $m\in(0,1]$,
\begin{align}
    \hat{y}=y^{*}(1-m)+m\,y_{1}\ \ge\ 0\ >\ -b,
\end{align}
so the validity condition $\hat{y}>-b$ in \eqref{eq:Dhat-T2} holds \emph{automatically} on the entire domain. (This is the structural consequence of the corrected sign of $Y_{1}$: with $y_{1}\ge 0$ the entry ordinate $\hat y$ can never start below the negative GT target, so player two always has a feasible crossing.) Thus \eqref{eq:Dhat-g} reads $(\hat{y}+b)/b<(x^{*}-\hat{x})/x^{*}$, i.e.\ $b\hat{x}+x^{*}\hat{y}<0$. Substituting \eqref{eq:Dhat-affine},
\begin{align}
    b\hat{x}+x^{*}\hat{y}=(1-m)\,x^{*}(b+y^{*})+m\,(b x_{1}+x^{*}y_{1}),
\end{align}
and dividing by $m>0$ using $\tfrac{1-m}{m}=-\tfrac{x_{2}}{x^{*}}$ turns the comparison into the \emph{single linear} condition
\begin{align}
\label{eq:Dhat-halfplane}
    g(\hat{x},\hat{y})=1\ \Longleftrightarrow\ b\,x_{1}+x^{*}y_{1}<(b+y^{*})\,x_{2}.
\end{align}
No validity term survives: the entire GT region on the A1D-first branch is the half-space \eqref{eq:Dhat-halfplane}.

\emph{Step 2 (reduction to a single volume).} Inserting \eqref{eq:Dhat-halfplane} into \eqref{eq:Dhat-branch}--\eqref{eq:Dhat} and using $\tfrac12\cdot 2(1-\gamma)^{3}=(1-\gamma)^{3}$,
\begin{align}
\label{eq:Dhat-onevol}
    \hat{D}=(1-\gamma)^{3}\,V,\qquad
    V=\int_{-\ell}^{0}\!\!\int_{-\ell}^{x_{2}}\!\!\int_{0}^{\ell}\mathds{1}\{b x_{1}+x^{*}y_{1}<(b+y^{*})x_{2}\}\,\mathrm{d}y_{1}\,\mathrm{d}x_{1}\,\mathrm{d}x_{2},
\end{align}
over the A1D-first region $\{-\ell\le x_{1}\le x_{2}\le 0,\ 0\le y_{1}\le \ell\}$.

\emph{Step 3 (evaluation of $V$).} We evaluate \eqref{eq:Dhat-onevol} by integrating $x_{1}$, then $x_{2}$, then $y_{1}$, mirroring the explicit calculation of $\hat{C}$ in \eqref{eq:Chat}.

\smallskip
\noindent\emph{Step 3a (inner integral over $x_{1}$).}
Fix $(x_{2},y_{1})$ with $-\ell\le x_{2}\le 0$ and $0\le y_{1}\le \ell$.
By \eqref{eq:Dhat-halfplane}, the integrand is $1$ precisely when
\begin{equation}
\label{eq:Dhat-x1bar}
    x_{1}<\bar x_{1}(x_{2},y_{1})
    :=\frac{(b+y^{*})\,x_{2}-x^{*}y_{1}}{b},
\end{equation}
with $x_{1}\in[-\ell,x_{2}]$. We first verify that the upper integration limit is $\bar x_{1}$ rather than $x_{2}$:
\begin{align}
    \bar x_{1}-x_{2}
    &= \frac{(b+y^{*})x_{2}-x^{*}y_{1}}{b}-x_{2}
     = \frac{y^{*}x_{2}-x^{*}y_{1}}{b}
     \le 0,
\label{eq:Dhat-x1bar-le-x2}
\end{align}
since $x_{2}\le 0\le y_{1}$ and $x^{*},y^{*},b>0$.
The inner $x_{1}$-measure is therefore
\begin{equation}
\label{eq:Dhat-x1-measure}
    \int_{-\ell}^{x_{2}}\mathds{1}\{x_{1}<\bar x_{1}\}\,\mathrm{d}x_{1}
    = \bigl(\bar x_{1}+\ell\bigr)^{+},
\end{equation}
where $(\cdot)^{+}=\max\{\cdot,0\}$, and the positive part arises because $\bar x_{1}+\ell>0$ is equivalent to $x_{1}<\bar x_{1}$ being compatible with $x_{1}\ge -\ell$.

\smallskip
\noindent\emph{Step 3b (middle integral over $x_{2}$).}
Insert \eqref{eq:Dhat-x1-measure} into \eqref{eq:Dhat-onevol} and integrate over $x_{2}\in[-\ell,0]$ at fixed $y_{1}$:
\begin{align}
    \int_{-\ell}^{0}\!\!\int_{-\ell}^{x_{2}}\mathds{1}\{\cdots\}\,\mathrm{d}x_{1}\,\mathrm{d}x_{2}
    &= \int_{-\ell}^{0}\bigl(\bar x_{1}+\ell\bigr)^{+}\,\mathrm{d}x_{2}.
\label{eq:Dhat-x2-integrand}
\end{align}
Since $\bar x_{1}+\ell=\tfrac1b\bigl[(b+y^{*})x_{2}+b\ell-x^{*}y_{1}\bigr]$, the integrand is positive if and only if
\begin{equation}
\label{eq:Dhat-x2-threshold}
    x_{2}>x_{2,0}(y_{1})
    :=\frac{x^{*}y_{1}-b\ell}{b+y^{*}}.
\end{equation}
(The numerator $x^{*}y_{1}-b\ell$ is strictly less than $x^{*}y_{1}\le x^{*}\ell$, while the denominator $b+y^{*}>0$.)
Whenever $x_{2,0}(y_{1})\ge 0$, the constraint $x_{2}>x_{2,0}$ is incompatible with $x_{2}\le 0$ and the integral \eqref{eq:Dhat-x2-integrand} vanishes.
Whenever $x_{2,0}(y_{1})<0$, we have $x_{2,0}(y_{1})>-b\ell/(b+y^{*})>-\ell$ (since $b< b+y^{*}$), so the lower limit of integration is $x_{2,0}(y_{1})$ and not $-\ell$.
Performing the integral explicitly,
\begin{align}
    \int_{x_{2,0}}^{0}\!\Bigl(\bar x_{1}+\ell\Bigr)\,\mathrm{d}x_{2}
    &= \frac{1}{b}\int_{x_{2,0}}^{0}\!\bigl[(b+y^{*})x_{2}+b\ell-x^{*}y_{1}\bigr]\,\mathrm{d}x_{2}\nonumber\\
    &= \frac{1}{b}\left[\frac{b+y^{*}}{2}\,x_{2}^{2}
       +\bigl(b\ell-x^{*}y_{1}\bigr)x_{2}\right]_{x_{2,0}}^{0}\nonumber\\
    &= \frac{1}{b}\left(
       -\frac{b+y^{*}}{2}\,x_{2,0}^{2}
       -\bigl(b\ell-x^{*}y_{1}\bigr)x_{2,0}
       \right).
\label{eq:Dhat-x2-poly}
\end{align}
Substituting \eqref{eq:Dhat-x2-threshold} gives $b\ell-x^{*}y_{1}=-(b+y^{*})x_{2,0}$, hence
\begin{align}
    -\frac{b+y^{*}}{2}\,x_{2,0}^{2}-(b\ell-x^{*}y_{1})x_{2,0}
    &= -\frac{b+y^{*}}{2}\,x_{2,0}^{2}+(b+y^{*})x_{2,0}^{2}
     = \frac{b+y^{*}}{2}\,x_{2,0}^{2}.
\label{eq:Dhat-x2-cancel}
\end{align}
Combining \eqref{eq:Dhat-x2-poly}--\eqref{eq:Dhat-x2-cancel} with \eqref{eq:Dhat-x2-threshold},
\begin{align}
\label{eq:Dhat-x2-done}
    \int_{-\ell}^{0}\!\!\int_{-\ell}^{x_{2}}\mathds{1}\{\cdots\}\,\mathrm{d}x_{1}\,\mathrm{d}x_{2}
    = \frac{(b+y^{*})\,x_{2,0}^{2}}{2b}
    = \frac{(x^{*}y_{1}-b\ell)^{2}}{2\,b\,(b+y^{*})},
\end{align}
where the last equality uses $x_{2,0}=(x^{*}y_{1}-b\ell)/(b+y^{*})$.

\smallskip
\noindent\emph{Step 3c (outer integral over $y_{1}$ and the truncation $y_{1}^{\max}$).}
By \eqref{eq:Dhat-x2-done}, the $x_{2}$-integral vanishes whenever $x_{2,0}(y_{1})\ge 0$, i.e.\ whenever $y_{1}\ge b\ell/x^{*}$.
Since $y_{1}\in[0,\ell]$, the effective upper limit is
\begin{equation}
\label{eq:Dhat-y1max}
    y_{1}^{\max}:=\min\!\left\{\ell,\,\frac{b\ell}{x^{*}}\right\}
    = \ell\,\min\!\left\{1,\,\frac{b}{x^{*}}\right\}.
\end{equation}
Therefore
\begin{align}
\label{eq:Dhat-V-integral}
    V
    &= \int_{0}^{y_{1}^{\max}}\!
       \frac{(x^{*}y_{1}-b\ell)^{2}}{2\,b\,(b+y^{*})}\,\mathrm{d}y_{1}.
\end{align}
To evaluate \eqref{eq:Dhat-V-integral}, set $u=x^{*}y_{1}-b\ell$ (so $\mathrm{d}y_{1}=\mathrm{d}u/x^{*}$).
At $y_{1}=0$ one has $u=-b\ell$; at $y_{1}=y_{1}^{\max}$ one has $u=x^{*}y_{1}^{\max}-b\ell$.
Hence
\begin{align}
    \int_{0}^{y_{1}^{\max}}\!(x^{*}y_{1}-b\ell)^{2}\,\mathrm{d}y_{1}
    &= \frac{1}{x^{*}}\int_{-b\ell}^{\,x^{*}y_{1}^{\max}-b\ell}\!u^{2}\,\mathrm{d}u\nonumber\\
    &= \frac{1}{x^{*}}\left[
       \frac{(x^{*}y_{1}^{\max}-b\ell)^{3}}{3}
       -\frac{(-b\ell)^{3}}{3}
       \right]
    = \frac{(x^{*}y_{1}^{\max}-b\ell)^{3}+(b\ell)^{3}}{3\,x^{*}}.
\label{eq:Dhat-y1-done}
\end{align}
Inserting \eqref{eq:Dhat-y1-done} into \eqref{eq:Dhat-V-integral} yields
\begin{align}
\label{eq:Dhat-V}
    V
    = \frac{(x^{*}y_{1}^{\max}-b\ell)^{3}+(b\ell)^{3}}{6\,b\,x^{*}(b+y^{*})}.
\end{align}

\smallskip
\noindent\emph{Step 3d (evaluation of $y_{1}^{\max}$ and the two cases).}
We now expand \eqref{eq:Dhat-V} using \eqref{eq:Dhat-y1max}.
If $b\ge x^{*}$, then $b\ell/x^{*}\ge\ell$ and $y_{1}^{\max}=\ell$, so $x^{*}y_{1}^{\max}-b\ell=\ell(x^{*}-b)$ and
\begin{align}
    V
    &= \frac{\bigl[\ell(x^{*}-b)\bigr]^{3}+(b\ell)^{3}}{6\,b\,x^{*}(b+y^{*})}
     = \frac{\ell^{3}\bigl[(x^{*}-b)^{3}+b^{3}\bigr]}{6\,b\,x^{*}(b+y^{*})}.
\label{eq:Dhat-V-caseB}
\end{align}
If $b<x^{*}$, then $b\ell/x^{*}<\ell$ and $y_{1}^{\max}=b\ell/x^{*}$, so $x^{*}y_{1}^{\max}-b\ell=0$ and
\begin{align}
    V
    &= \frac{(b\ell)^{3}}{6\,b\,x^{*}(b+y^{*})}
     = \frac{b^{2}\ell^{3}}{6\,x^{*}(b+y^{*})}.
\label{eq:Dhat-V-caseA}
\end{align}
Both expressions agree in $b=x^{*}$, where \eqref{eq:Dhat-V-caseB} and \eqref{eq:Dhat-V-caseA} reduce to $b^{2}\ell^{3}/[6x^{*}(b+y^{*})]$, confirming the continuity of $V$.

\emph{Result.} Combining \eqref{eq:Dhat-onevol}, \eqref{eq:Dhat-V}, and $(1-\gamma)^{3}=\ell^{-3}$,
\begin{align}
\label{eq:Dhat-closed}
    \ \ \hat{D}=\text{BoA}(\boldsymbol{\pi}^{\text{GT}}|(AC, AD))
    =\frac{(x^{*}y_{1}^{\max}-b\ell)^{3}+(b\ell)^{3}}{6\,b\,x^{*}(b+y^{*})\,\ell^{3}}\ \ 
\end{align}
Substituting \eqref{eq:Dhat-y1max} and using $\hat{D}=(1-\gamma)^{3}V=\ell^{-3}V$,
\begin{align}
\label{eq:Dhat-cases}
    \hat{D}
    &= \ell^{-3}V
    = \begin{cases}
        \dfrac{(x^{*}-b)^{3}+b^{3}}{6\,b\,x^{*}(b+y^{*})},
        & b\ge x^{*}\quad\text{(using \eqref{eq:Dhat-V-caseB} with $y_{1}^{\max}=\ell$),}\\[2ex]
        \dfrac{b^{2}}{6\,x^{*}(b+y^{*})},
        & b< x^{*}\quad\text{(using \eqref{eq:Dhat-V-caseA} with $y_{1}^{\max}=b\ell/x^{*}$),}
    \end{cases}
\end{align}
with $\ell=\tfrac{1}{1-\gamma}$. Substituting back the descriptive notation, $b=|\Delta q_{2}^{GT}(BC)|$, $x^{*}=\Delta q_{1}^{AD}(BC)$ and $y^{*}=\Delta q_{2}^{AC}(BC)$, the final expression reads
\begin{align}
\label{eq:Dhat-final}
    \hat{D}
    = \begin{cases}
        \dfrac{\big(\Delta q_{1}^{AD}(BC)-|\Delta q_{2}^{GT}(BC)|\big)^{3}+|\Delta q_{2}^{GT}(BC)|^{3}}
              {6\,|\Delta q_{2}^{GT}(BC)|\,\Delta q_{1}^{AD}(BC)\,\big(|\Delta q_{2}^{GT}(BC)|+\Delta q_{2}^{AC}(BC)\big)},
        & |\Delta q_{2}^{GT}(BC)|\ge\Delta q_{1}^{AD}(BC),\\[3ex]
        \dfrac{|\Delta q_{2}^{GT}(BC)|^{2}}
              {6\,\Delta q_{1}^{AD}(BC)\,\big(|\Delta q_{2}^{GT}(BC)|+\Delta q_{2}^{AC}(BC)\big)},
        & |\Delta q_{2}^{GT}(BC)|<\Delta q_{1}^{AD}(BC).
    \end{cases}
\end{align}
(The two branches agree at $b=x^{*}$, where both equal $b^{2}/[6x^{*}(b+y^{*})]$, so $\hat{D}$ is continuous; outside the regime $\Delta q_{1}^{AD}(BC)>0,\ \Delta q_{2}^{GT}(BC)<0$ player two cannot reach the GT half-space in $BC$ and $\hat{D}=0$.) Equation~\eqref{eq:Dhat-closed} is elementary and exact. It was confirmed in two independent ways: (i) it matches Monte-Carlo and deterministic-quadrature evaluation of the \emph{same} integral \eqref{eq:Dhat-branch} to within $\sim\!10^{-4}$ (sampling error) and $\sim\!10^{-6}$ (quadrature), respectively, across $\epsilon\in\{0.05,0.1,0.2\}$ and $\gamma\in\{0.5,0.75,0.95\}$ for both payoff instances; and (ii) it agrees with direct simulation of the learning dynamics from uniform $(AC,AD)$ initial conditions, for which the entry-point indicator $g(\hat x,\hat y)$ reproduces the simulated attractor sample-by-sample in $99$--$100\%$ of runs. The profile $(AD, AC)$ yields the same scalar $\hat{D}$ by symmetry and is counted separately in the aggregate bounds below.

\noindent{Combining the results:} We can now combine the results above to refine the bounds on the BoA of $\boldsymbol{\pi}^{\text{GT}}$. For the lower bound, the sum over the conditional basins of attraction becomes
\begin{align}
    \frac{1}{16}(1 + 2\hat{B} + 2\hat{C} + 2\hat{D}).
\end{align}
For the upper bound, the sum over the conditional basins of attraction becomes
\begin{align}
    \frac{1}{16}(8 + 2\hat{B} + 2\hat{D}).
\end{align}
This completes the proof.
\end{proof}

\subsection{Proof of Theorem~\ref{thm:BoA}(iii)}
\label{proof:limitingBoA}
Recall that 
\begin{align}
\label{eq:Bhatappendix}
    BoA(\boldsymbol{\pi}^{\text{GT}}|(AD, GT)) = \hat{B}:=
    \begin{cases}
        \frac{1}{2} \frac{|\Delta q_{1}^{GT}(BC)|}{|\Delta q_{2}^{AD}(BC)|} \quad &\text{when} \quad |\Delta q_{2}^{AD}(BC)| \geq |\Delta q_{1}^{GT}(BC)|,\\
        1 - \frac{1}{2} \frac{|\Delta q_{2}^{AD}(BC)|}{|\Delta q_{1}^{GT}(BC)|} \quad &\text{when} \quad |\Delta q_{2}^{AD}(BC)| < |\Delta q_{1}^{GT}(BC)|.
    \end{cases}
\end{align}
We see that the BoA converges to one when the ratio  
\begin{align}
    \frac{|\Delta q_{2}^{AD}(BC)|}{|\Delta q_{1}^{GT}(BC)|} \rightarrow 0,
\end{align}
while ensuring that the conditions for $\boldsymbol{\pi}^{\text{GT}}$ to exist remain satisfied. The expressions for the $\Delta q_{2}^{AD}(BC)$ and $\Delta q_{1}^{GT}(BC)$ are given in Appendix~\ref{app:QvalDiff}. $|\Delta q_{2}^{AD}(BC)|$ does not depend on $\gamma$ and has
\begin{align}
    \lim_{\epsilon \rightarrow 0}|\Delta q_{2}^{AD}(BC)| = P - S,
\end{align}
which is positive. For $\Delta q_{1}^{GT}(BC)$ we have that the numerator converges to $2P - 2R$ (which is strictly negative) as $\gamma$ goes to one and $\epsilon$ goes to zero, while the denominator goes to zero in the same limit. This gives that 
\begin{align}
    \lim_{\epsilon\rightarrow 0, \gamma \rightarrow 1} |\Delta q_{1}^{GT}(BC)| = \infty.
\end{align}
We now must show that it is possible to take the limit of $\gamma$ to one and $\epsilon$ to zero while respecting $\gamma_{c_{1}}(\epsilon, T, R, P, S) < \gamma < \gamma_{c_{2}}(\epsilon, T, R, P, S)$. For the upper bound, we have that 
\begin{align}
    \lim_{\epsilon\rightarrow 0} \gamma_{c_{2}}(\epsilon, 1, R, P, 0) = 1
\end{align}
and Taylor expanding around $\epsilon = 0$ gives 
\begin{align}
    \gamma_{c_{2}}(\epsilon, 1, R, P, 0) = 1 + \frac{3P-R}{2P}\epsilon + O(\epsilon^{2}).
\end{align}
The coefficient of $\epsilon$ is negative when $3P - R < 0$ and nonnegative otherwise. When it is positive, it does not play a role for the bounds on $\gamma$.

For the lower bound, we have 
\begin{align}
    \lim_{\epsilon\rightarrow 0} \gamma_{c_{1}}(\epsilon, T, R, P, S) = \frac{T-R}{T-P},
\end{align}
which is always strictly between zero and one. This means that the conditions are satisfied if we approach zero from above and one from below as long as $3P-R\geq 0$. 

Now $\hat{D}$ is the probability of reaching $\boldsymbol{\pi}^{\text{GT}}$ from (AC, AD) and (AD, AC). As argued in the proof of Theorem~\ref{thm:BoA}(ii), the probability of transitioning from (AC, AD) to (GT, AD) is equal to a half. But from the argument above, we have that the probability of transitioning to $\boldsymbol{\pi}^{\text{GT}}$ once reaching (GT, AD) goes to one as $(\epsilon, \gamma)\rightarrow (0, 1)$, so that 
\begin{align}
    \lim_{\epsilon\rightarrow 0, \gamma \rightarrow 1} \hat{D} = \frac{1}{2}.
\end{align}

Regarding $\hat{C}$, we obtain
\begin{align}
    \lim_{\epsilon\rightarrow 0, \gamma\rightarrow 1} \Delta q_{1}^{GT}(A1D) = \frac{1}{2}(3P - R),
\end{align}
and
\begin{align}
    \lim_{\epsilon\rightarrow 0, \gamma\rightarrow 1} \Delta q_{2}^{AC}(BC) = T-R.
\end{align}
These are both always positive (as is required) when $3P - R > 0$. We thus have
\begin{align}
     \lim_{\epsilon\rightarrow 0, \gamma \rightarrow 1} \hat{C} = 
     \begin{cases}
         \frac{3P-1}{3P-R} \quad \text{when } &3P+R>2\\
         \frac{3P-R}{4(1-R)}\; &\text{otherwise}.
     \end{cases}
\end{align}
Note that $3P+R>2$ implies that $3P - 1 > 1-R>0$ so that the numerator of the first case is always positive when its condition is satisfied. In addition, note that $3P+R>2$ implies $3P-R>2(1-R)>0$ so that the condition $3P+R>2$ ensures that the condition $3P-R>0$ is satisfied. Substituting these expressions into the expressions for the bounds of the BoA gives the result.

\section{Value Iteration Algorithm}
\begin{algorithm}[H]
\caption{Value Iteration}
\label{alg:value_iteration}
\KwIn{Number of samples $M$, tolerance $\varepsilon$, learning rate $\alpha$}
\KwOut{Histories $q1\_hist, q2\_hist$, equilibrium indicators eqC, eqD}

\For{$m \gets 1$ \KwTo $M$}{
    Initialize Q-values: $(q_1, q_2) \gets \text{initialize\_q\_values}()$\;
    Set $(q_1^{val}, q_2^{val}) \gets (q_1, q_2)$\;
    Initialize histories: $q1\_hist \gets [\ ], \ q2\_hist \gets [\ ]$\;
    Set $\text{error} \gets 1$\;

    \While{$\text{error} > 0$}{
        $(\pi_1, \pi_2) \gets \text{update\_policies}(q_1, q_2)$\;
        $q\_error \gets 1$\;

        \While{$q\_error > \varepsilon$}{
            $(q_1^{val}, q_2^{val}, q\_error) \gets \text{update\_q\_values}(\pi_1, \pi_2, q_1^{val}, q_2^{val})$\;
        }

        $\text{error} \gets \mathbf{1}\{ \, |q_1^{val} - q_1| + |q_2^{val} - q_2| > \varepsilon \, \}$\;

        $q_1 \gets q_1 + \alpha \cdot (q_1^{val} - q_1)$\;
        $q_2 \gets q_2 + \alpha \cdot (q_2^{val} - q_2)$\;
        Append $q_1, q_2$ to histories\;
    }

    Save histories: $q1\_hist, q2\_hist$ for sample $m$\;\\
    $\text{eqC}[m] \gets (\pi_1 = \pi^{eqC} \ \wedge \ \pi_2 = \pi^{eqC})$\;\\
    $\text{eqD}[m] \gets (\pi_1 = \pi^{eqD} \ \wedge \ \pi_2 = \pi^{eqD})$\;
}
\end{algorithm}
\newpage

\section{Case by case analysis, vector fields and graphs}
\label{appendix:case_by_case}

In this appendix, we illustrate the dynamics for each case, emphasizing the role of the BoA Boundary (BoAB) where relevant. To avoid redundancy, we do not plot trees that are identical to others,these typically correspond to the same pair of strategies but in reversed order. For example, ADGT and GTAD yield identical trees, as do GTAGT and AGTGT, and similar cases follow. Therefore, if certain trees appear to be missing, it is because their identical counterparts (with inverted strategies) have already been plotted.
\vspace{1em}
\newline
We present the analysis in the following order to facilitate the reader’s understanding. We begin with the vector fields and trees corresponding to initializations in which convergence to equilibrium is inevitable,namely, cases where, regardless of the starting point, the dynamics always lead to either the ADAD or GTGT equilibrium. These configurations result in a tree consisting of a single node with a self-loop edge. Subsequently, we show cases that require one step to reach equilibrium, then those that require two steps, and so on.

% ==== First 2 plots ====
\begin{figure}[H]
    \centering
    \begin{subfigure}{0.49\textwidth}
        \includegraphics[width=\linewidth]{images/vf_adad.png}
        \caption{\textsc{AD} vs \textsc{AD}}
    \end{subfigure}
    \begin{subfigure}{0.49\textwidth}
        \includegraphics[width=\linewidth]{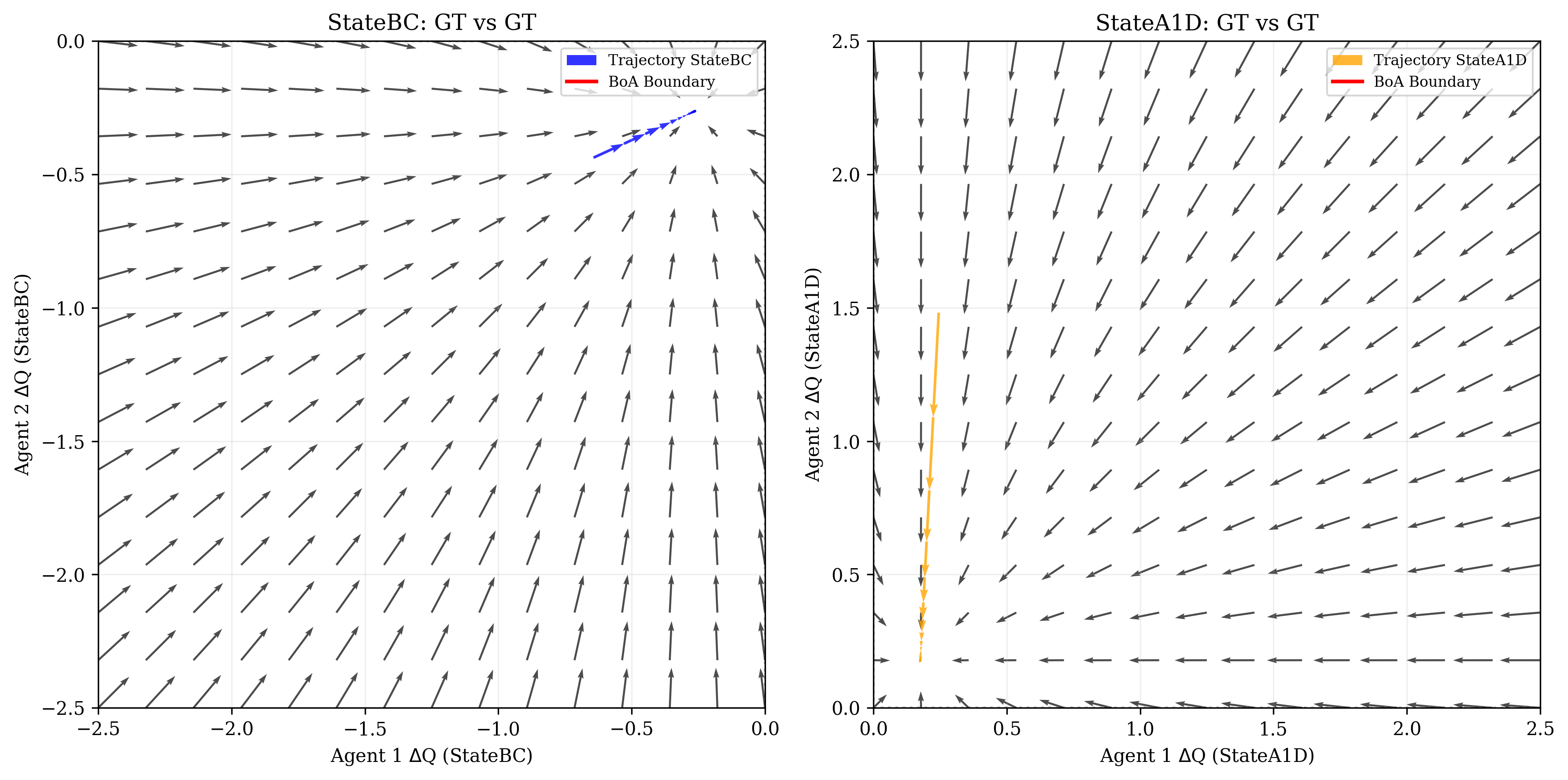}
        \caption{\textsc{GT} vs \textsc{GT}}
    \end{subfigure}
    \caption{Vector fields corresponding to cases where all trajectories lead directly to the equilibrium point (ADAD or GTGT).}
    \label{fig:vectorfields1eq}
    
\end{figure}

\begin{figure}[H]
\centering
\includegraphics[width=0.4\textwidth]{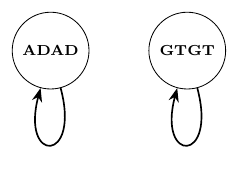}
\label{fig:eq_tree}
\caption{Trees showing the possible paths to an equilibrium when starting in (AD, AD) and (GT, GT). Here the resulting tree is a node with a self-loop.}
\end{figure}

% ==== Next 4 plots ====
\begin{figure}[H]
    \centering
    \begin{subfigure}{0.49\textwidth}
        \includegraphics[width=\linewidth]{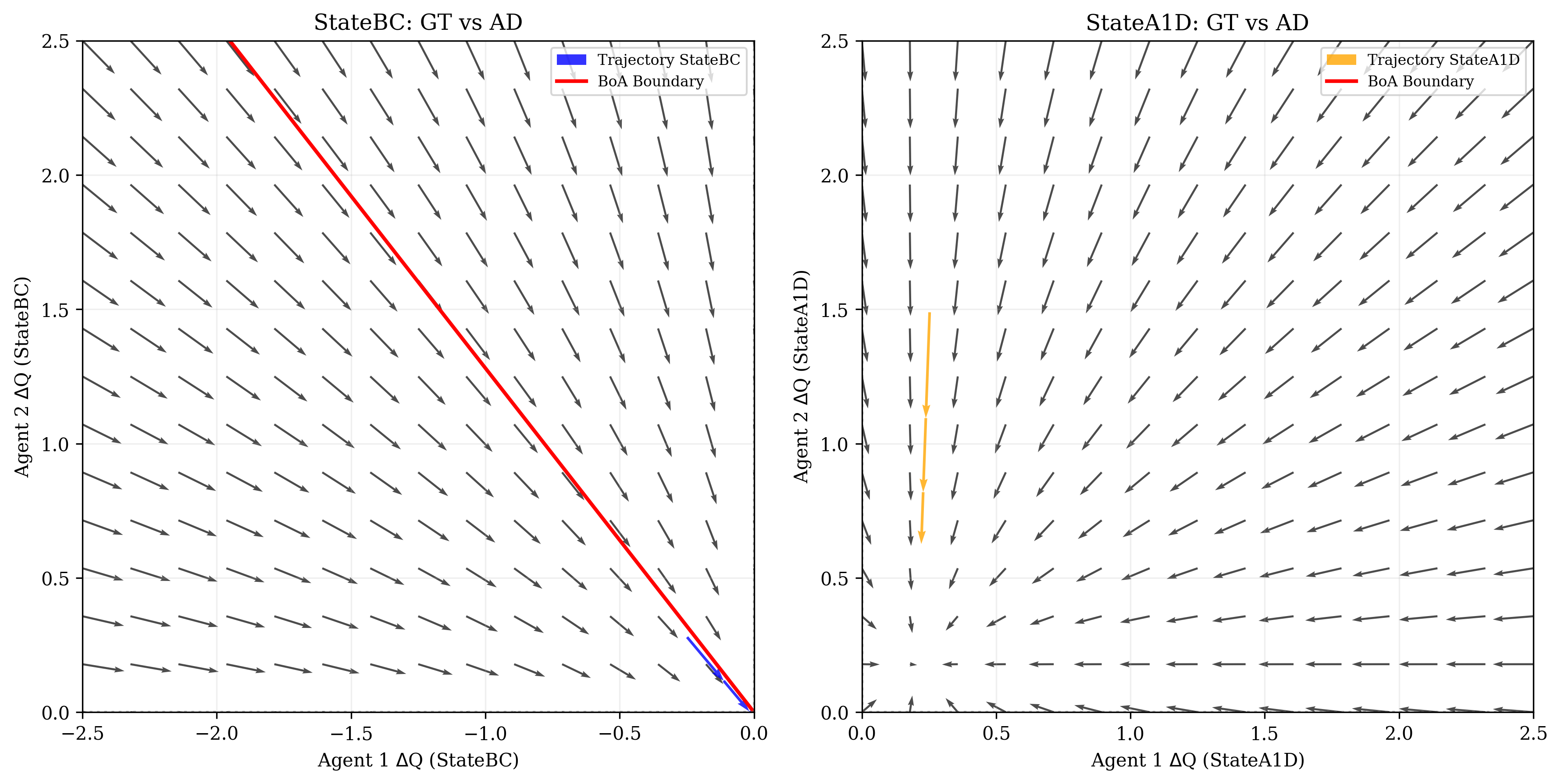}
        \caption{\textsc{GT} vs \textsc{AD}}
    \end{subfigure}
    \hfill
    \begin{subfigure}{0.49\textwidth}
        \includegraphics[width=\linewidth]{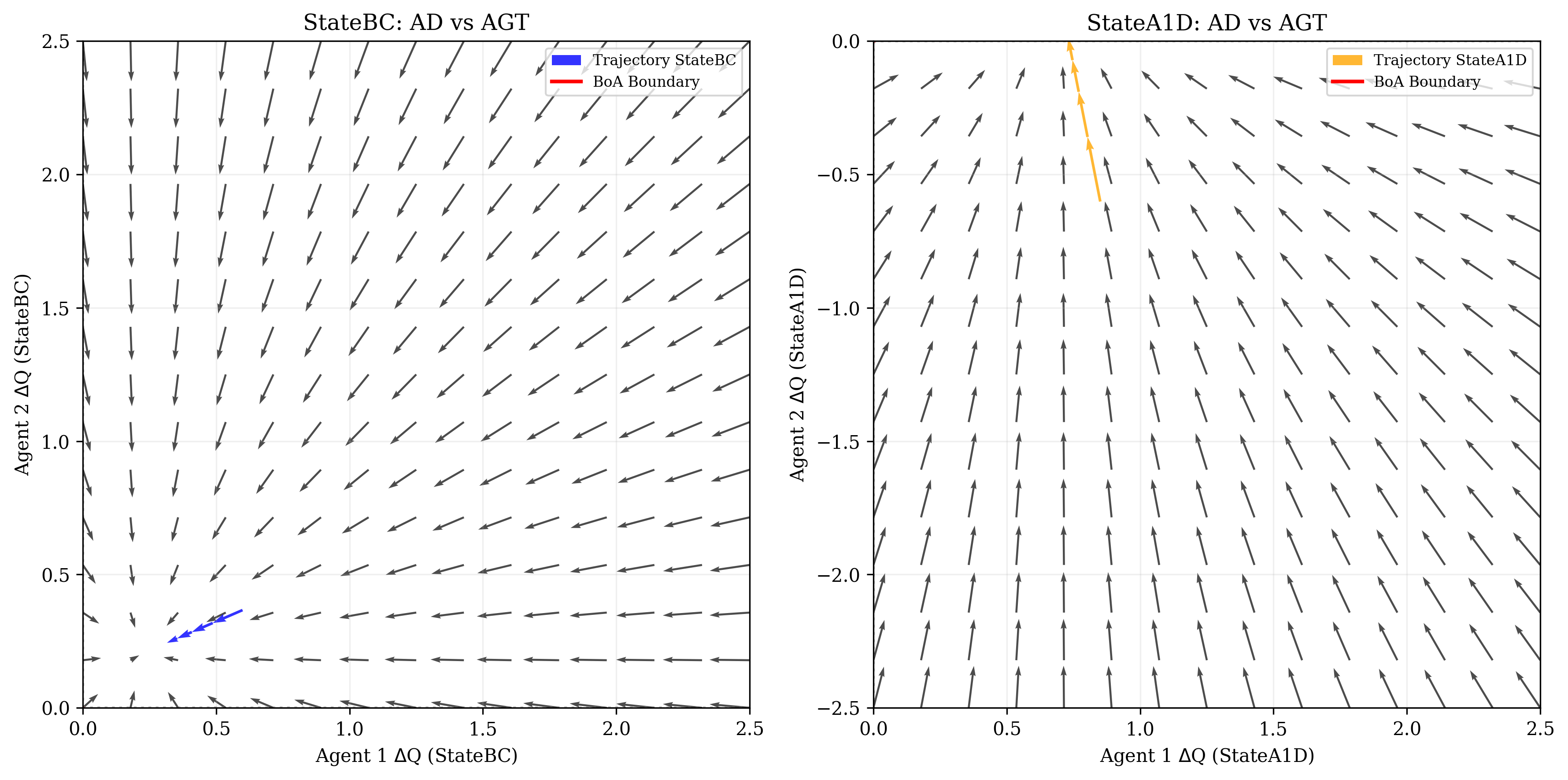}
        \caption{\textsc{AD} vs \textsc{AGT}}
    \end{subfigure}
    \caption{Vector fields showing dynamics that reach equilibrium in a single step: GTAD and ADAGT.}
    \label{fig:vectorfields2}
\end{figure}

\begin{figure}[H]
\centering
\includegraphics[width=0.6\textwidth]{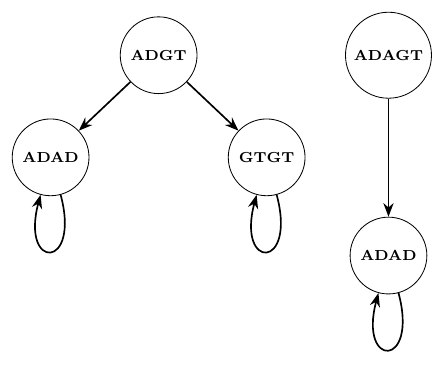}
\label{fig:onestep_tree}
\caption{Trees showing the possible paths to an equilibrium when starting in (AD, GT) and (AD, AGT). Here the maximum path length is one. }
\end{figure}

% ==== Next 4 plots ====
\begin{figure}[H]
    \centering
    \begin{subfigure}{0.49\textwidth}
        \includegraphics[width=\linewidth]{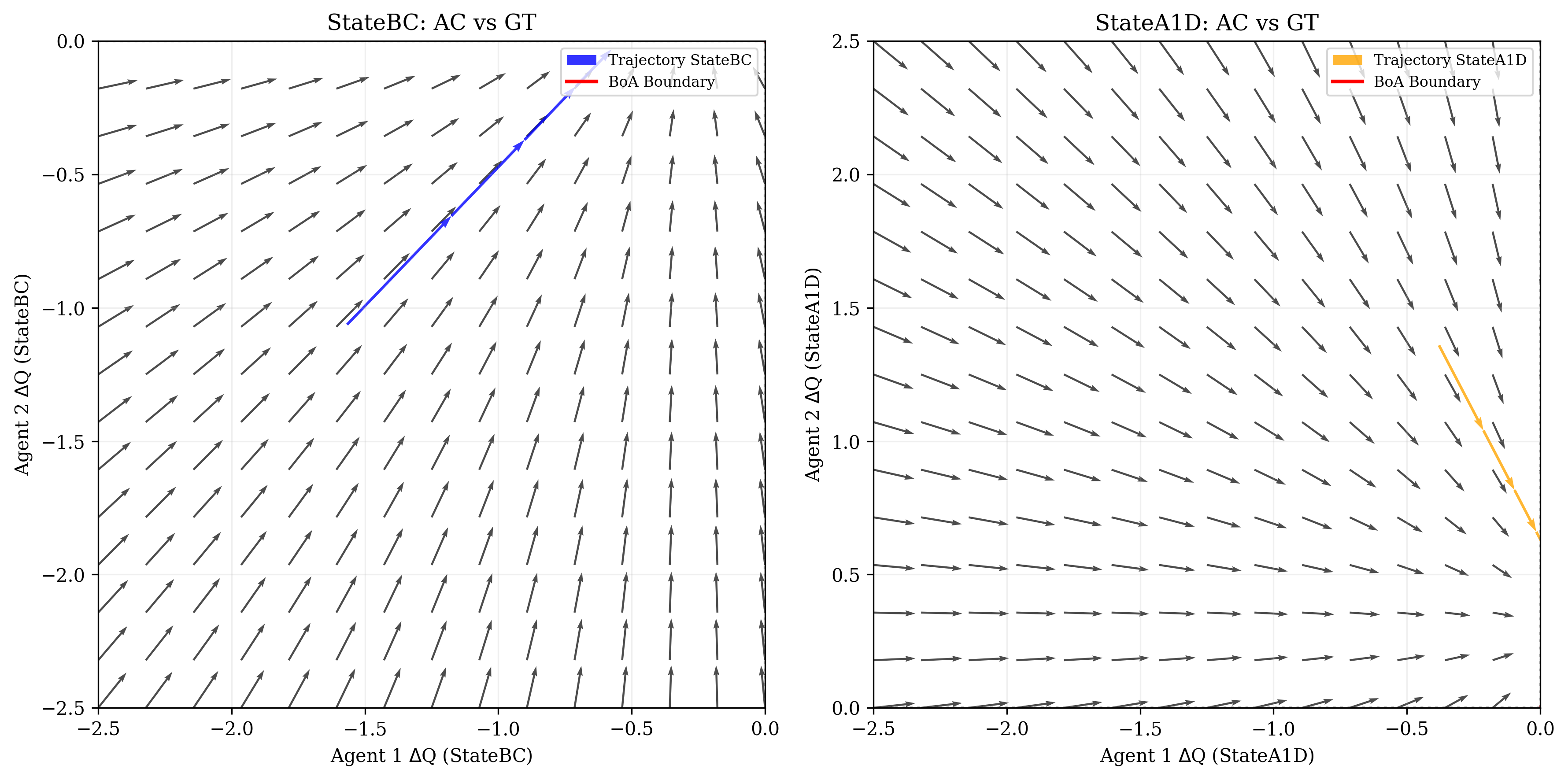}
        \caption{\textsc{AC} vs \textsc{GT}}
    \end{subfigure}
        \begin{subfigure}{0.49\textwidth}
        \includegraphics[width=\linewidth]{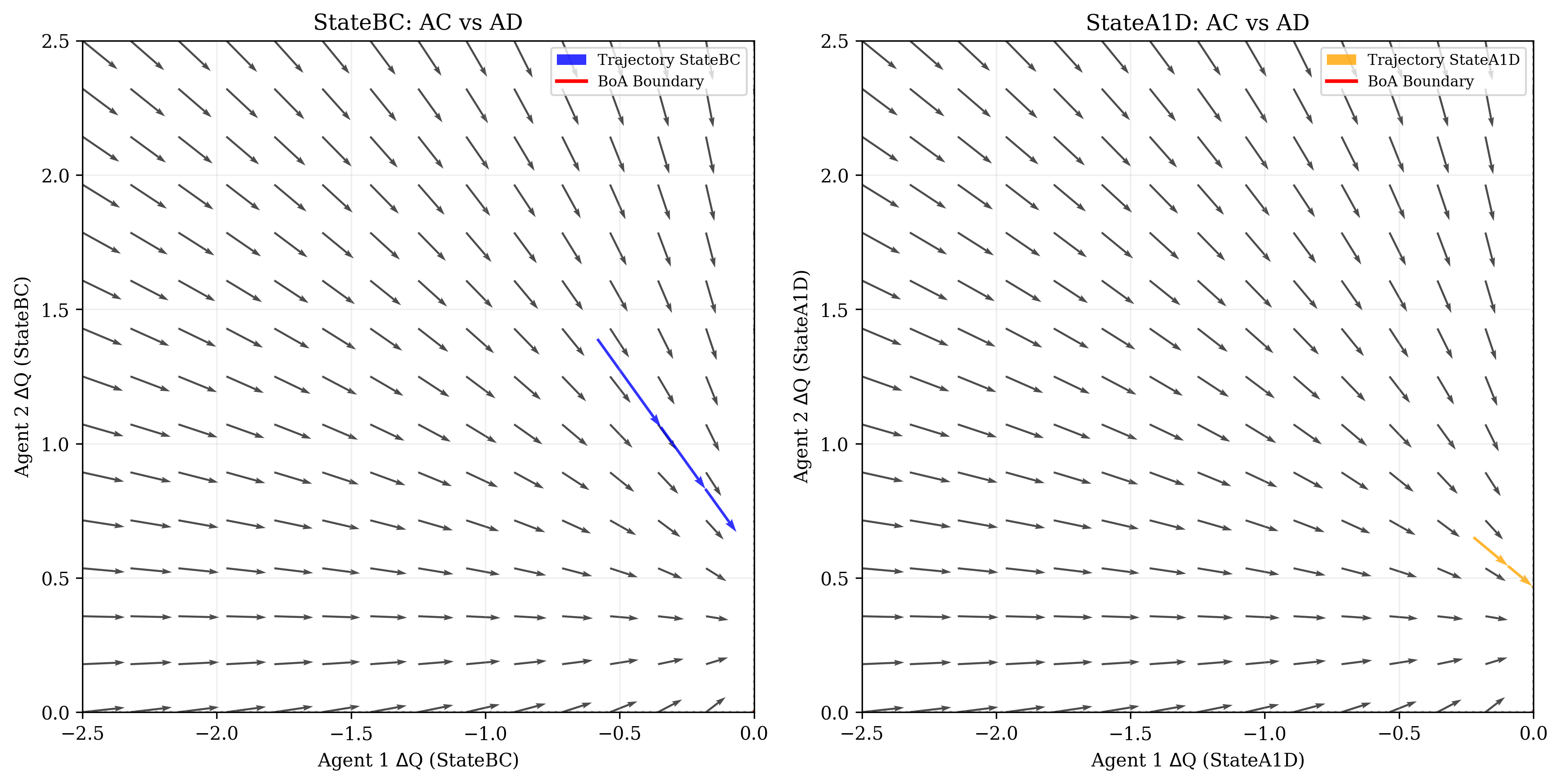}
        \caption{\textsc{AC} vs \textsc{AD}}
    \end{subfigure}
    \begin{subfigure}{0.49\textwidth}
        \includegraphics[width=\linewidth]{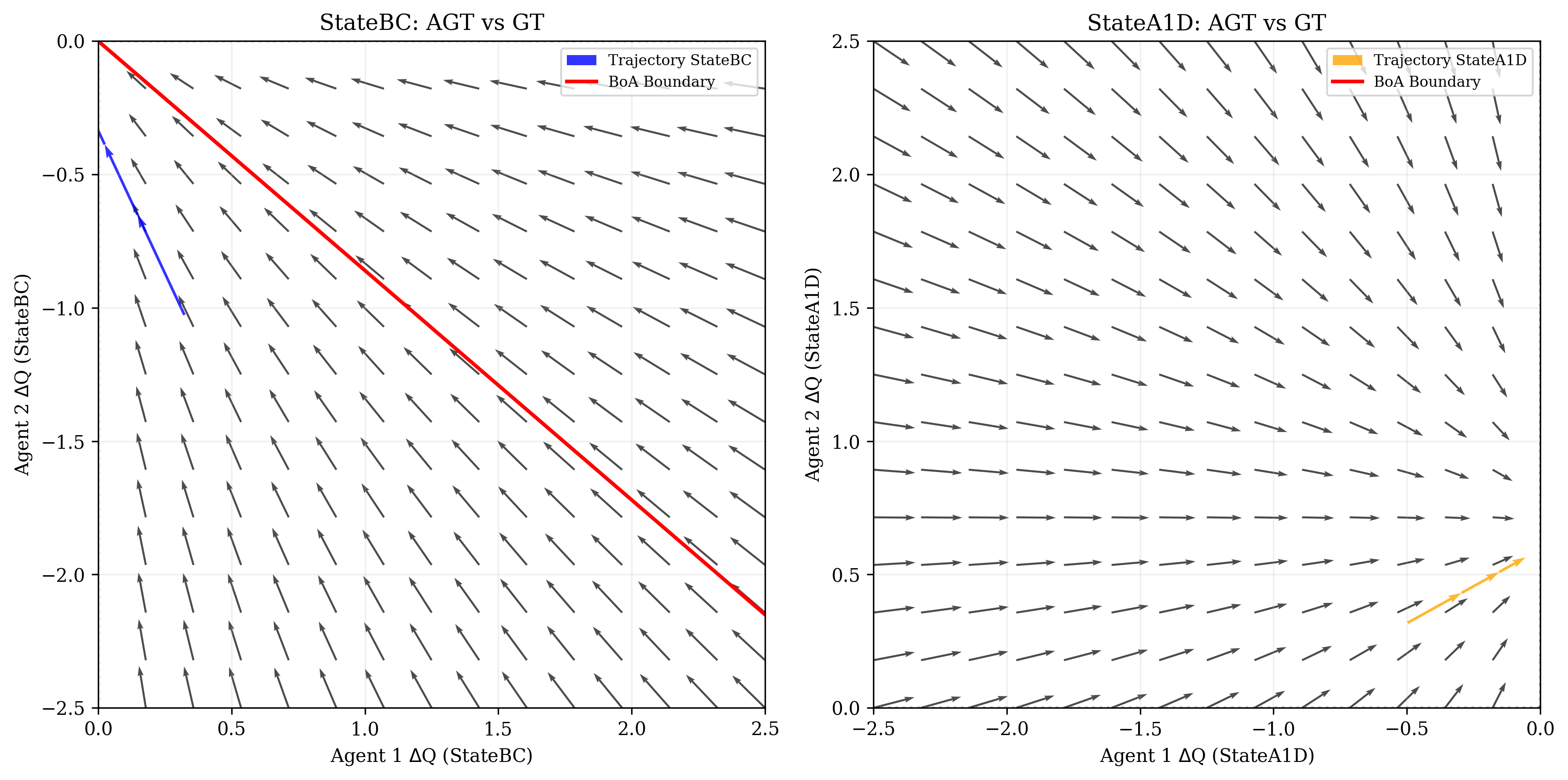}
        \caption{\textsc{AGT} vs \textsc{GT}}
    \end{subfigure}
    \begin{subfigure}{0.49\textwidth}
        \includegraphics[width=\linewidth]{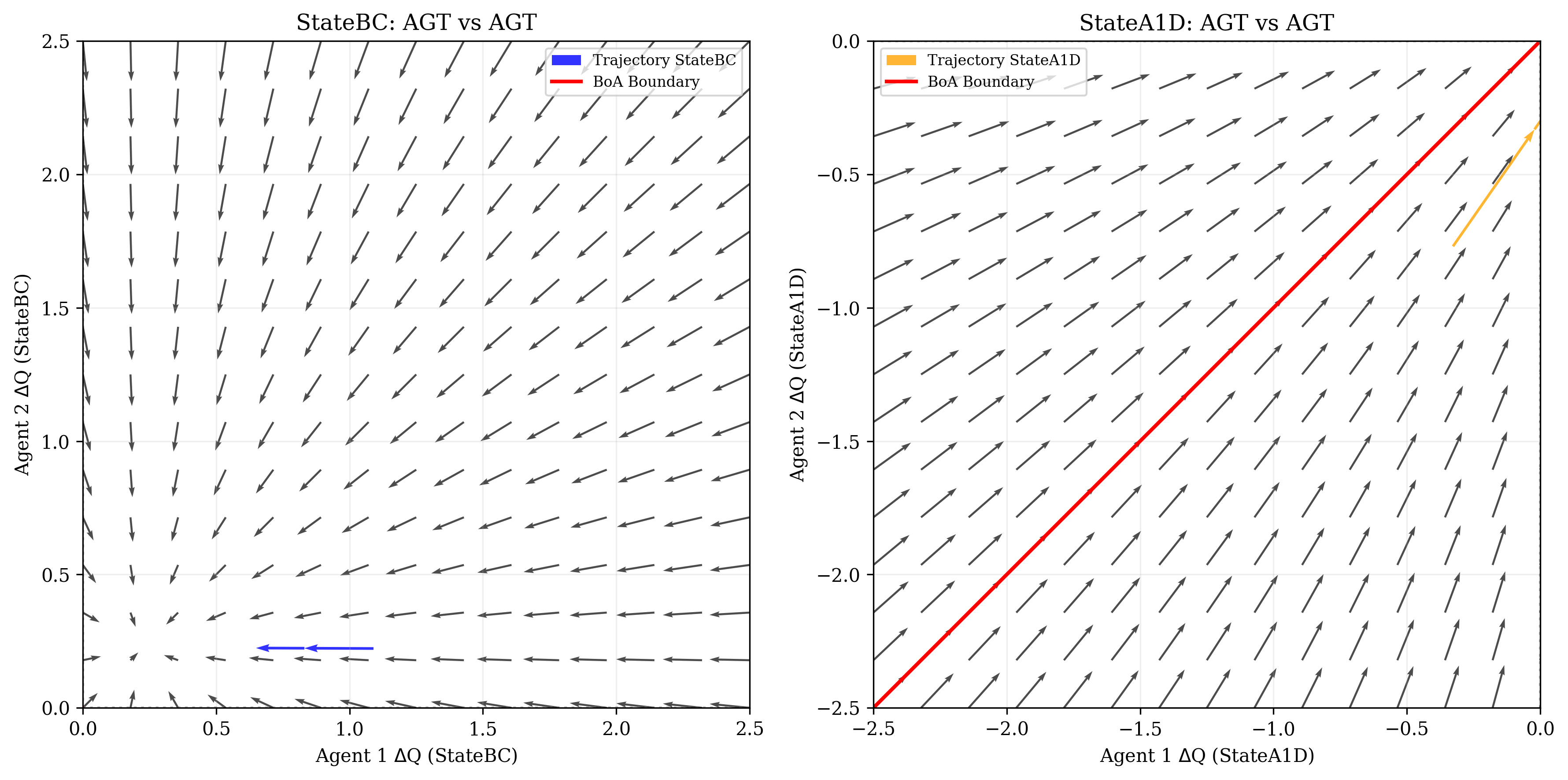}
        \caption{\textsc{AGT} vs \textsc{AGT}}
    \end{subfigure}

    \caption{Vector fields illustrating trajectories that require two steps to reach the equilibrium: ACGT, ACAD,  AGTGT and AGTAGT}
    \label{fig:twosteps}
\end{figure}
\begin{figure}[H]
\centering
\includegraphics[width=1\textwidth]{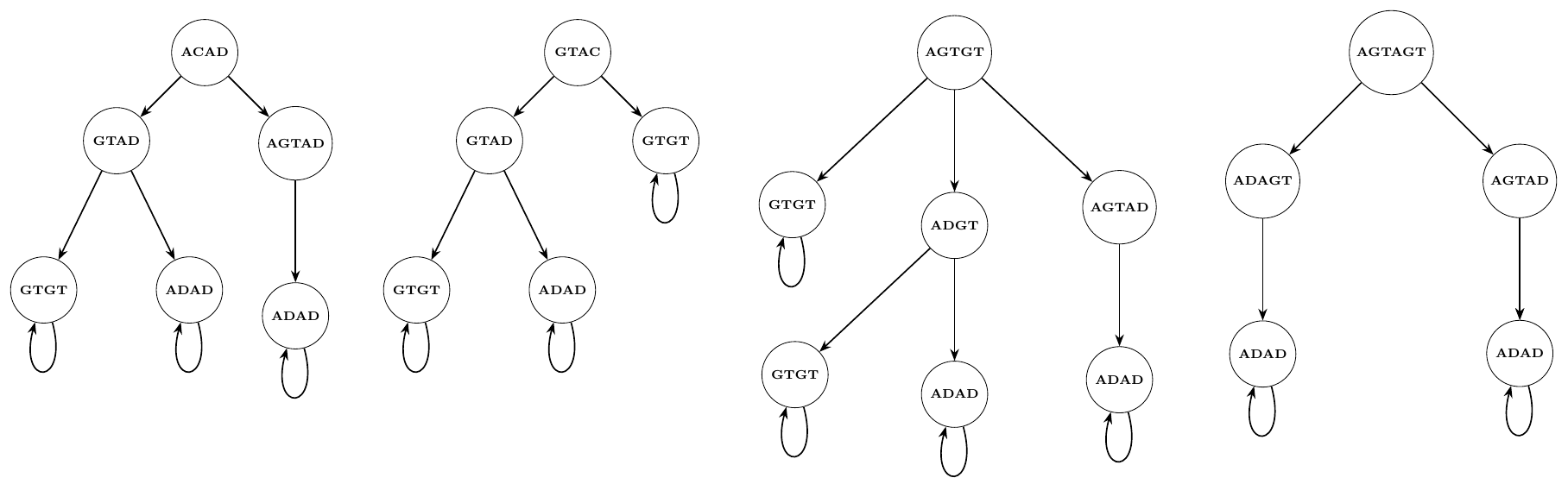}
\label{fig:twostepstree}
\caption{Trees showing the possible paths to reach an equilibrium when starting in (AC, AD), (GT, AC), (AGT, GT) and (AGT, AGT). Here the maximum path length is two. }
\label{fig:AGT_tree}
\end{figure}
% ==== Last 4 plots ====
\begin{figure}[H]
    \centering
    \begin{subfigure}{0.49\textwidth}
        \includegraphics[width=\linewidth]{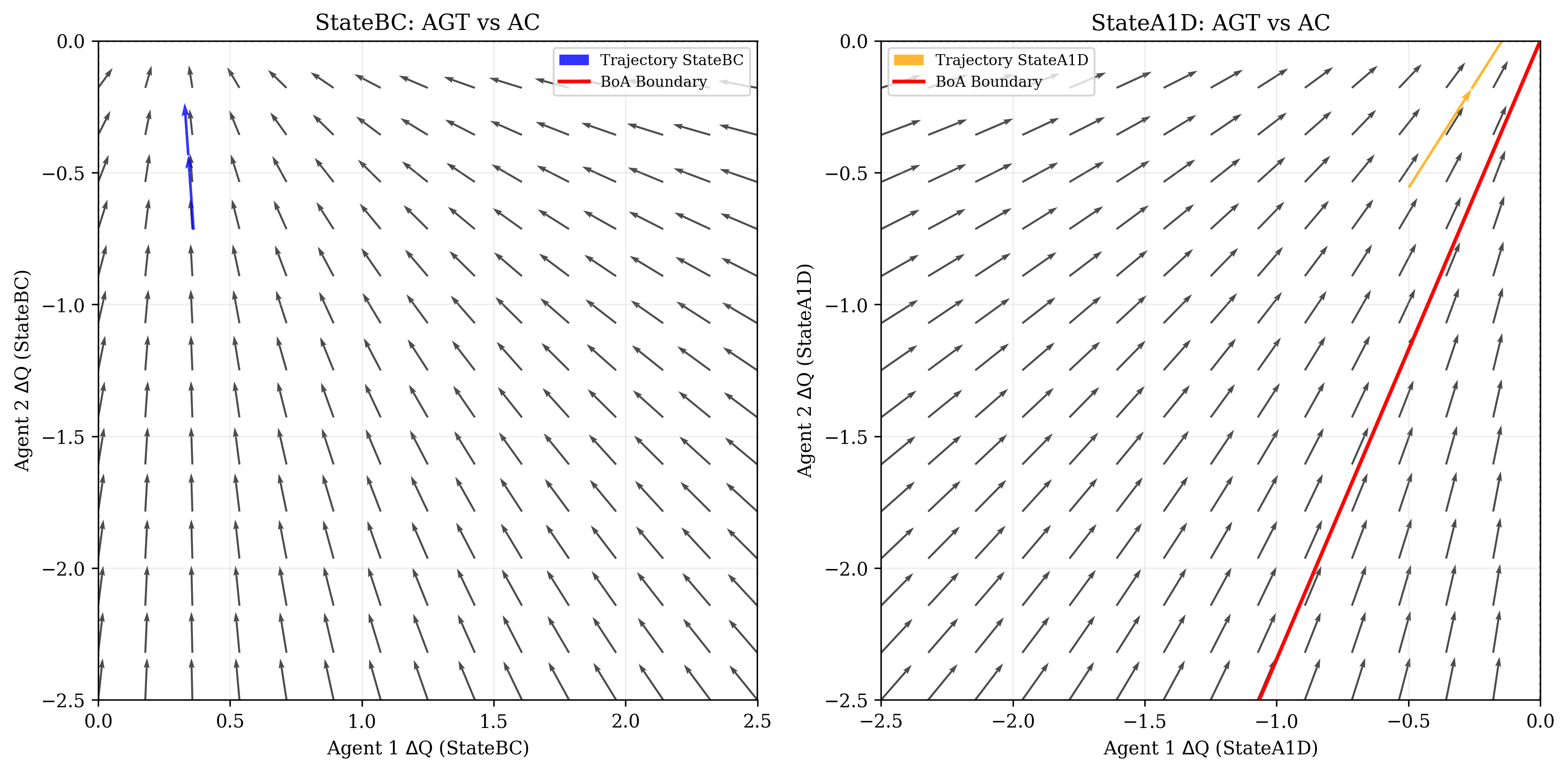}
        \caption{\textsc{AGT} vs \textsc{AC}}
    \end{subfigure}
    \begin{subfigure}{0.49\textwidth}
        \includegraphics[width=\linewidth]{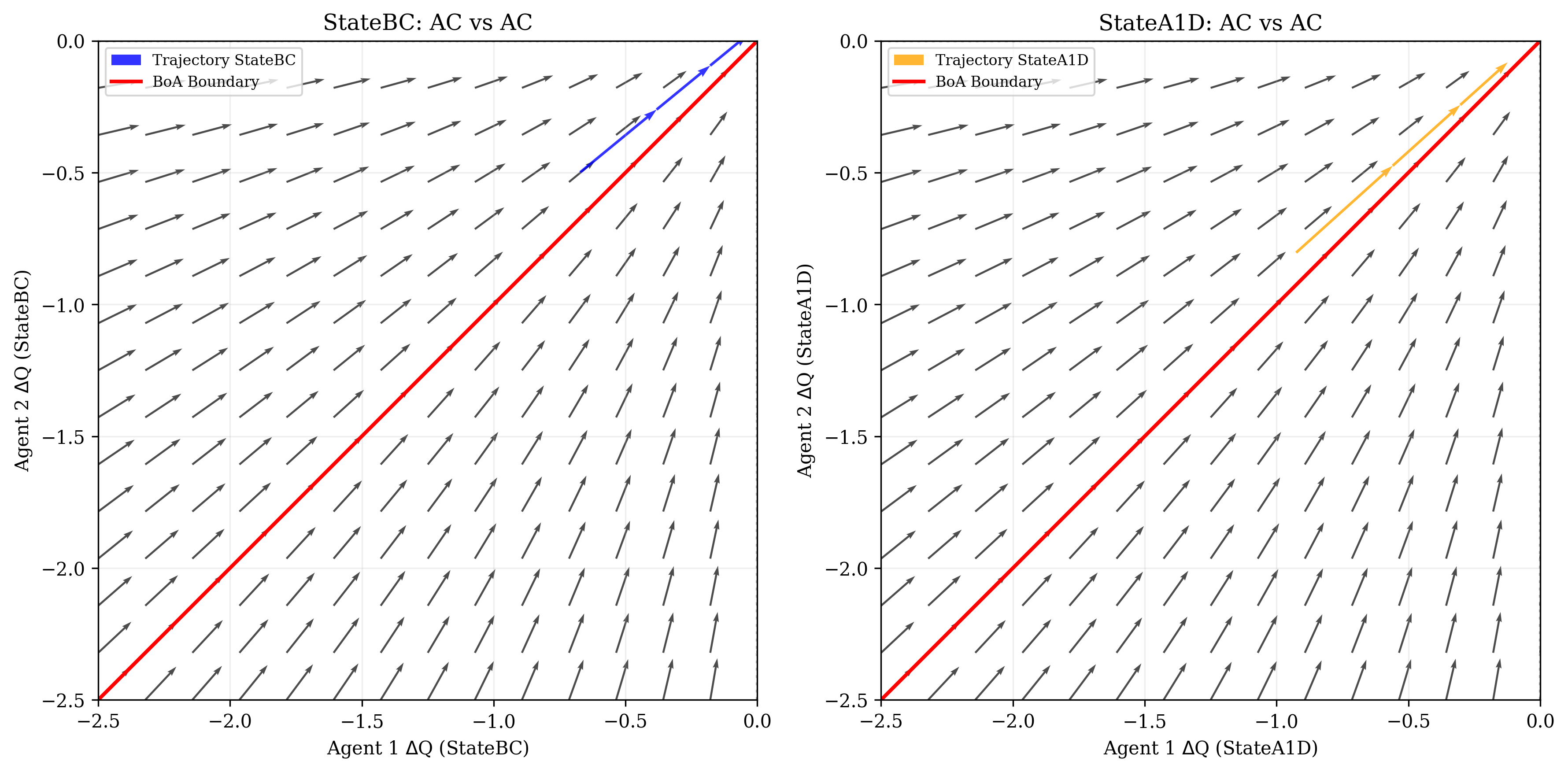}
        \caption{\textsc{AC} vs \textsc{AC}}
    \end{subfigure}
    \caption{Vector fields depicting more complex dynamics, where trajectories converge to equilibrium after three or more steps: AGTAC and ACAC}
    \label{fig:vectorfields4}
\end{figure}
\begin{figure}[H]
\centering
\includegraphics[width=1\textwidth]{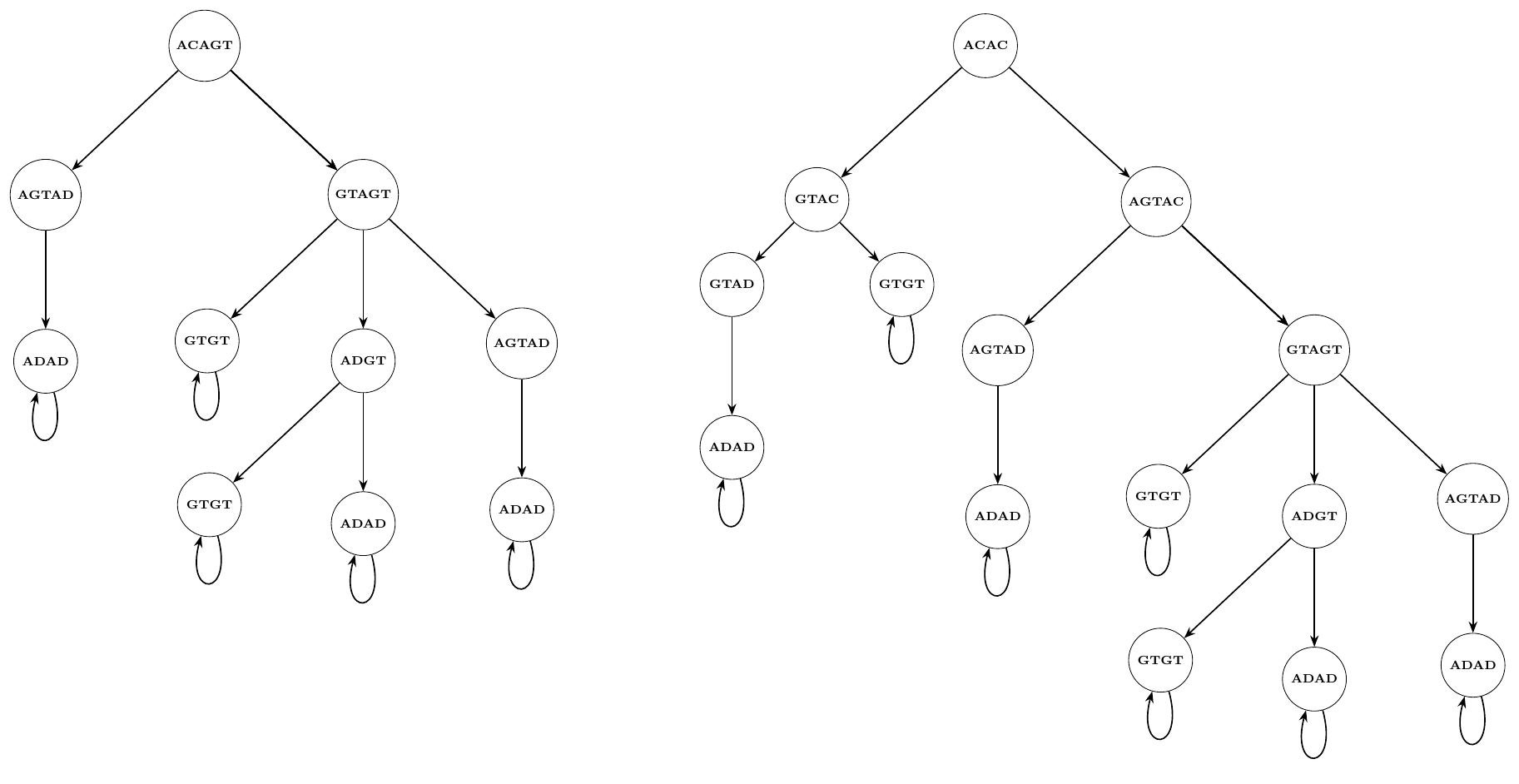}
\caption{Trees showing the possible paths to an equilibrium when starting in (AC, AGT) and (AC, AC). Here the maximum path length is three for (AC, AGT) and four for (AC, AC).}

\end{figure}

\newpage
\section{BoA conditioned on starting in $(AC,\cdot)$}
\label{app:acgt}

This appendix complements the $(AD,GT)$ validation in Figure~\ref{fig:alpha_effect_ADGT} with simulations conditioned on $(AC,GT)$ and $(AC,AD)$. We use the same hyperparameters as in Section~\ref{sec:simulations} ($M=2000$, $\alpha\in\Theta_{\alpha}$, $\lambda\in\Theta_{\lambda}$, $\varepsilon\in\Theta_{\epsilon}$, $\gamma\in\Theta_{\gamma}$) and the same two payoff configurations: (1)~$R=0.7$, $P=0.2$ and (2)~$R=0.8$, $P=0.5$. In each run the initial Q-values are drawn so that the realised strategy profile matches the conditioning event (fixed-profile initialization).

\subsection{Profile $(AC,GT)$: lower bound $\hat{C}$ and full BoA}
\label{app:acgt-profile}

\begin{figure}[H]
    \centering
    \begin{subfigure}[b]{0.48\textwidth}
        \centering
        \includegraphics[width=\textwidth]{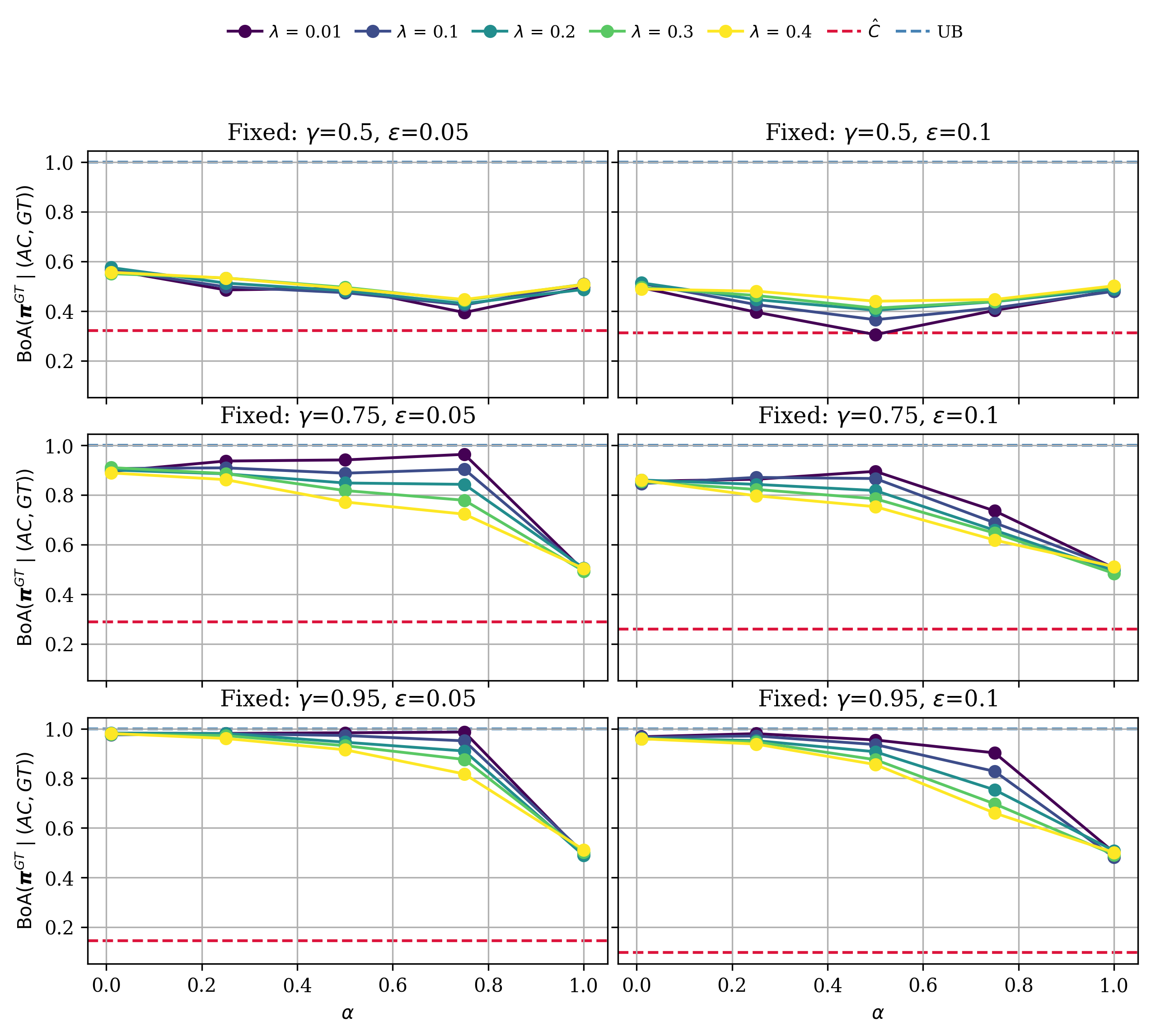}
        \caption{$R = 0.7,\ P = 0.2$}
        \label{fig:alpha_effect_ACGT_R07}
    \end{subfigure}
    \hfill
    \begin{subfigure}[b]{0.48\textwidth}
        \centering
        \includegraphics[width=\textwidth]{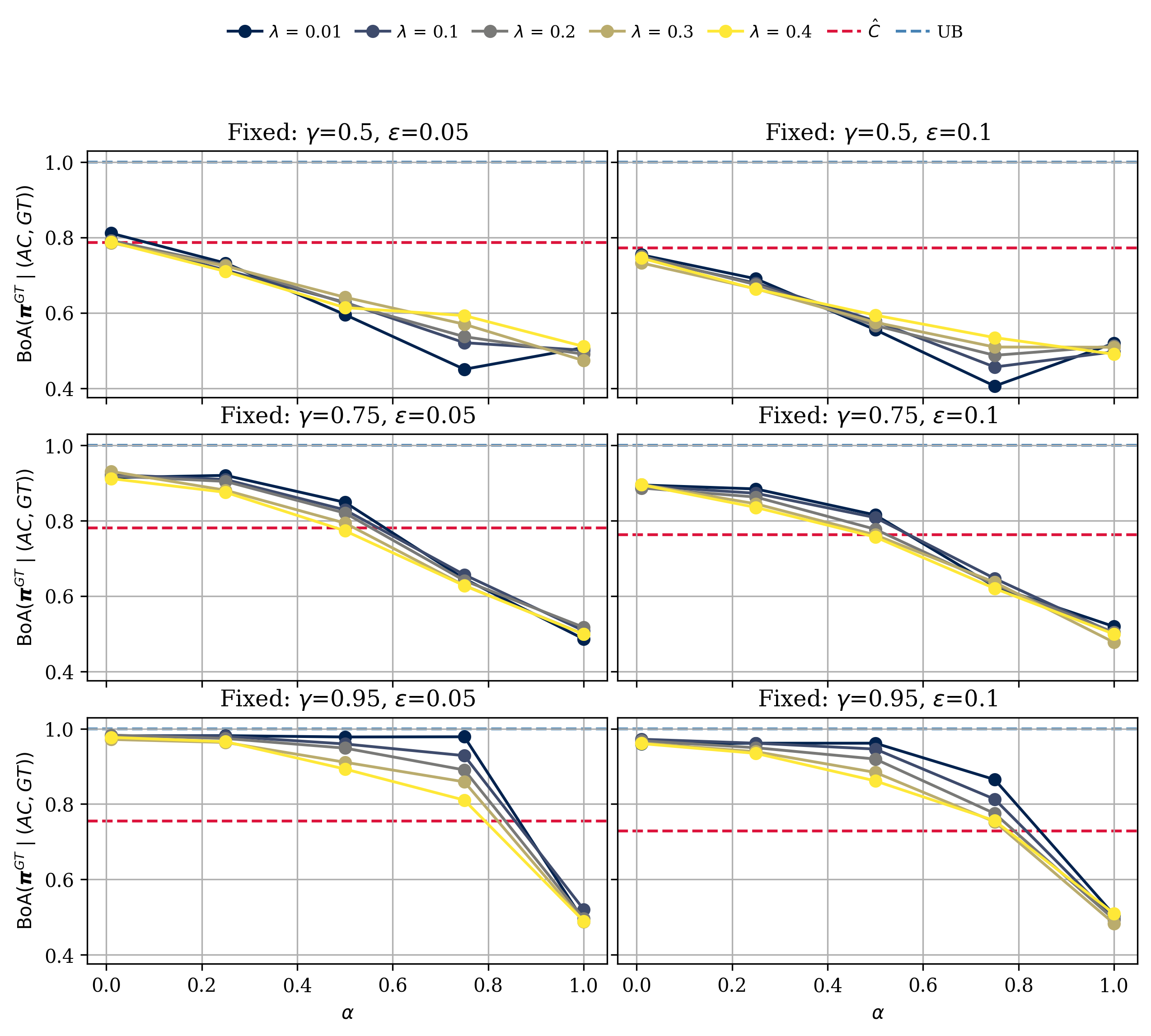}
        \caption{$R = 0.8,\ P = 0.5$}
        \label{fig:alpha_effect_ACGT_R08}
    \end{subfigure}
    \caption{Empirical $\text{BoA}(\boldsymbol{\pi}^{\text{GT}}\mid(AC,GT))$ (markers) versus the theoretical lower bound $\hat{C}$ (dash-dotted line, \eqref{eq:Chat}), as a function of $\alpha$ for several $\lambda$. Each subplot fixes $(\gamma,\varepsilon)$.}
    \label{fig:alpha_effect_ACGT}
\end{figure}

From $(AC,GT)$ the first profile change is either to $\boldsymbol{\pi}^{\text{GT}}$ (player~1's A1D component crosses zero first) or to $(AC,AD)$ (player~2's BC component crosses first). In the proof of Theorem~\ref{thm:BoA}(ii) (Appendix~\ref{proof:PropBoA}) we showed that the probability of the direct GT path in the $\alpha\to 0$ limit is
\begin{equation}
\label{eq:Chat-bound}
    \hat{C}
    \;\;\leq\;\;
    \text{BoA}(\boldsymbol{\pi}^{\text{GT}} \mid (\text{AC},\text{GT}))
    \;\;\leq\;\; 1,
\end{equation}
where $\hat{C}$ is the closed form in \eqref{eq:Chat}. The dash-dotted line in Figure~\ref{fig:alpha_effect_ACGT} plots $\hat{C}$; the panels report the \emph{full} empirical BoA $\text{BoA}(\boldsymbol{\pi}^{\text{GT}}\mid(AC,GT))$, which can exceed $\hat{C}$ because trajectories that first reach $(AC,AD)$ may still converge to $\boldsymbol{\pi}^{\text{GT}}$.

Figure~\ref{fig:alpha_effect_ACGT} shows that the full BoA lies above $\hat{C}$ as expected and approaches one as $\gamma\to 1$ and $\varepsilon\to 0$. For small $\alpha$ and $\lambda$ the panels are close to the $\alpha\to 0$ limit; $\lambda$ has limited effect near $\alpha\in\{0.01,1\}$.

\subsection{First-crossing validation of $\hat{C}$}
\label{app:acgt-boundary}

To test $\hat{C}$ directly,rather than the full BoA,we record for each trajectory which boundary is crossed first: $\Delta q_{1}(A1D)=0$ (A1D-first, direct path toward $\boldsymbol{\pi}^{\text{GT}}$) or $\Delta q_{2}(BC)=0$ (BC-first, via $(AC,AD)$). Let $P_{\text{direct}}$ denote the fraction of A1D-first crossings along the simulated trajectory, and let $P_{\text{analytic}}$ denote the mean over trajectories of the hitting-time indicator $\mathbb{P}[T_{A1D}<T_{BC}]$ evaluated at the \emph{initial} $\Delta q$ values (the $\alpha\to 0$ limit). By construction $P_{\text{analytic}}$ targets $\hat{C}$; $P_{\text{direct}}$ approximates it when $\alpha$ and $\lambda$ are small.

\begin{figure}[H]
    \centering
    \begin{subfigure}[b]{0.48\textwidth}
        \centering
        \includegraphics[width=\textwidth]{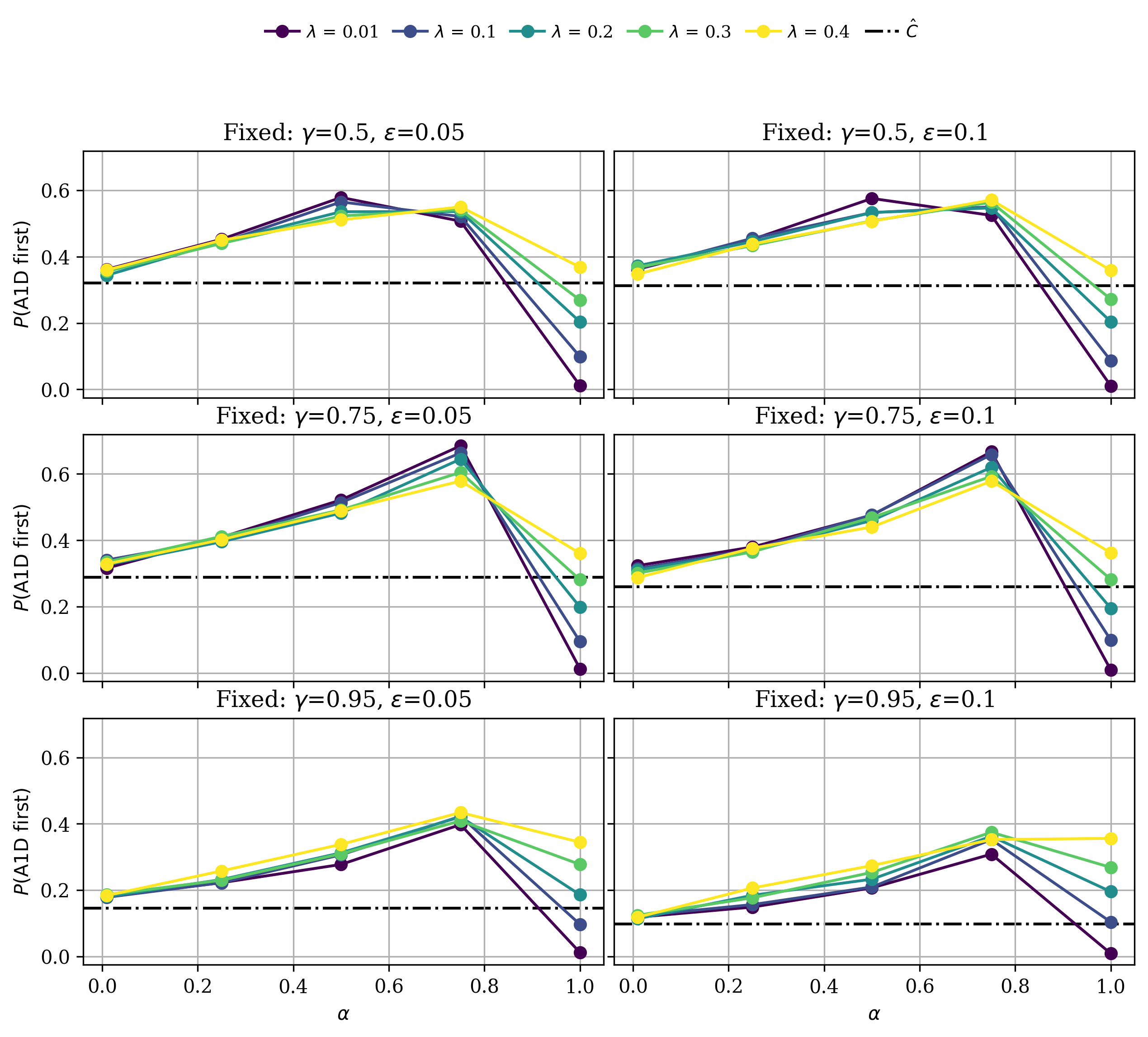}
        \caption{$R = 0.7,\ P = 0.2$}
        \label{fig:alpha_effect_ACGT_boundary_R07}
    \end{subfigure}
    \hfill
    \begin{subfigure}[b]{0.48\textwidth}
        \centering
        \includegraphics[width=\textwidth]{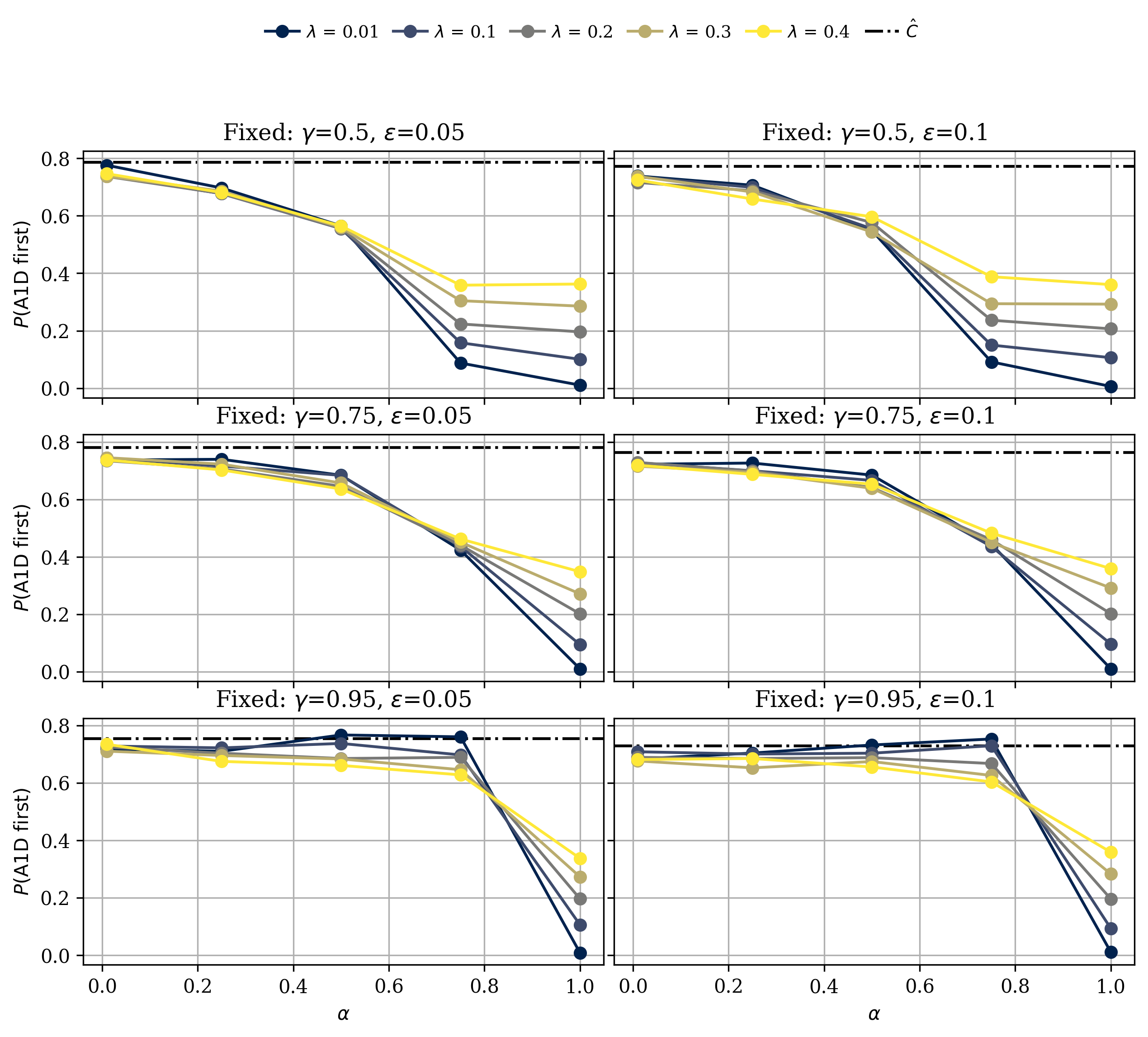}
        \caption{$R = 0.8,\ P = 0.5$}
        \label{fig:alpha_effect_ACGT_boundary_R08}
    \end{subfigure}
    \caption{Validation of $\hat{C}$: empirical A1D-first fraction $P_{\text{direct}}$ (markers) and mean init-time indicator $P_{\text{analytic}}$ (solid) versus $\hat{C}$ (dash-dotted, \eqref{eq:Chat}), for $(AC,GT)$ initialization. Same axes as Figure~\ref{fig:alpha_effect_ACGT}.}
    \label{fig:alpha_effect_ACGT_boundary}
\end{figure}

Figure~\ref{fig:alpha_effect_ACGT_boundary} reports $P_{\text{direct}}$ and $P_{\text{analytic}}$ against $\hat{C}$. Across both payoff configurations we find mean absolute errors of order $0.04$ at $\alpha=\lambda=0.01$, consistent with the hitting-time derivation in Appendix~\ref{proof:PropBoA}. As $\alpha$ grows, $P_{\text{direct}}$ can deviate from $\hat{C}$ because discrete Q-updates depart from the continuous flow; $P_{\text{GT}}$ (full convergence to $(GT,GT)$) is not a valid comparator for $\hat{C}$ since it includes successful paths through $(AC,AD)$.

\subsection{Profile $(AC,AD)$: bound $\hat{D}$}
\label{app:acad-profile}

\begin{figure}[H]
    \centering
    \includegraphics[width=0.55\textwidth]{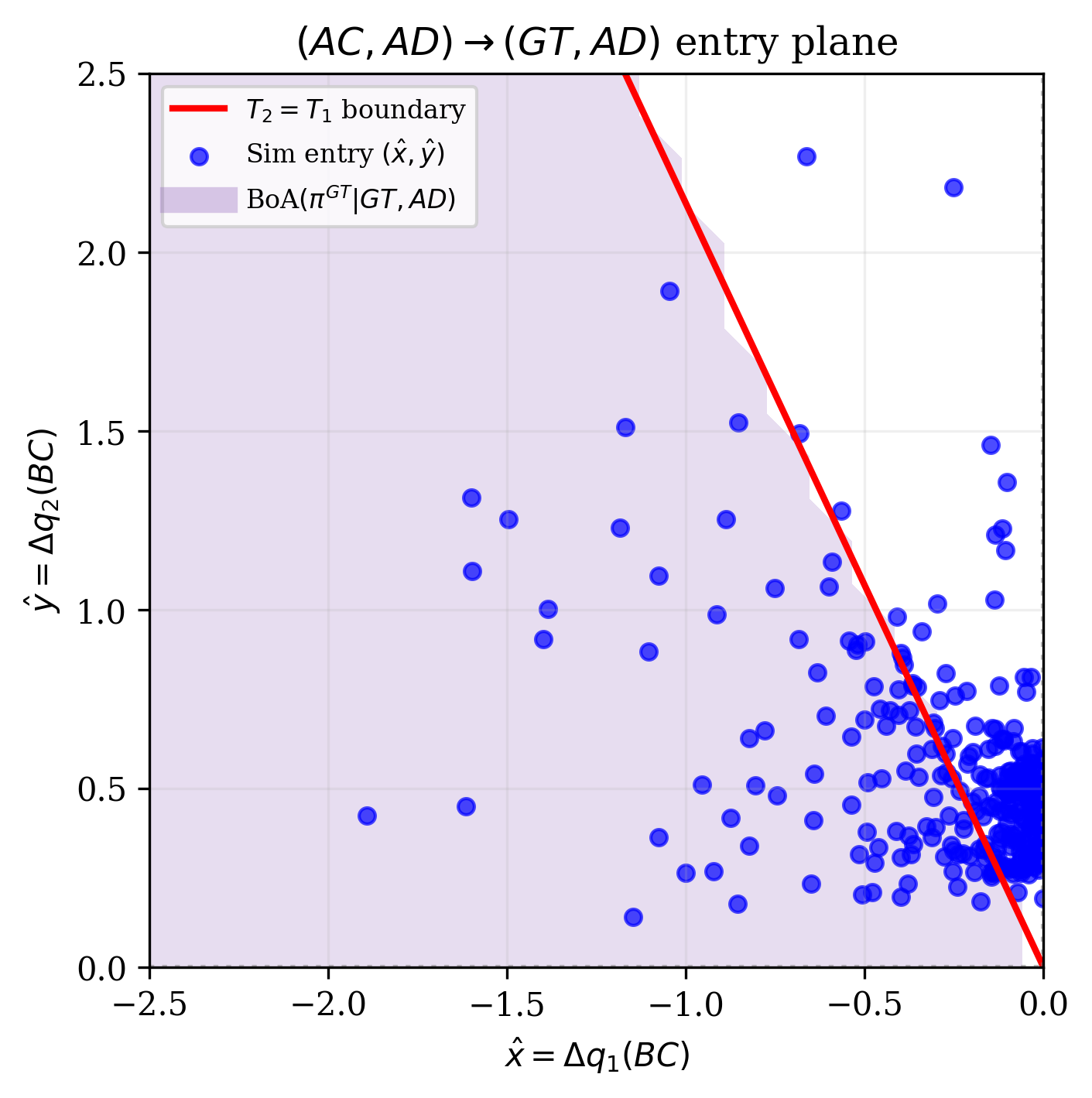}
    \caption{Entry plane $(\hat{x},\hat{y})$ after an A1D-first crossing from $(AC,AD)$: shaded region where the subsequent $(GT,AD)$ dynamics reach $\boldsymbol{\pi}^{\text{GT}}$ (indicator $g(\hat{x},\hat{y})=1$ in \eqref{eq:Dhat-g}). The integral \eqref{eq:Dhat-branch} averages $g$ over the A1D-first law of $(\hat{x},\hat{y})$. Basically it shows where trajectories land after A1D-first, and which landing points lie in the GT basin on (GT,AD).}
    \label{fig:acad-entry}
\end{figure}

For $(AC,AD)$, player~1's first boundary hit determines the next profile: A1D-first ($|X_{1}|>|X_{2}|$) leads to $(GT,AD)$; BC-first leads to $(AGT,AD)$, which has zero BoA for $\boldsymbol{\pi}^{\text{GT}}$. The proof in Appendix~\ref{proof:PropBoA} gives
\begin{equation}
\label{eq:Dhat-bound}
    \hat{D}
    \;=\;
    \text{BoA}(\boldsymbol{\pi}^{\text{GT}} \mid (\text{AC},\text{AD}))
    \;=\;
    \frac{1}{2}\,\hat{D}_{\text{A1D}},
\end{equation}
where $\hat{D}_{\text{A1D}}$ is the integral in \eqref{eq:Dhat-branch}--\eqref{eq:Dhat} over the entry plane $(\hat{x},\hat{y})$ (Figure~\ref{fig:acad-entry}). This integral admits the closed form \eqref{eq:Dhat-closed}--\eqref{eq:Dhat-cases} derived in Appendix~\ref{proof:PropBoA}; the Monte-Carlo and quadrature routines of Subsection~\ref{app:acgt-numerics} are retained only as numerical cross-checks of that formula.

\begin{figure}[H]
    \centering
    \begin{subfigure}[b]{0.48\textwidth}
        \centering
        \includegraphics[width=\textwidth]{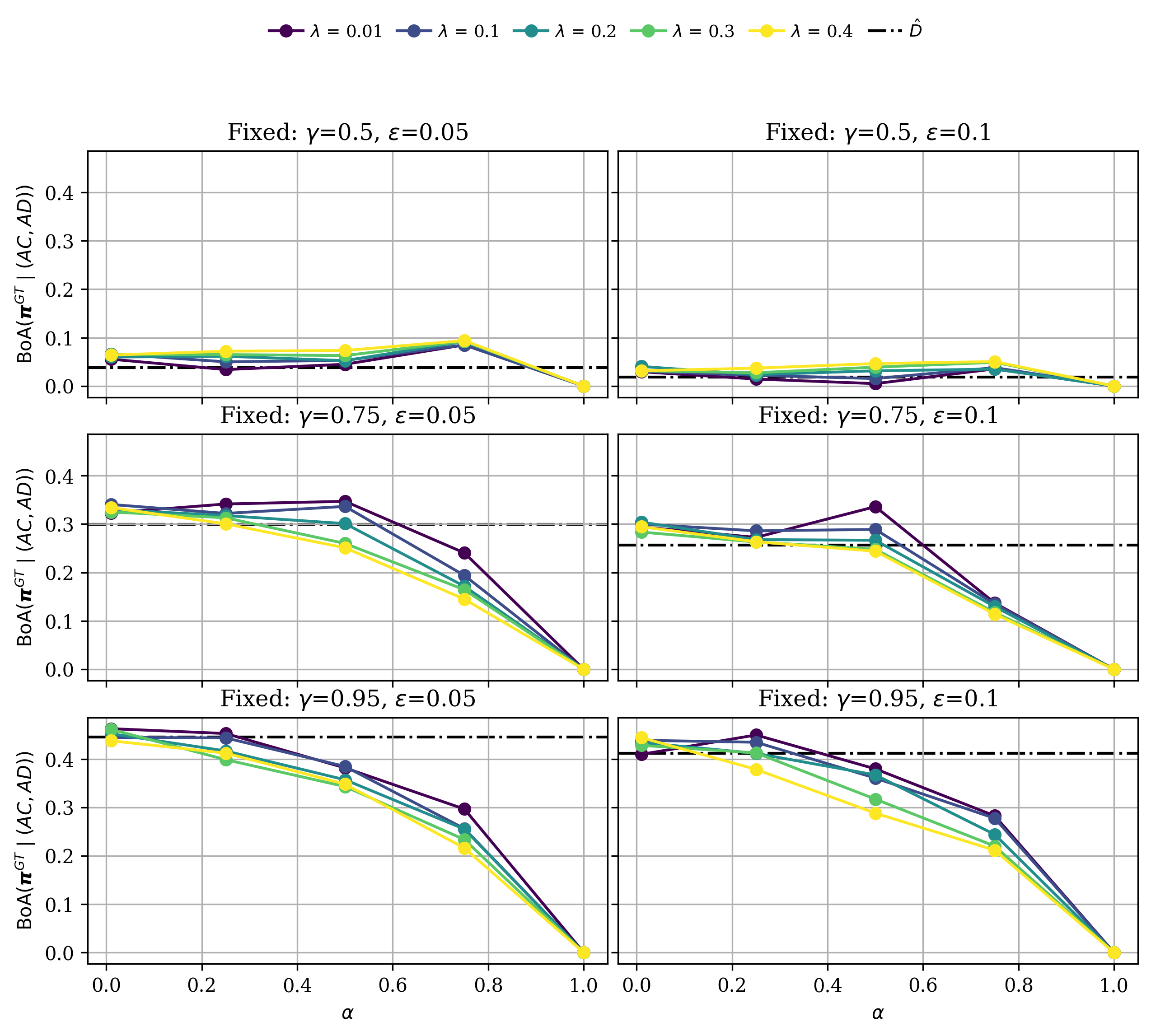}
        \caption{$R = 0.7,\ P = 0.2$}
        \label{fig:alpha_effect_ACAD_R07}
    \end{subfigure}
    \hfill
    \begin{subfigure}[b]{0.48\textwidth}
        \centering
        \includegraphics[width=\textwidth]{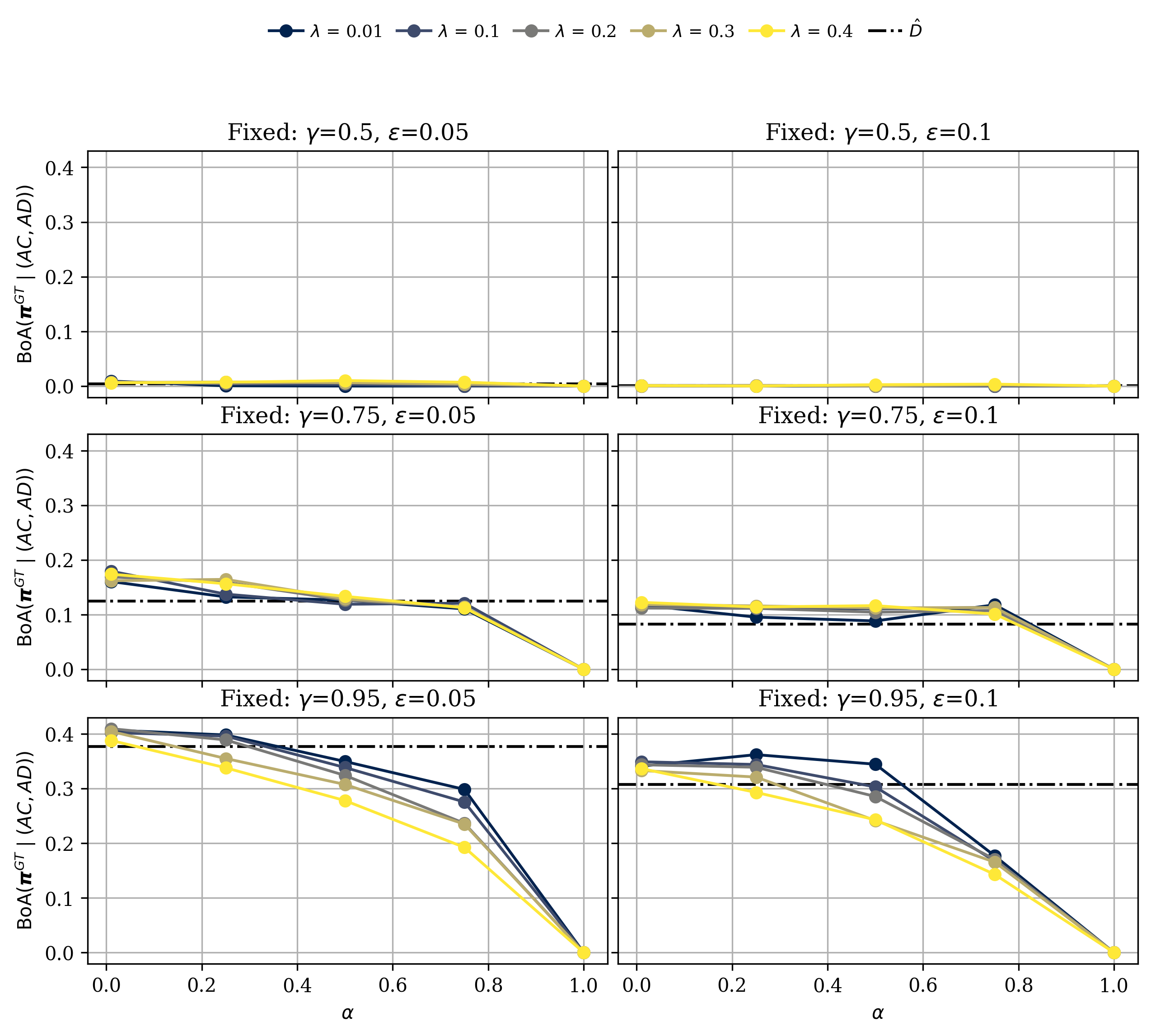}
        \caption{$R = 0.8,\ P = 0.5$}
        \label{fig:alpha_effect_ACAD_R08}
    \end{subfigure}
    \caption{Empirical $\text{BoA}(\boldsymbol{\pi}^{\text{GT}}\mid(AC,AD))$ (markers) versus $\hat{D}$ (dash-dotted line, \eqref{eq:Dhat}), as a function of $\alpha$ for several $\lambda$. Each subplot fixes $(\gamma,\varepsilon)$.}
    \label{fig:alpha_effect_ACAD}
\end{figure}

Figure~\ref{fig:alpha_effect_ACAD} compares the empirical $\text{BoA}(\boldsymbol{\pi}^{\text{GT}}\mid(AC,AD))$ with $\hat{D}$. Agreement is strongest for configuration~(1) at small $\alpha$ and $\lambda$ (mean absolute error $\approx 0.03$); configuration~(2) shows somewhat larger deviations ($\approx 0.10$), consistent with fixed-profile initialization that does not exactly match the uniform $\Delta q$ law used in the proof.

\subsection{Numerical evaluation of $\hat{D}$}
\label{app:acgt-numerics}

The production value of $\hat{D}$ is the closed form \eqref{eq:Dhat-closed}--\eqref{eq:Dhat-cases} (\texttt{calculate\_D\_hat\_closed}). As an independent check, the integral \eqref{eq:Dhat-branch} is also evaluated by \emph{Monte Carlo} (\texttt{calculate\_D\_hat\_mc}): draw $X_{1},X_{2}\sim\mathrm{Unif}[-L,0]$ independently and keep the A1D-first samples $\{X_{1}<X_{2}\}$ (this reproduces the uniform-on-triangle conditional law, whose $X_{2}$-marginal is $\propto(X_{2}+L)$, rather than the biased $X_{1}\sim\mathrm{Unif}[-L,X_{2}]$), draw $Y_{1}\sim\mathrm{Unif}[0,L]$ (player two defects in $BC$, so $\Delta q_{2}^{0}(BC)>0$), propagate to $(\hat{x},\hat{y})$, apply $g(\hat{x},\hat{y})$, and return $\hat{D}=\tfrac12$ times the sample mean. A nested \emph{quadrature} implementation (\texttt{calculate\_D\_hat\_quadrature}) provides a third, deterministic cross-check. All three agree to within sampling/discretisation error. Marginal densities $p(\hat{x})$ and $p(\hat{y})$ must not be multiplied: $\hat{x}$ and $\hat{y}$ are correlated through the shared stopping time $T_{A1D}(X_{2})$ (see discussion after \eqref{eq:pBCx}).

\end{document}